\documentclass[10pt,conference,letterpaper]{IEEEtran}                    
\usepackage{graphicx}
\usepackage{verbatim}
\usepackage{multicol}
\usepackage{cases}
\usepackage{mathptmx}
\usepackage{setspace}
\usepackage{cite}
\usepackage{url}
\usepackage{hyperref}
\usepackage{tfrupee}
\usepackage{soul}
\usepackage{array,etoolbox}
\usepackage{spreadtab}
\usepackage{multirow}
\usepackage{subcaption}

\preto\tabular{\setcounter{magicrownumbers}{0}}
\newcounter{magicrownumbers}

\IEEEoverridecommandlockouts

\def\hb{\hbox to 10.7 cm{}}
\newcommand{\fref}[1]{Fig.~\ref{#1}}

\newcommand{\cref}[1]{Chapter~\ref{#1}}

\newcommand{\junk}[1]{}

\begin{document}

\title{Estimating Time to Clear Pendency of Cases in High Courts in India using Linear Regression\thanks{Note: As of January 01, 2019, there are 25 high courts in India, however, for consistency with the earlier data, we have merged the High Court of Andhra Pradesh and the High Court of Telangana as Telangana and Andhra High Court.}}


%
\author{\IEEEauthorblockN{Kshitiz Verma\IEEEauthorrefmark{1},
Anshu Musaddi\IEEEauthorrefmark{2},
Ansh Mittal\IEEEauthorrefmark{2},
Anshul Jain\IEEEauthorrefmark{2}}
\IEEEauthorblockA{kshitiz.verma@jklu.edu.in, \{17ucs185, 17ucs028, 17ucs029\}@lnmiit.ac.in}
\IEEEauthorblockA{\IEEEauthorrefmark{1} JK Lakshmipat University, Jaipur, India}
\IEEEauthorblockA{\IEEEauthorrefmark{2} The LNM Institute of Information Technology, Jaipur, India}
}
\thispagestyle{plain}
\pagestyle{plain}

\maketitle
\begin{abstract}

Indian Judiciary is suffering from burden of millions of cases that are lying pending in its courts at all the levels. The High Court National Judicial Data Grid (HC-NJDG) indexes all the cases pending in the high courts and publishes the data publicly. In this paper, we analyze the data that we have collected from the HC-NJDG portal on 229 randomly chosen days between August 31, 2017 to March 22, 2020, including these dates. Thus, the data analyzed in the paper spans a period of more than two and a half years. We show that:
\begin{itemize}
\item the pending cases in most of the high courts is increasing linearly with time.
\item the case load on judges in various high courts is very unevenly distributed, making judges of some high courts hundred times more loaded than others.
\item for some high courts it may take even a hundred years to clear the pendency cases if proper measures are not taken.
\end{itemize}
We also suggest some policy changes that may help clear the pendency within a fixed time of either five or fifteen years. Finally, we find that the rate of institution of cases in high courts can be easily handled by the current sanctioned strength. However, extra judges are needed only to clear earlier backlogs.
 

\end{abstract}

\section{Introduction}
\label{sec:intro}

\par Justice delayed is justice denied. Delays in a justice system don't just violate the fundamental, constitutional and human rights of a victim but they also have an adverse effects on the rights of the accused as well as those who are convicted. Recently, Supreme Court of India acquitted two persons, accused of a gang rape after 28 years of the incident \cite{twenty_eight}. In an another case, a litigant had to fight for more than four decades to retrieve possession\cite{unfortunate}. Such unfortunate examples are not exceptions in India. They contribute towards the disrepute of the judicial system, lower the faith of the people in judiciary and also impact the economic growth. The judicial delays may potentially translate to a loss of 0.48\% of the national Domestic Gross Product (GDP) of India\cite{daksh_report}. Hence, the government has all reasons to implement policies that help removing the pendency. We define \emph{pendency} and \emph{delay} as they are defined in The Report No. 245 of Law Commission of India\cite{lci_report245}:

\begin{enumerate}
\item \emph{Pendency: All cases instituted but not disposed of, regardless of when the case was instituted.}
\item \emph{Delay: A case that has been in the court/judicial system for longer than the normal time that it should take for a case of that type to be disposed of.}
\item Nullifying pendency: To make pendency equal to zero.
\end{enumerate}

Note that in our definition of pendency, even if a case was instituted just one day ago, it still counts as pendency. Hence, it is a very strict way of computing judicial backlogs. We will use this meaning of pendency throughout the paper. 

\par In the year 2009, it was accepted by the Prime Minister of India that the pendency in Indian courts is the maximum in the world \cite{max_india}. India does very poorly compared to the other major democracies of the world\cite{dushyant_doj}. As of October, 2020 more than 34 million cases were pending in all the levels in Indian courts \cite{NJDG}. A substantial percentage of them have been around for more than ten, twenty or even thirty years \cite{NJDG}\cite{94kcases}. Indian Judiciary has started digitization of courts, on the initiatives of Supreme Court of India, through e-Courts project to take help of the Information and Communications Technologies (ICT) in the judicial sector through its e-Committee \cite{sci_actionplan2005}\cite{sci_actionplan2014}. A great leap in providing free access to the judicial information was provided by the implementation of the National Judicial Data Grid (NJDG) \cite{NJDG}, an important outcome of the e-Courts projects, has data of more than 34 million cases pending in Indian courts at all the levels. The data is open and is available publicly.

Many attempts have been made to estimate the number of years required to clear the pendency of the cases. According to Justice V. V. Rao, a prediction made in 2010 \cite{justicerao320}, it would take 320 years to clear the pendency. A report of Delhi High Court stated that it would require 466 years to clear all the pendency in Delhi Courts \cite{delhi_466}. On the other hand, a commitment to clear all the pending cases in five years was made by the then Union Law Minister in 2011 \cite{thing_past} and later by the then CJI Justice H.L. Dattu in 2015\cite{fiveYearsJDattu}. Yet another commitment was made to make average disposal of cases as three years rather than then 15 years \cite{reforms_three}. Law Commission of India published a report discussing delays and pendency in July 2014\cite{lci_report245}. However, they primarily addressed the issue in lower judiciary. Moreover, the statistics and the situation has changed drastically in last six years as even though the number of judges have increased, the pendency has not decreased. National Judicial Data Grid for High Courts (HC-NJDG) \cite{njdg_hc}, provides data on pending cases in High Courts and was visioned to be a game changer\cite{njdg_gamechg}. 


\begin{figure*}[t!]
\begin{subfigure}{.5\textwidth}
  \centering
    \includegraphics[width=8.6cm]{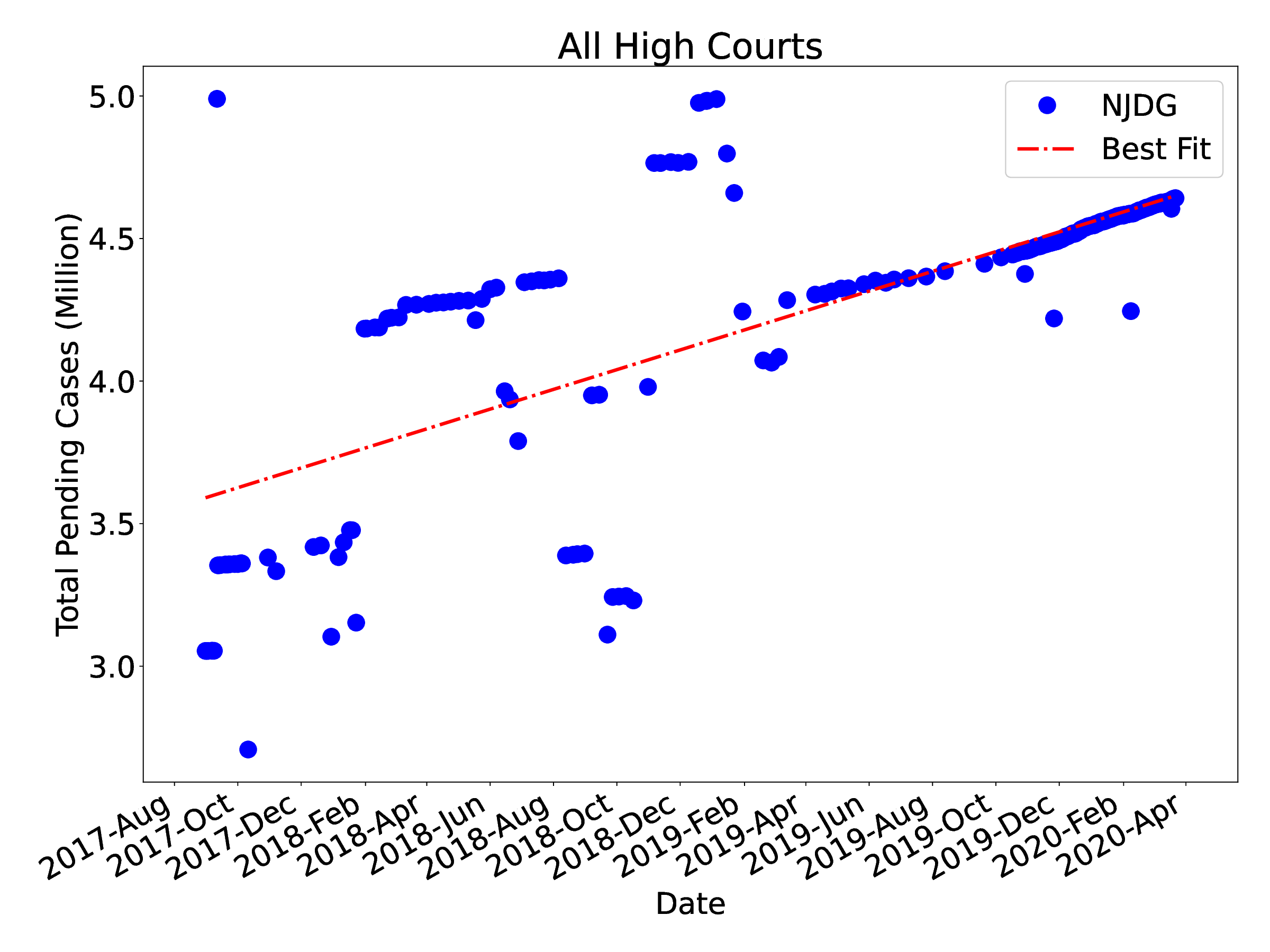}
    \caption{Evolution of pending cases in all the High Courts of India.} 
    \label{fig:hcpendency}
\end{subfigure}
\begin{subfigure}{.5\textwidth}
  \centering
    \includegraphics[width=8.6cm]{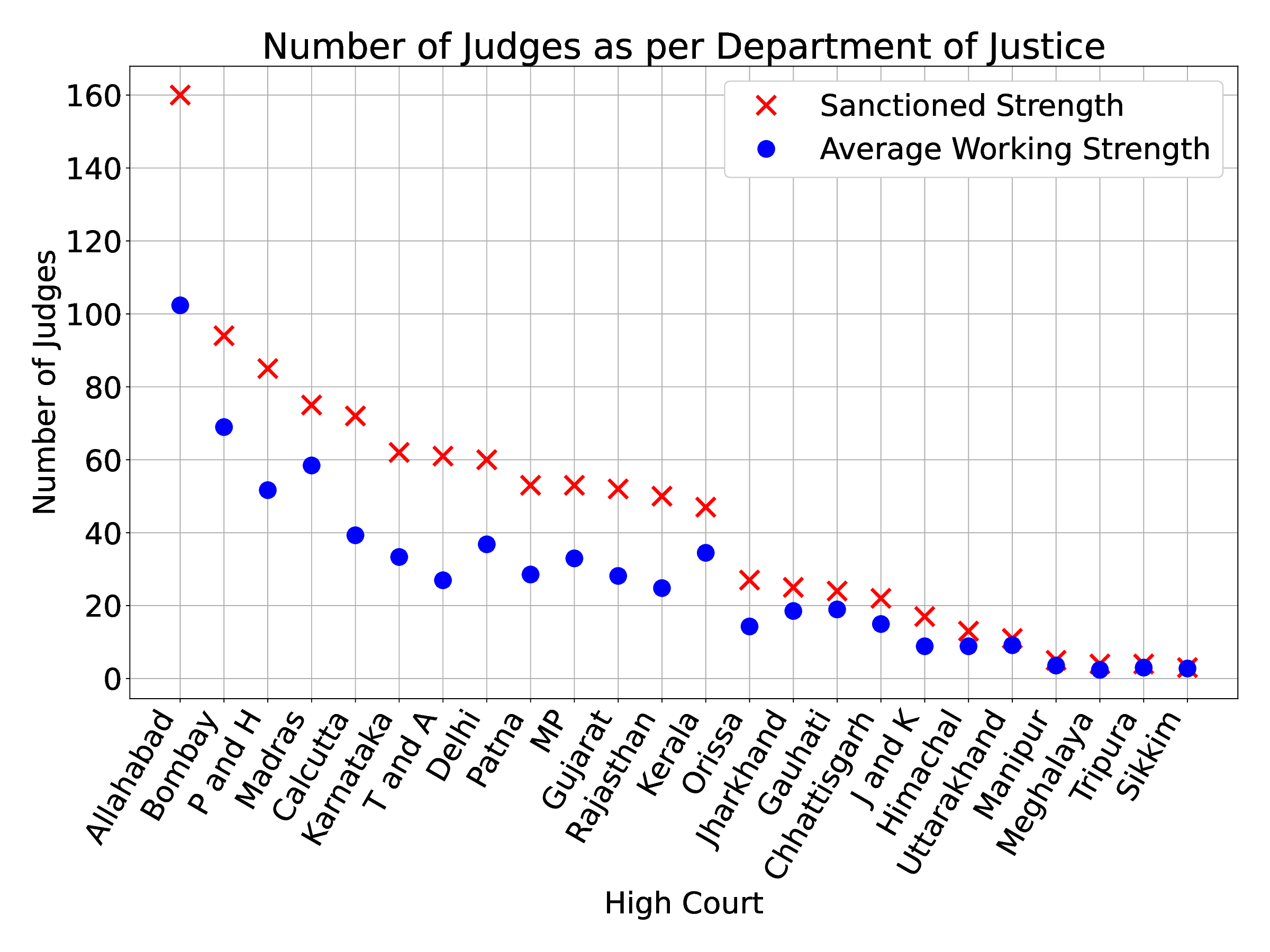}
    \caption{Sanctioned and average working strength of High Courts.} 
    \label{fig:sanctioned_working}
    \end{subfigure}
\caption{The number of cases in the high courts are rising pretty fast and the number of judges in the high courts is really low. }
\label{fig:fig}
\end{figure*}

In this paper, we exclusively study data from HC-NJDG and report results related to High Courts only. As of October 2020, there were more than 5 million cases pending in the High Courts of India\cite{HC_5M}. We have chosen to study pendency in high courts only because they are generally well equipped with resources to implement the suggested measures effectively. Also, since the number of high courts is only 25 (the 25$^{th}$ started on January 1, 2019), analysis and results are easy to interpret. 

\par The launching of NJDG also caused a sudden jump in India's rank in the Ease of Doing Business Report by the World Bank. It led to an improvement of 30 ranks in the year 2017 compared to the year 2016 \cite{WB_lauds}. However, it is not yet time to celebrate its success. This effort can be called successful only after all the stakeholders (e.g. the judges, court staff, advocates and litigants) find it useful in reducing their burden and NJDG helps improving the efficiency of the whole judicial system. It is still very far from that stage. Note that the success of portals like NJDG depends immensely on individual high courts updating their data regularly on the portal. We have seen positive signs and are hopeful that NJDG is indeed making progress towards an improvement that can make difference. 

This study, to the best of our knowledge, is the first one to analyze NJDG data over such a long period of time. Most of the existing studies consider the data from only one day on NJDG. Hence, we differ from the other studies in this basic premises itself. We also also try to answer the question, "How reliable is single day analysis of NJDG?". We answer this as negative, i.e., the NJDG data collected on just one day may not be taken as reliable for any reasonable analysis. There have been instances when the data on NJDG was very erroneous and such days are not rare. For example, a recent article on the pendency statistics of Bombay High Court claimed that 4.64 lakh cases are pending \cite{bombay_hc464}. Our finding is that throughout the data collection period, the Bombay High Court has updated the number of pending cases only once.

\subsection{Results in the paper}

Our work revolves around answering one central question. \emph{How long will it take to clear all the pending cases in the High Courts?} We summarize some of our results below:

\paragraph{Increasing pendency} Pending cases are increasing in most of the high courts rather than decreasing (Fig. \ref{fig:hcpendency}, \ref{fig:np_hc1}, \ref{fig:np_hc2_p2}). Hence, in the absence of clear policies and their implementation, pendency can never be cleared in most of the high courts.
\paragraph{Load on judges in different high courts} There is a huge difference in terms of average load of cases on judges of different high courts (\fref{fig:pend_date_r}). For example, a judge of Rajasthan High Court has almost 250 times more load than a judge in Sikkim High Court. 
\paragraph{Years to clear the pendency} Assuming a linear increase in the number of cases as well as in the number of judges so that the high courts operate at their sanctioned strength of judges, then for most of the high courts pendency can be nullified (Fig. \ref{fig:years_combat_real}, \ref{fig:years_combat_national_avg_lowest}). However, the number of years required to do so may vary significantly. 
\paragraph{Proposed policies for clearing the pendency} \fref{fig:req_individual} and \fref{fig:increase_judges} may help the government of India to take informed decisions on the number of judges that need to be increased. We find that the current sanctioned strength of the high courts is adequate to take care of the newly instituted cases. They are, however, insufficient for clearing the previous backlog of cases. Hence, the government may enact legislation to create some temporary positions in high courts, just to clear the pendency, without changing the actual current sanctioned strength.

\subsection{Organization of the paper}
\par The rest of the paper is organized as follows: Section \ref{sec:rw} encompasses the scope of the work and the related studies. Section \ref{sec:njdg} discusses data collection and explains the graphs used in the paper. Section \ref{sec:pending_hc} elaborates on pending cases in the high courts. Section \ref{sec:combat} is home to the most important result of the paper in which we estimate the time required to nullify the pendency in the high courts. Section \ref{sec:policy} focuses on forming policies to ensure that the pendency decreases. Section \ref{sec:conclusion} concludes the paper.


\section{Scope and Related Work}
\label{sec:rw}

Before we proceed any further to discuss the technical findings of the paper, we first understand the scope and the limitations of our work. There have been many news reports, articles and studies on pendency in Indian courts \cite{arrears}  \cite{over_22lakh} \cite{jkpile} \cite{verma2018courts}. In this paper, instead of studying the pending cases in all the courts of India, we decided to limit ourselves to only the high courts. This limits the number of court complexes in our study to just 39 without decreasing the complexity of the problem as the high courts are more clogged with cases than subordinate courts. Moreover, the improvements suggested in this paper are relatively easier to implement in high courts than in lower courts because of better infrastructure and budget available. Thus, we would concretely know where things can be improved. Hence, throughout this paper, we have concentrated on the pendency in high courts rather than the subordinate courts.

In our study, we have not taken input from any real person. No judge, advocate, litigant or court staff was interviewed. This may have both positive and negative impact. Intervention of court staff and those who are involved in updating NJDG may have provided more insights to interpret our results. On the other side of it, their views might have biased our results. So we decided to leave it for future because we wanted our assessment to be purely technical and statistical based only on the observations made from the data that we have collected from NJDG.

A study by Alok Prasanna Kumar has used number of the District and Magistrate courts, collected from the National Judicial Data Grid as of 18 March, 2016 \cite{comparative}. There are studies advocating to decrease the holidays available in judiciary\cite{funda_reason}. There are studies conducted by the Department of Justice as well\cite{dojeval}. While this study is very comprehensive, the results reported are different from ours and the parameters considered for evaluation are different as well. Various studies including \cite{vidhi1}, have conducted research on e-Court policies. The importance of data analysis of judicial data and the role of computer science is also suggested in \cite{judmess}. The Department of Justice also encourages research conducted on judicial reforms by means of funding \cite{doj_projects}. Another rich source of information on pending cases are the annual reports published by the Supreme Court of India \cite{SCAR}. 

The most relevant work that can be compared with our work is Law Commission of India report of July 2014 \cite{lci_report245}. The availability of data was a major concern for the authors of the report. They studied data on pendency at the end of years from 2002 to 2012. However, the primary focus of their analysis is the courts that are subordinate to the jurisdiction of high courts. Moreover, the pendency figures have more than doubled now compared to 2012. Hence an understanding of the rate of increase of cases is crucial in studying pendency. We make a clear distinction from the report by exclusively studing the high court data, collecting data over 229 days, and counting the contribution of each and every data point by using linear regression for the analysis.

Other relevant related work in this area is the Daksh report on the state of the Indian Judiciary \cite{daksh_report}. Their approach, however, is very different from ours. They have conducted a ground level research by surveying and obtaining the first hand experience of the litigants and other stake holders. Our work, on the other hand, relies completely on the data provided by the National Judicial Data Grid for High Courts (HC-NJDG). 

Hence, a lot of studies agree that the judicial throughput has to be increased. Either the number of vacations may be reduced or the number of judges may be increased. 

The issue has been of utmost importance to all the Chief Justices of India \cite{ranjan_plan}\cite{bobde_speedy}. Hence, a lot of research needs to be conducted in the area so as to help solve a crucial problem faced by Indian Judiciary.


\section{HC-NJDG Data Collection}
\label{sec:njdg}

High Court National Judicial Data Grid (HC-NJDG) was launched in July 2017 \cite{njdg_gamechg, njdg_hc}. We started collecting data from the portal on August 31, 2017. The last data used in this paper was collected on March 22, 2020. We have collected data on 229 randomly chosen dates, spanning a period of more than two and a half years.

To provide a glimpse of the data, some of the statistics, as collected on August 20, 2018, are provided in Table \ref{tab:njdg_stats}. The portal has more statistics available but we have chosen to present only the ones that are relevant for this paper. The data related to the number of pending cases in all the high courts in India is shown. The table also presents data on the number of monthly disposed and filed cases. 

\begin{table}[h]
\centering
\footnotesize
    \begin{tabular}{ | l | r | r | r | r |}
    \hline
    Title & Civil & Criminal & Writs & Total \\ \hline
    Pending Cases & 1506780 & 769754 & 1114448 & 3390982 \\ \hline
    Cased Filed (monthly) & 27663 & 42404 & 32009 & 102063 \\ \hline
    Cased Disposed (monthly) & 28080 & 47368 & 36548 & 111996 \\ \hline
    \end{tabular}
    \caption{Some of the statistics available on the HC-NJDG portal\cite{njdg_hc}.}
    \label{tab:njdg_stats}
\end{table}

Four high courts, namely, Allahabad High Court, Gauhati High Court, High Court of Jammu and Kashmir and High Court of Madhya Pradesh have joined HC-NJDG after we started collecting the data. Hence, they appear fewer number of times. However, majority of the high courts -- 20 to be precise -- had their presence on HC-NJDG when we started collecting the data, i.e., on August 31, 2017. 

The $25^{th}$ High Court for the state of Telangana was formed on January 01, 2019. For consistency with the previous data, we have continuted to consider Telangana and Andhra High Court as one in our analysis. For this reason, only 24 high courts appear in our study. Note that the last day for data collection for this paper was March 22, 2020, just before the nation-wide lockdown was announced in India due to Covid-19 pandemic. We hypothesize that data pre-lockdown and during the lockdown would be very different and hence not comparable for our work.


We have plotted many graphs in the paper. In order to maintain coherence and simplicity, and to have a reach to wider audience, we have restricted ourselves to only two kinds of graphs, as explained below. 
\paragraph{Temporal data graphs (Dates on horizontal axis)} The horizontal axis (also referred to as X-axis in the paper), consists of dates beginning August 31, 2017 to March 22, 2020 from left to right. The title of each graph is present on the top stating the name of the high court that plot corresponds to. If on some date data was not collected, then data is not shown against that date but the date is still present on the X-axis in the all cases. \fref{fig:hcpendency} is an example of this kind of graph.

\paragraph{Spatial data graphs (High Courts on horizontal axis)} In these graphs, the data from the high courts is plotted. The horizontal axis, or X-axis, in these graphs have 24 points, each representing one of the 24 high courts. Y-axis plots the value of the considered parameter. If a high court does not have a valid data for that parameter, its name still appears on the X-axis but have no value on the Y-axis. The title of the graph is present at the top. \fref{fig:sanctioned_working} is an example of this kind of graph.

\section{Pending Cases in High Courts}
\label{sec:pending_hc}

As discussed before, pendency in high courts is more than 10\% of the total pendency in India. Hence, concentrating on high courts capture the problem of pendency really well and offer much better quality data that can be studied to deduce meaningful conclusions.

\begin{figure*}[h!]
\includegraphics[width=4.45cm]{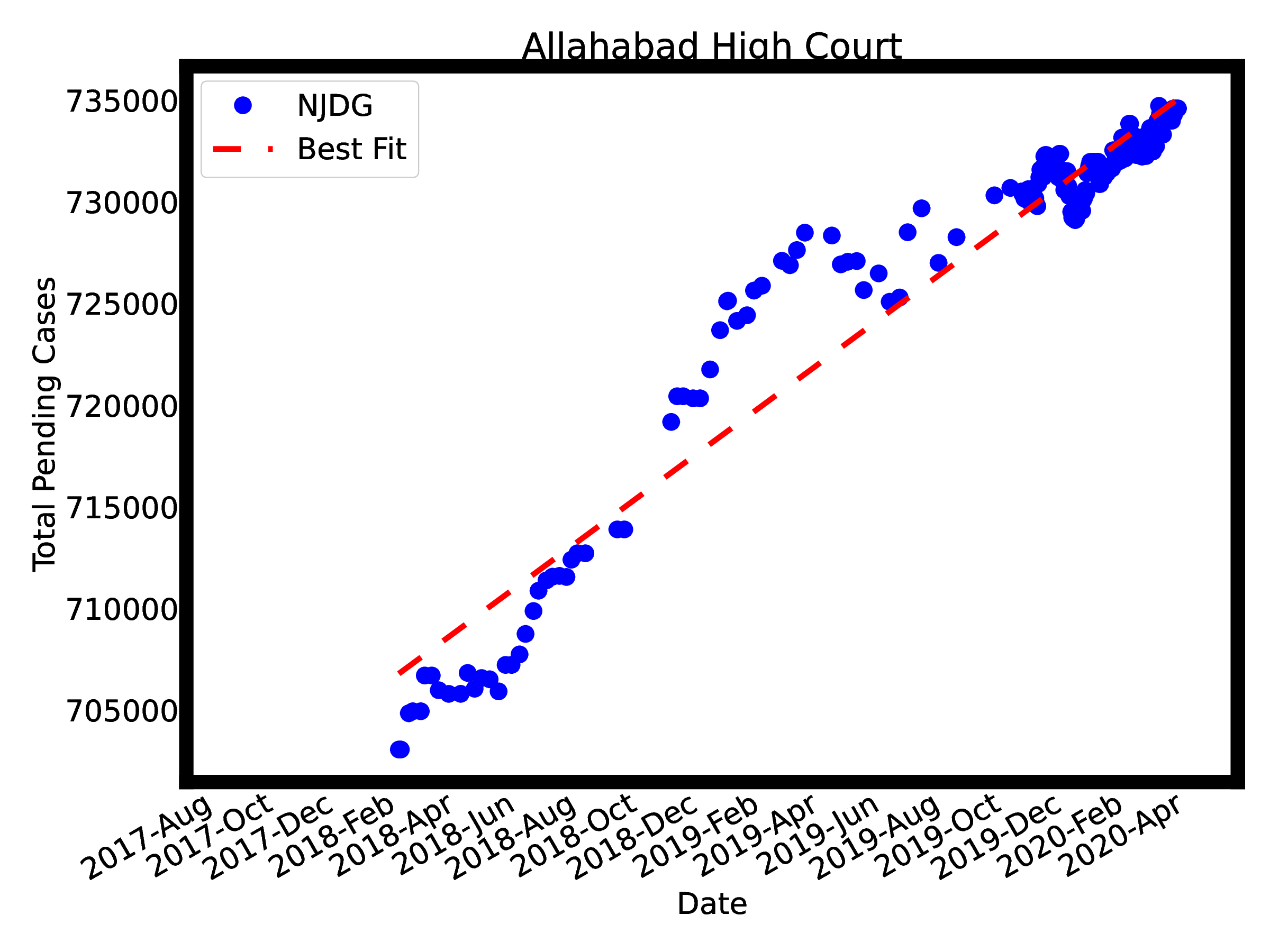}
\includegraphics[width=4.45cm]{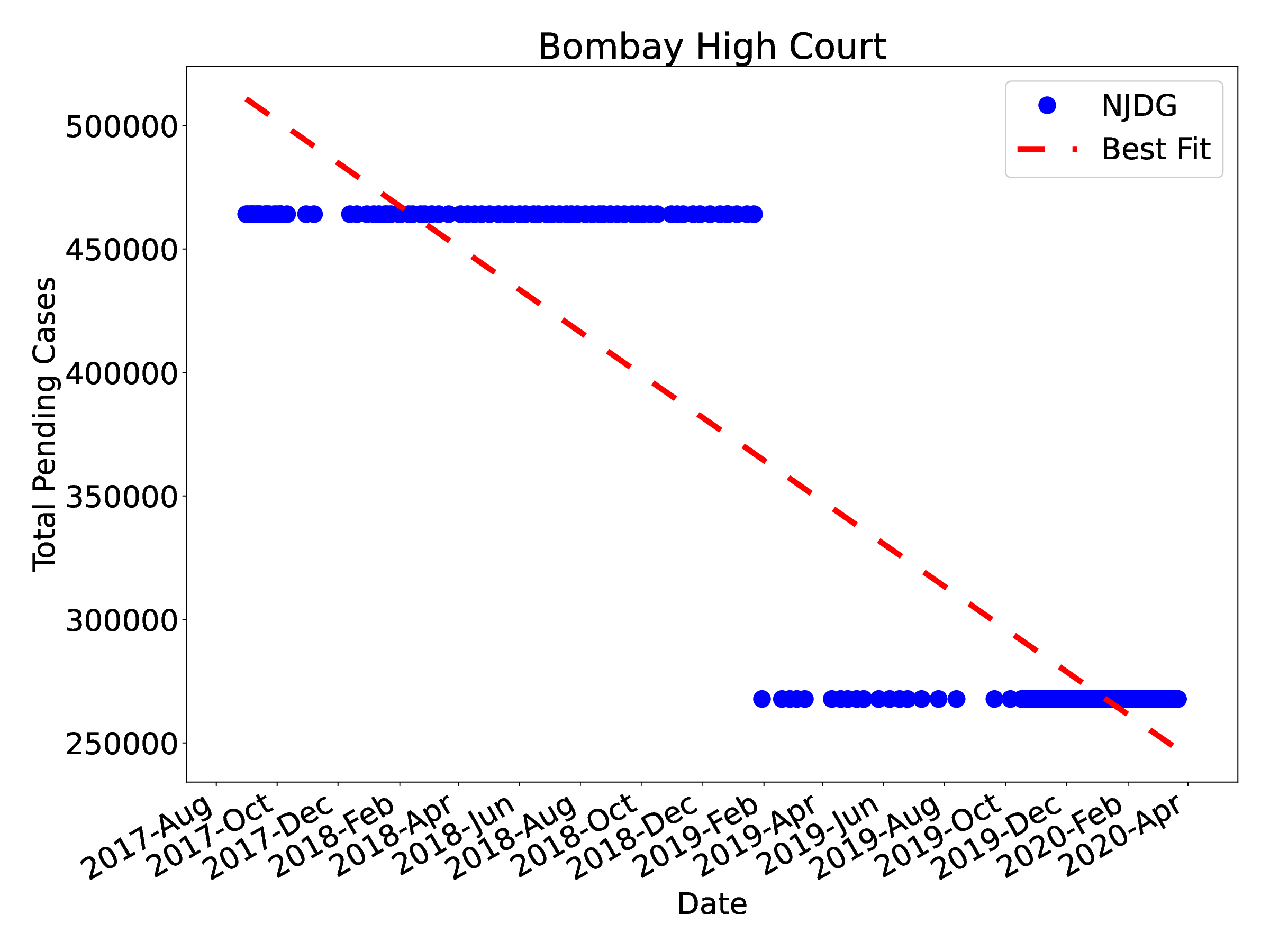}
\includegraphics[width=4.45cm]{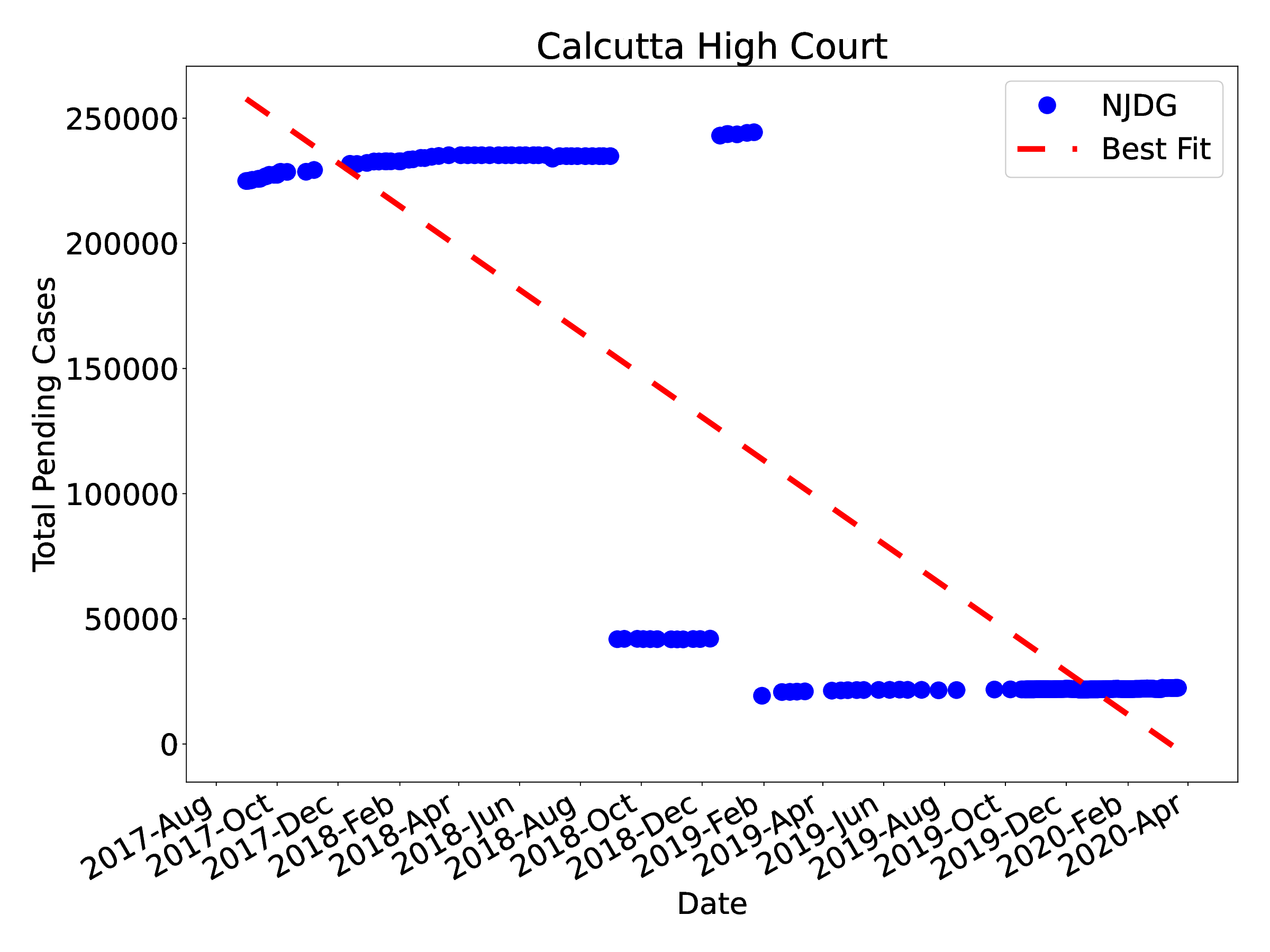}
\includegraphics[width=4.45cm]{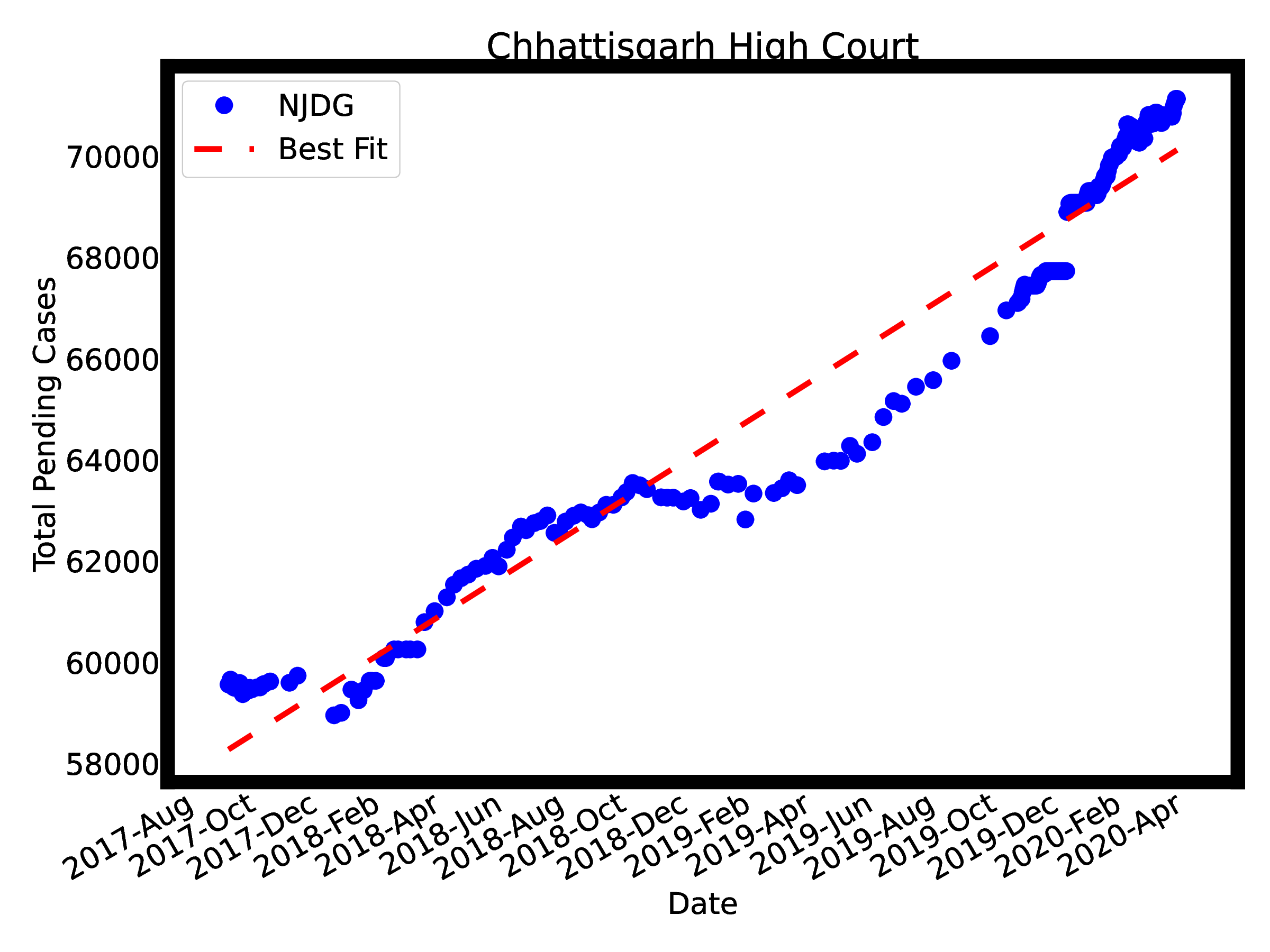}
\includegraphics[width=4.45cm]{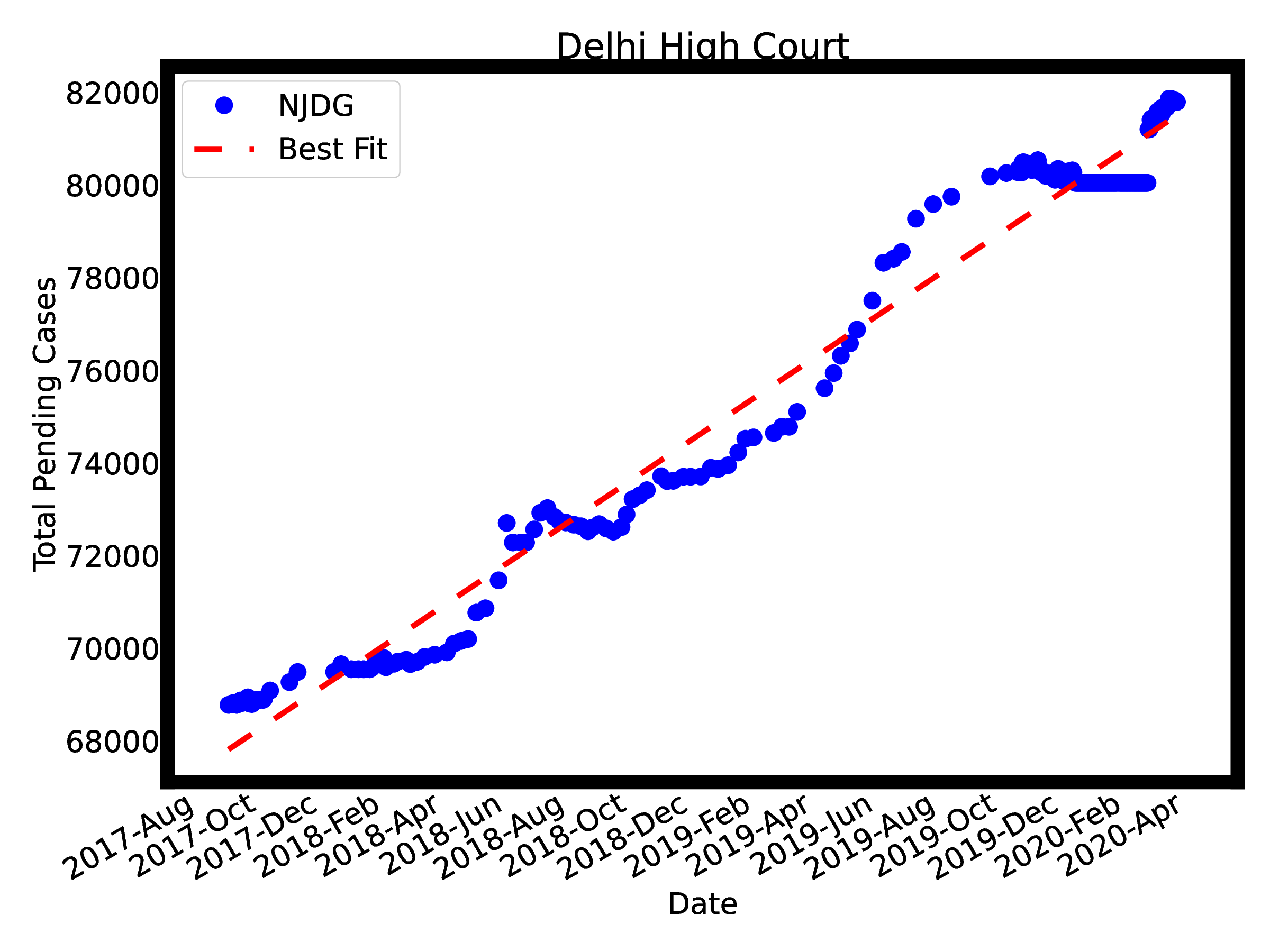}
\includegraphics[width=4.45cm]{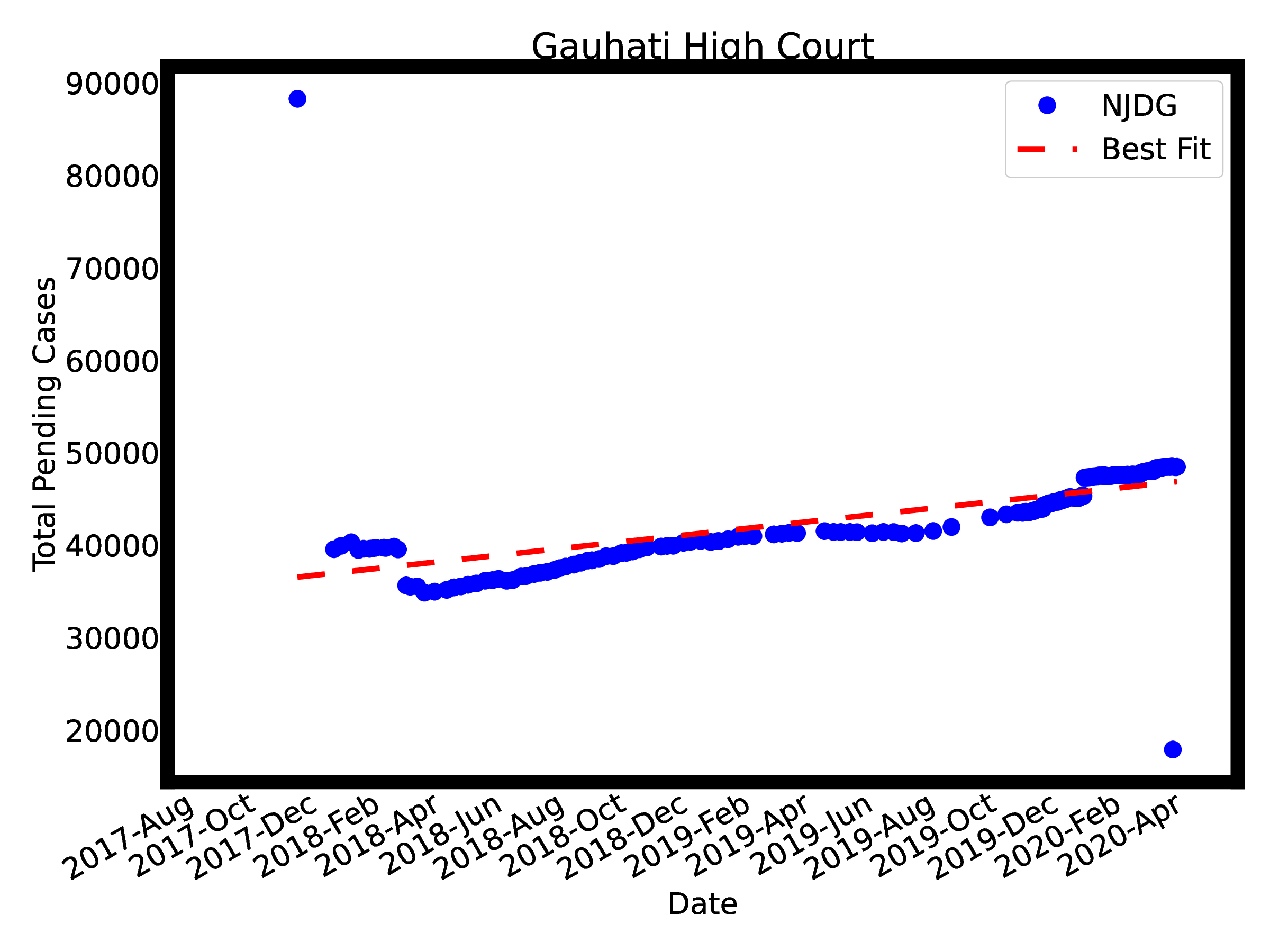}
\includegraphics[width=4.45cm]{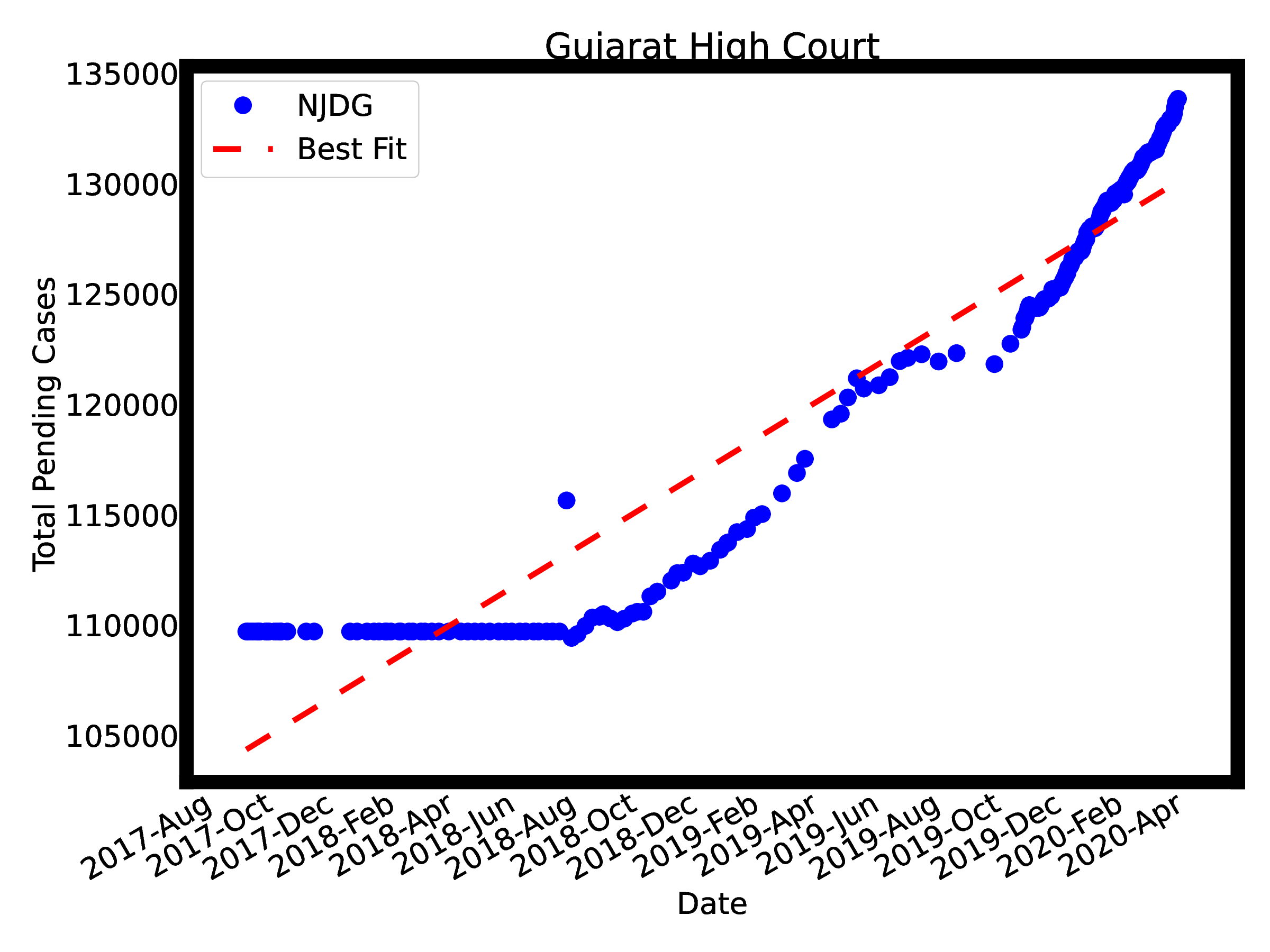}
\includegraphics[width=4.45cm]{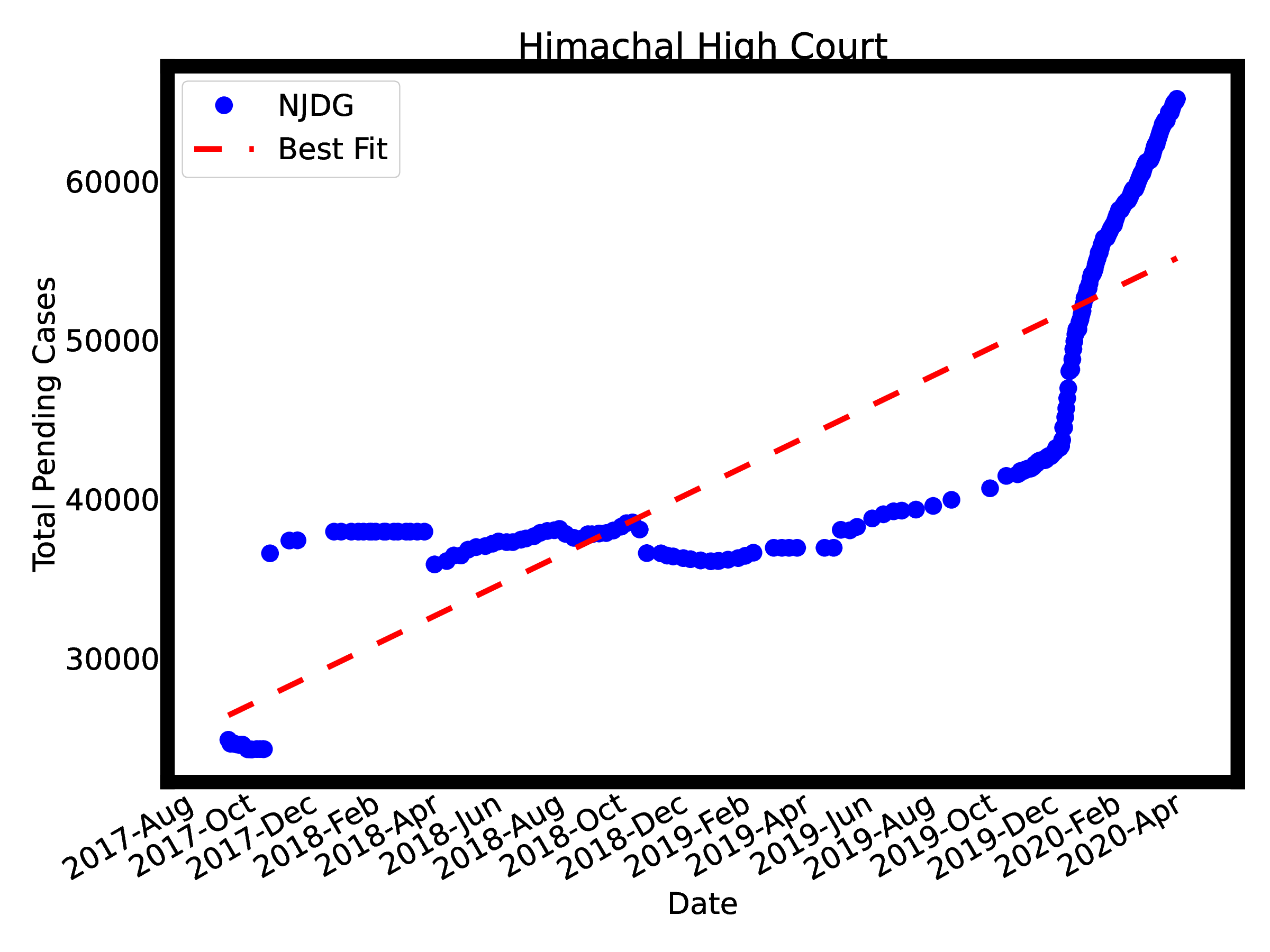}
\includegraphics[width=4.45cm]{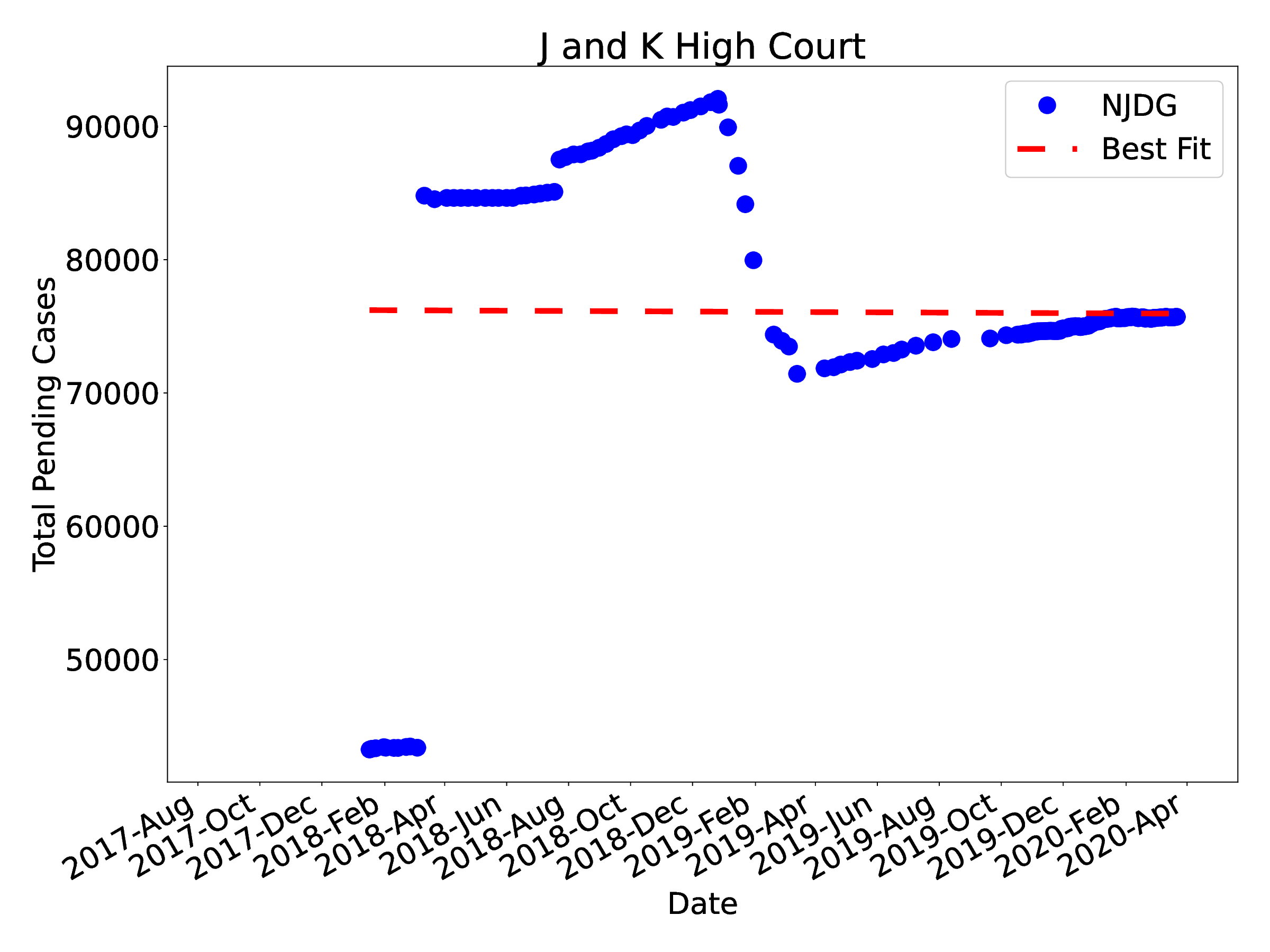}
\includegraphics[width=4.45cm]{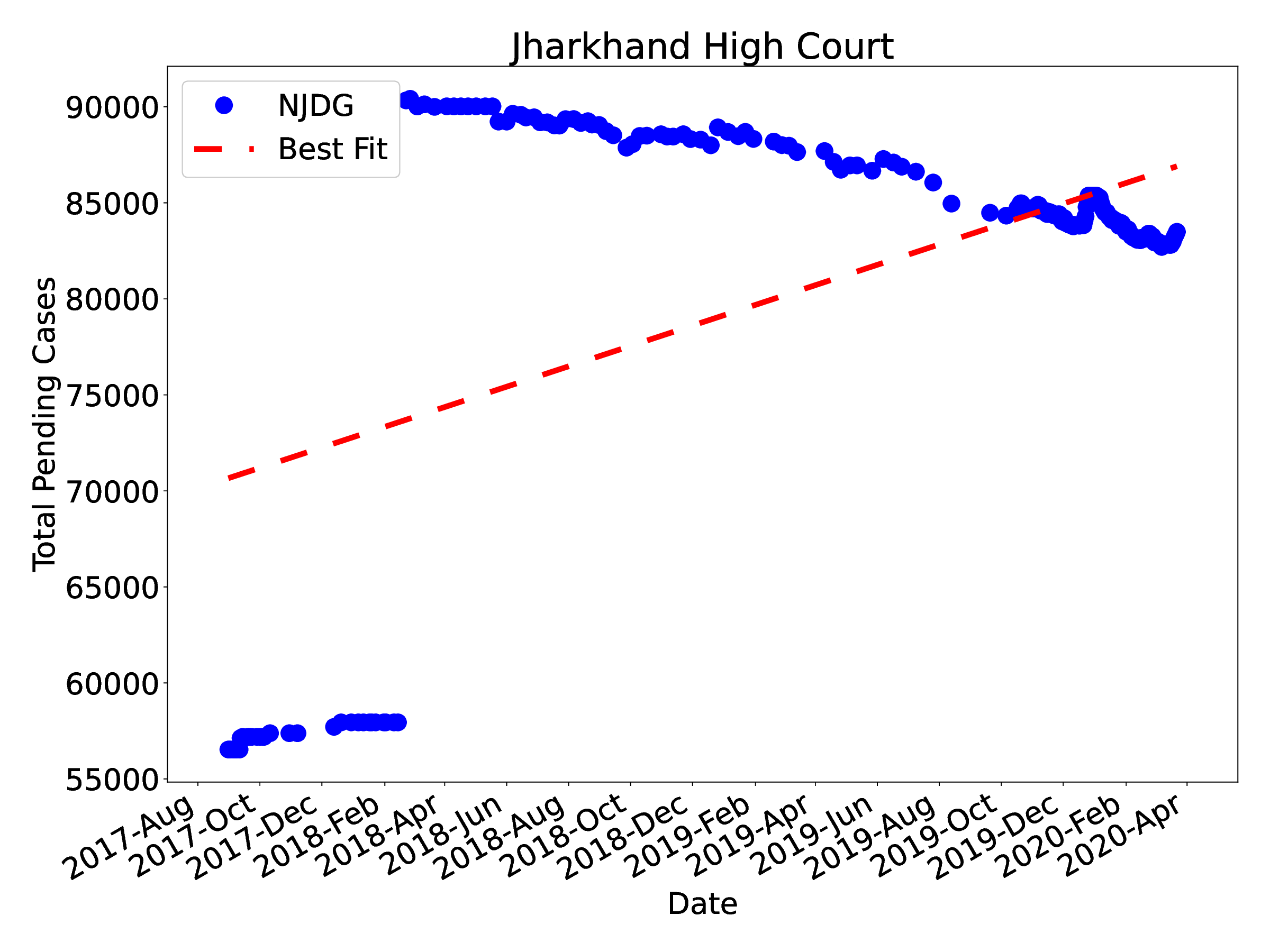}
\includegraphics[width=4.45cm]{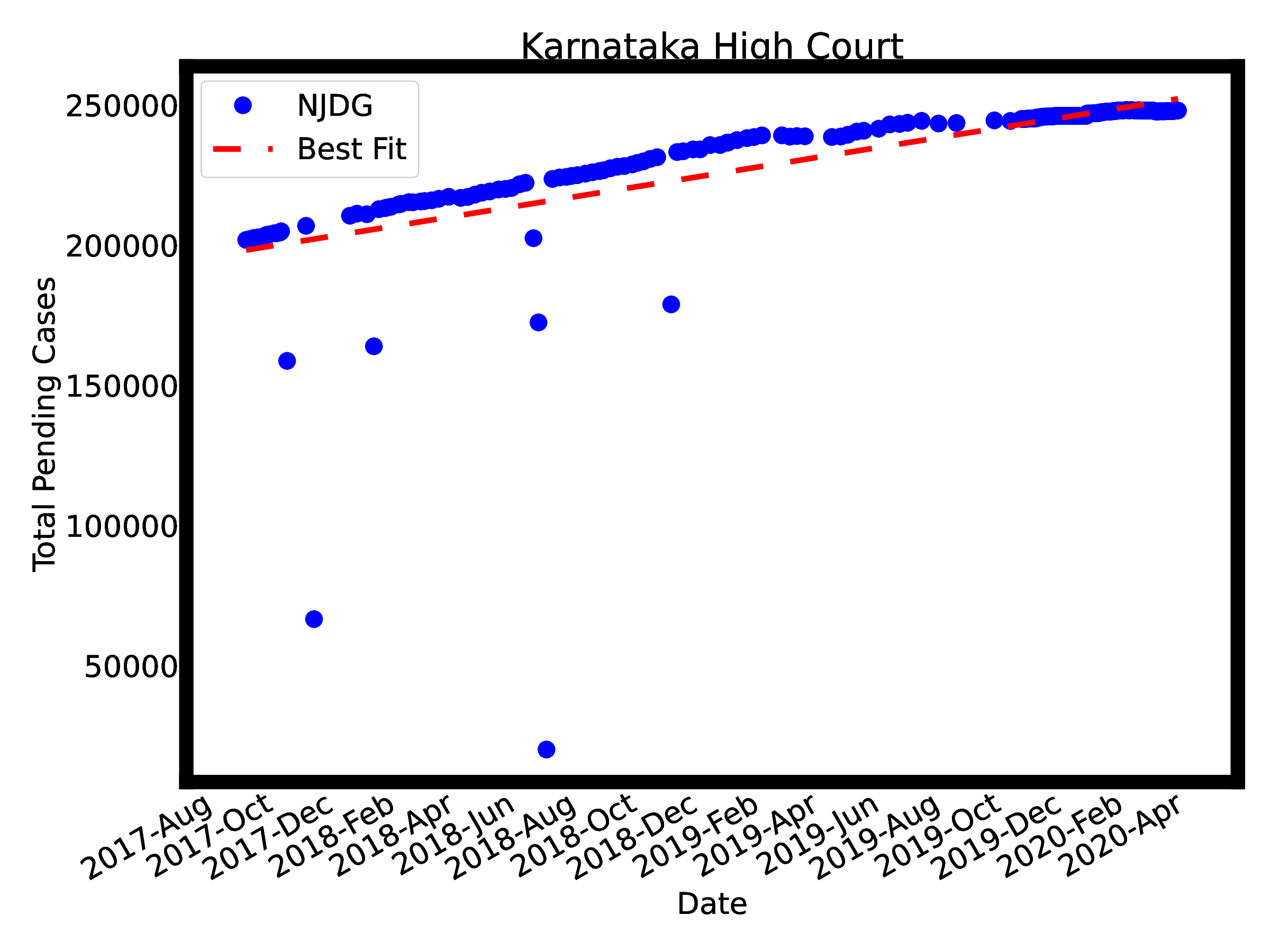}
\includegraphics[width=4.45cm]{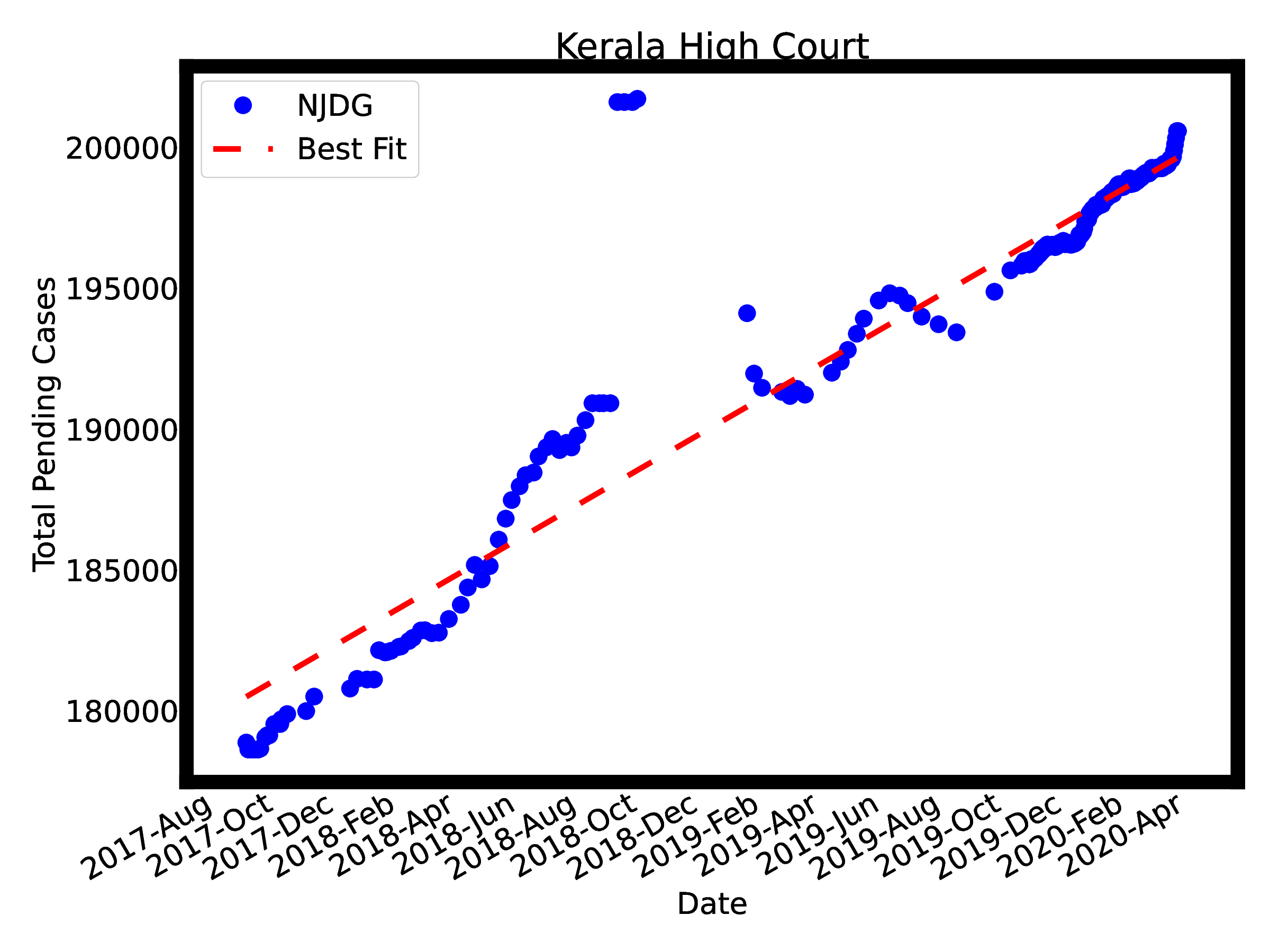}
\includegraphics[width=4.45cm]{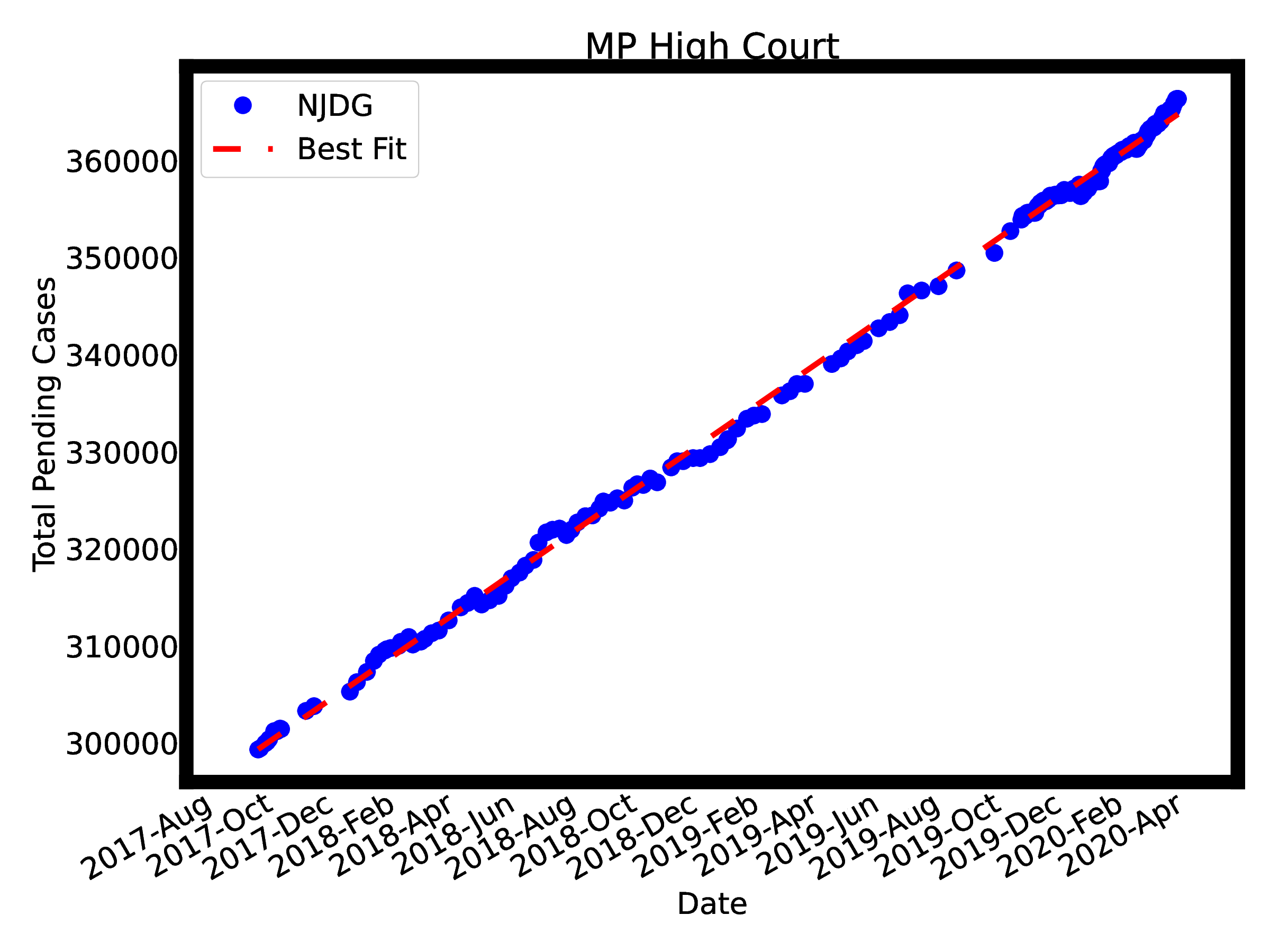}
\includegraphics[width=4.45cm]{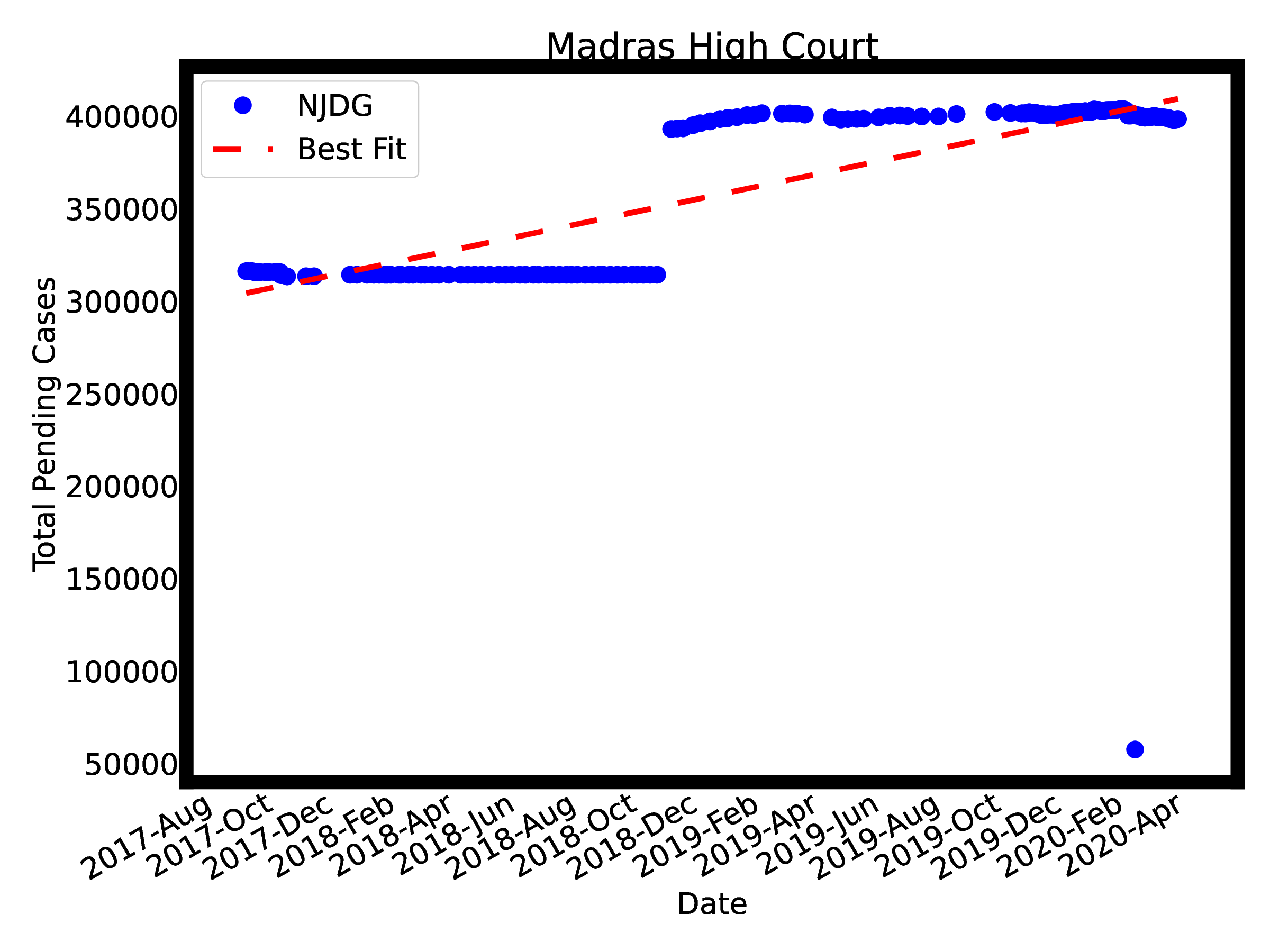}
\includegraphics[width=4.45cm]{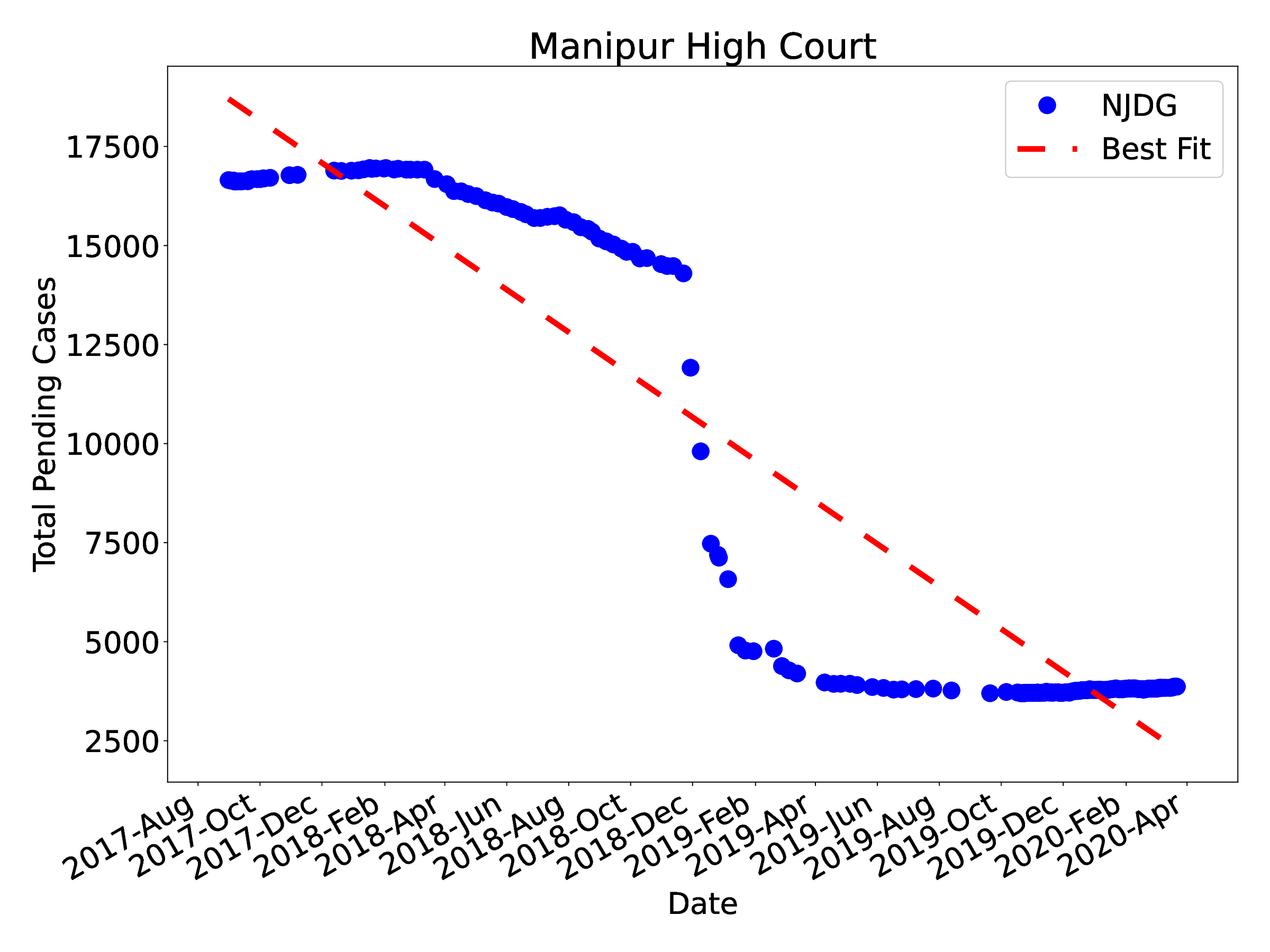}
\includegraphics[width=4.45cm]{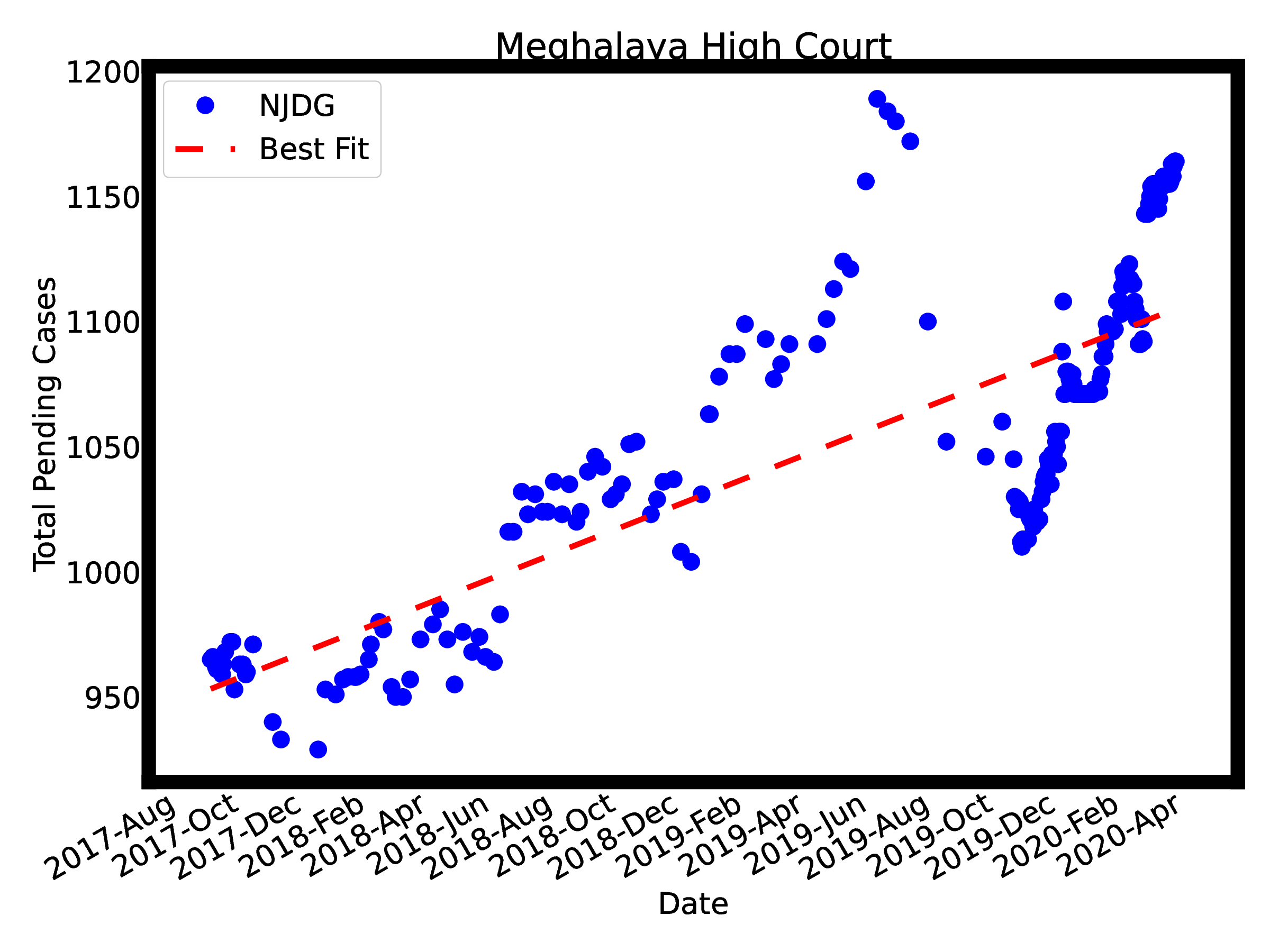}
\includegraphics[width=4.45cm]{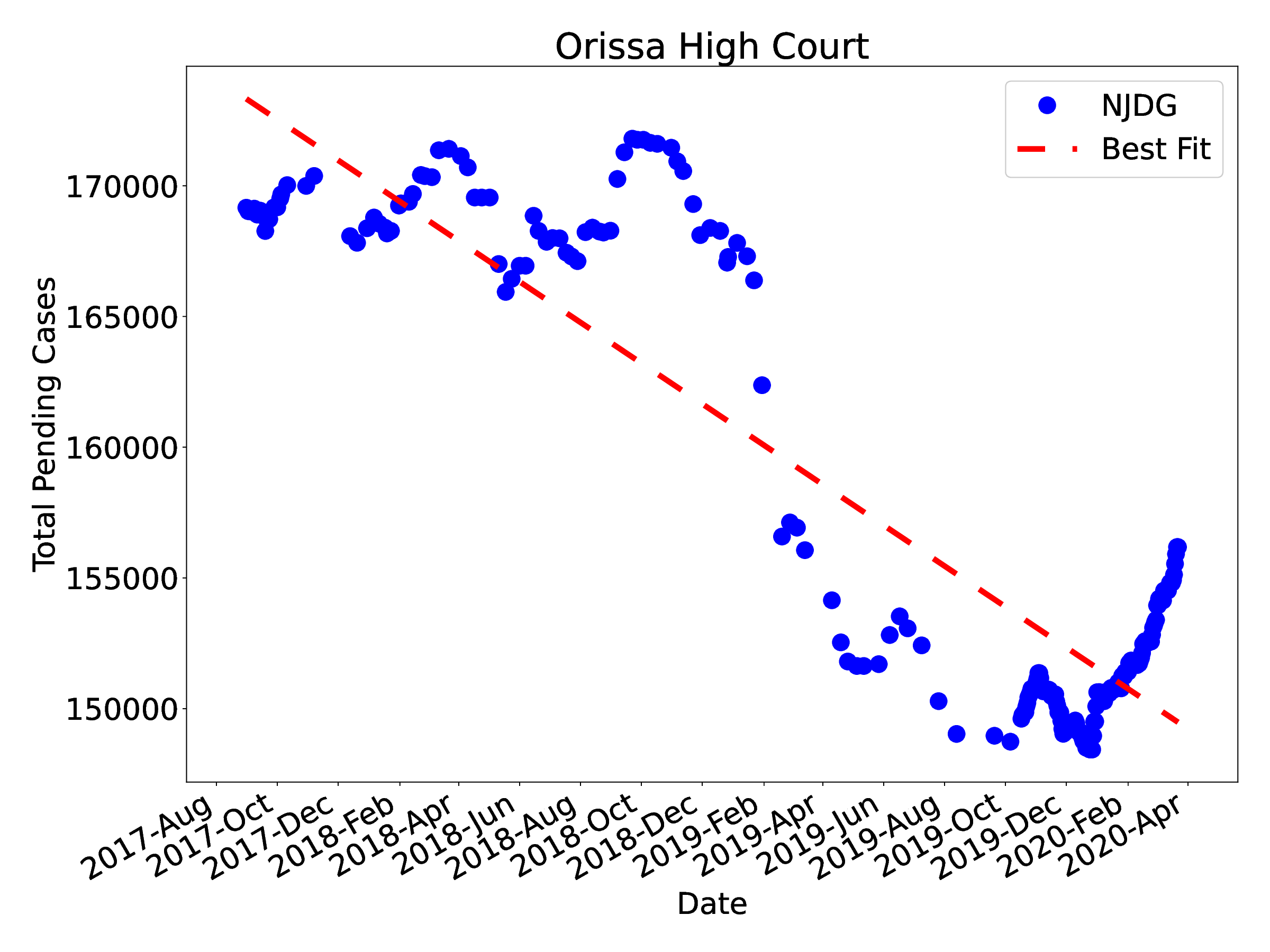}
\includegraphics[width=4.45cm]{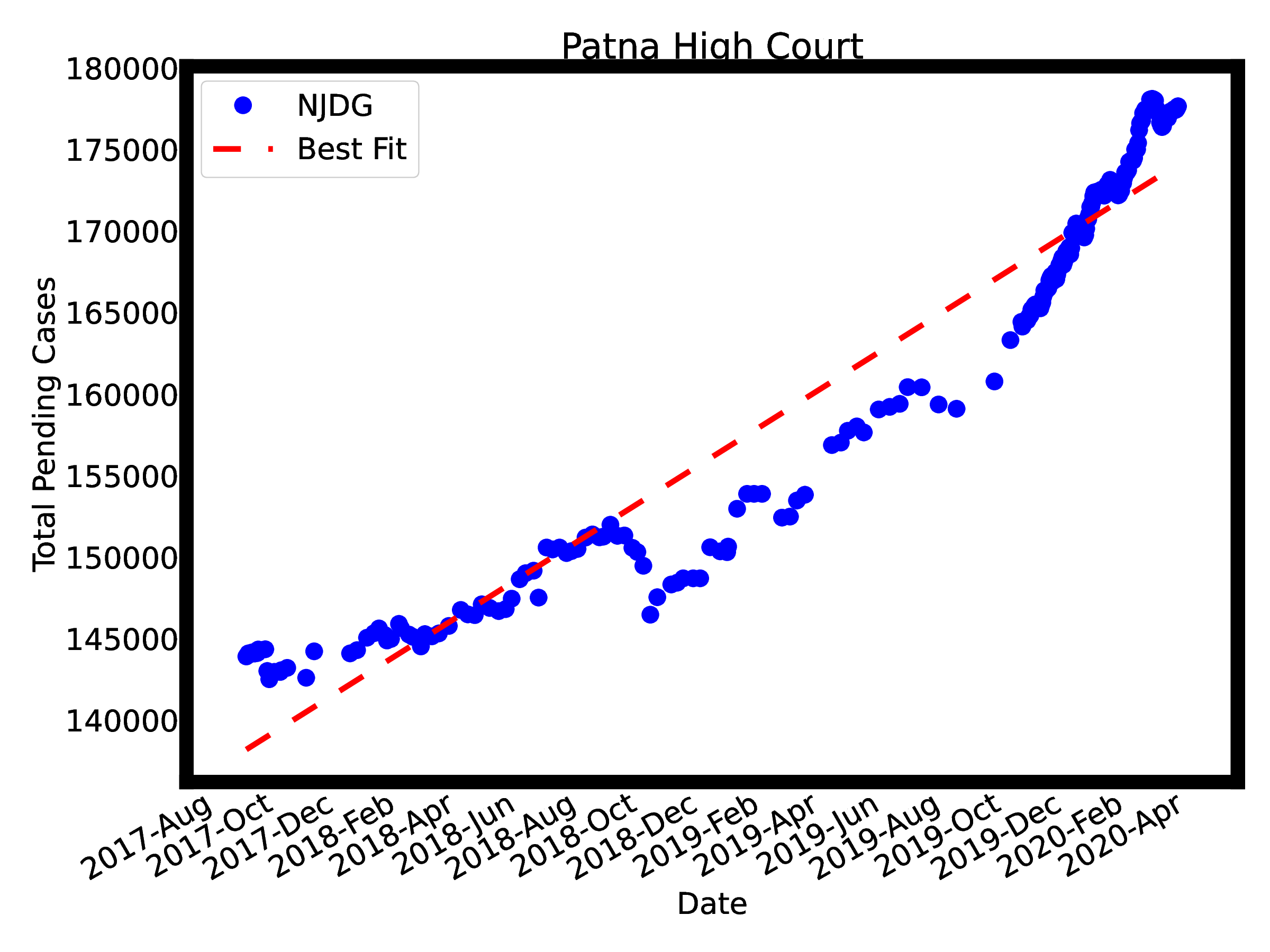}
\includegraphics[width=4.45cm]{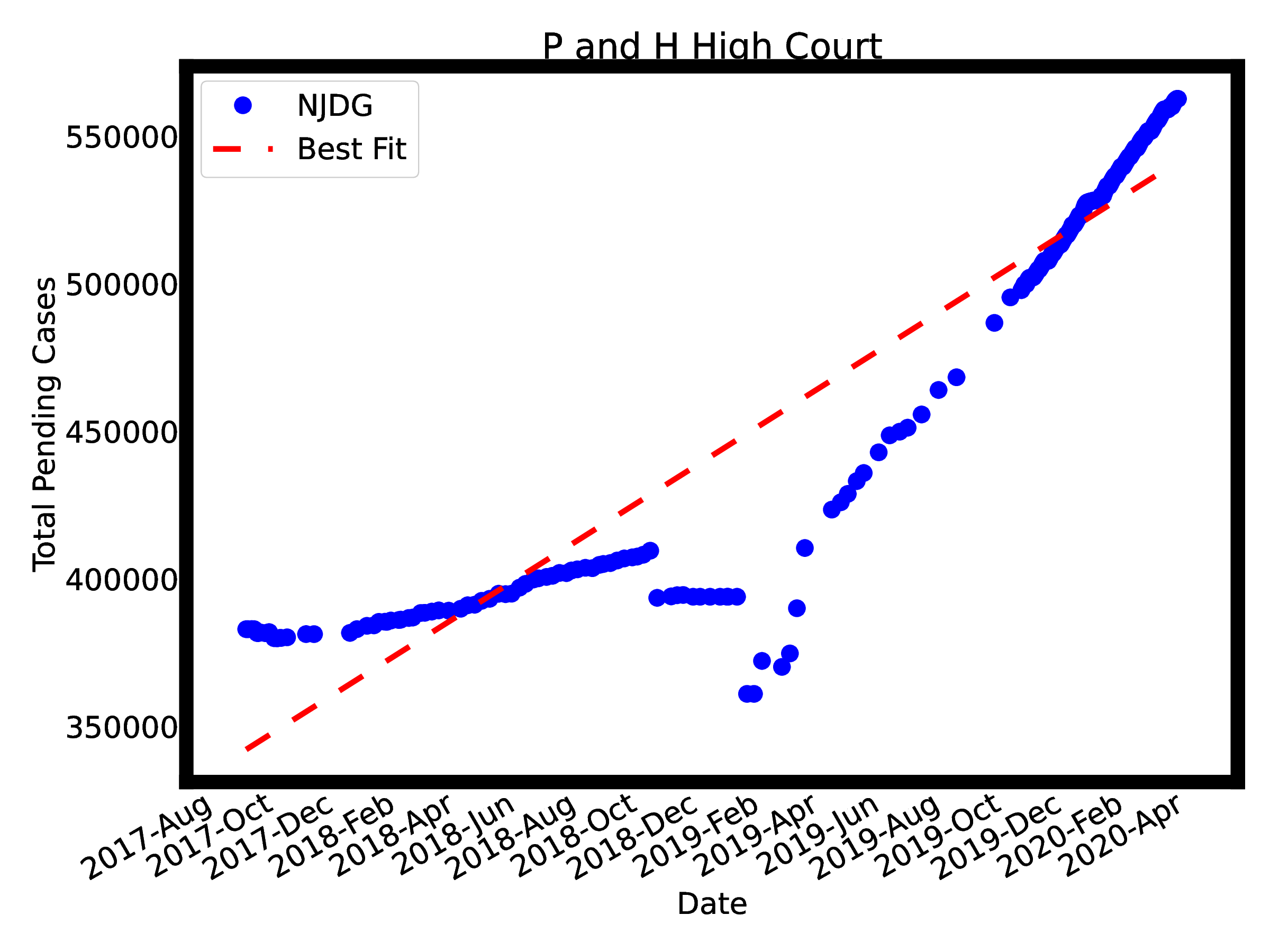}
\includegraphics[width=4.45cm]{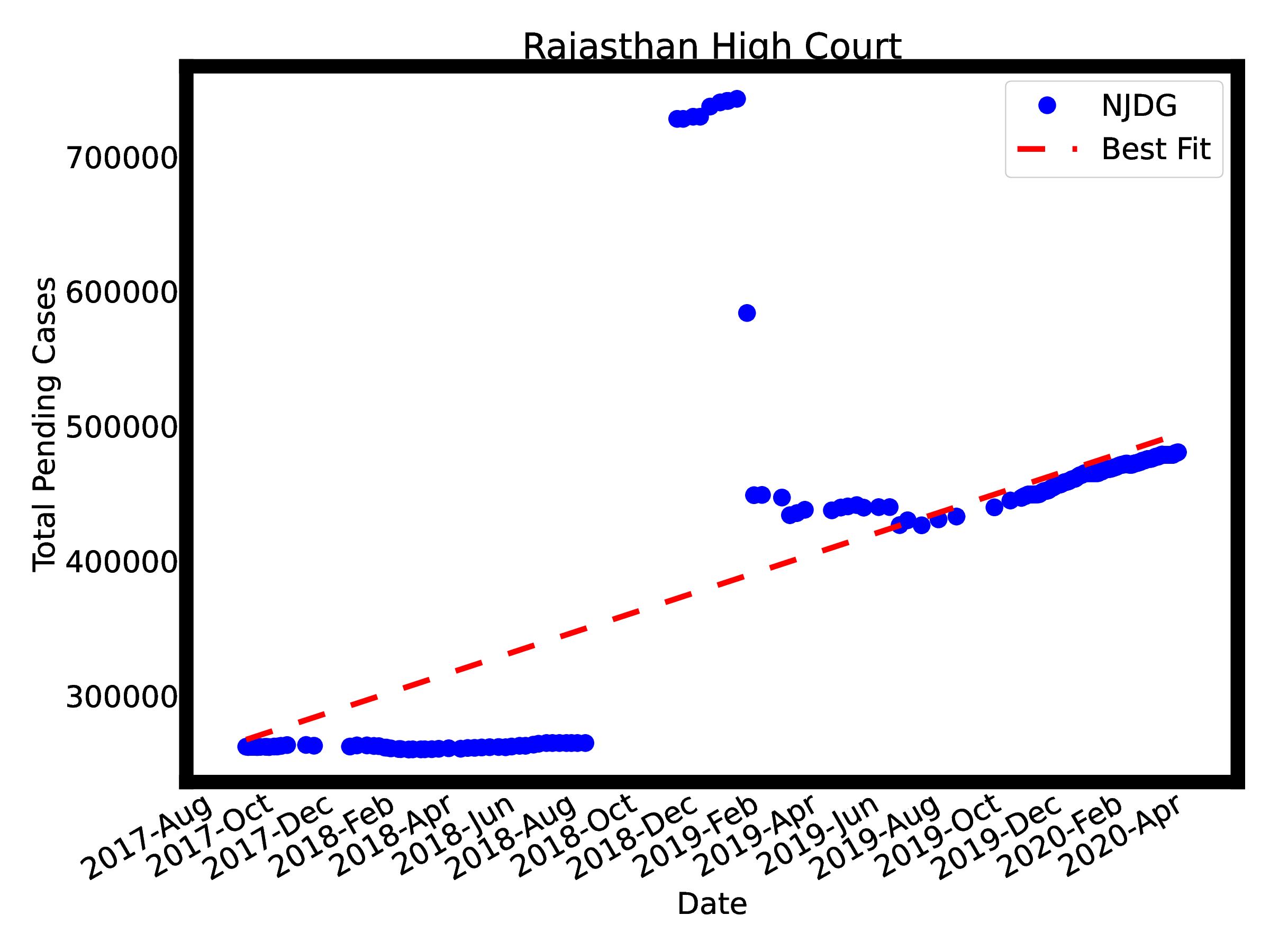}
\includegraphics[width=4.45cm]{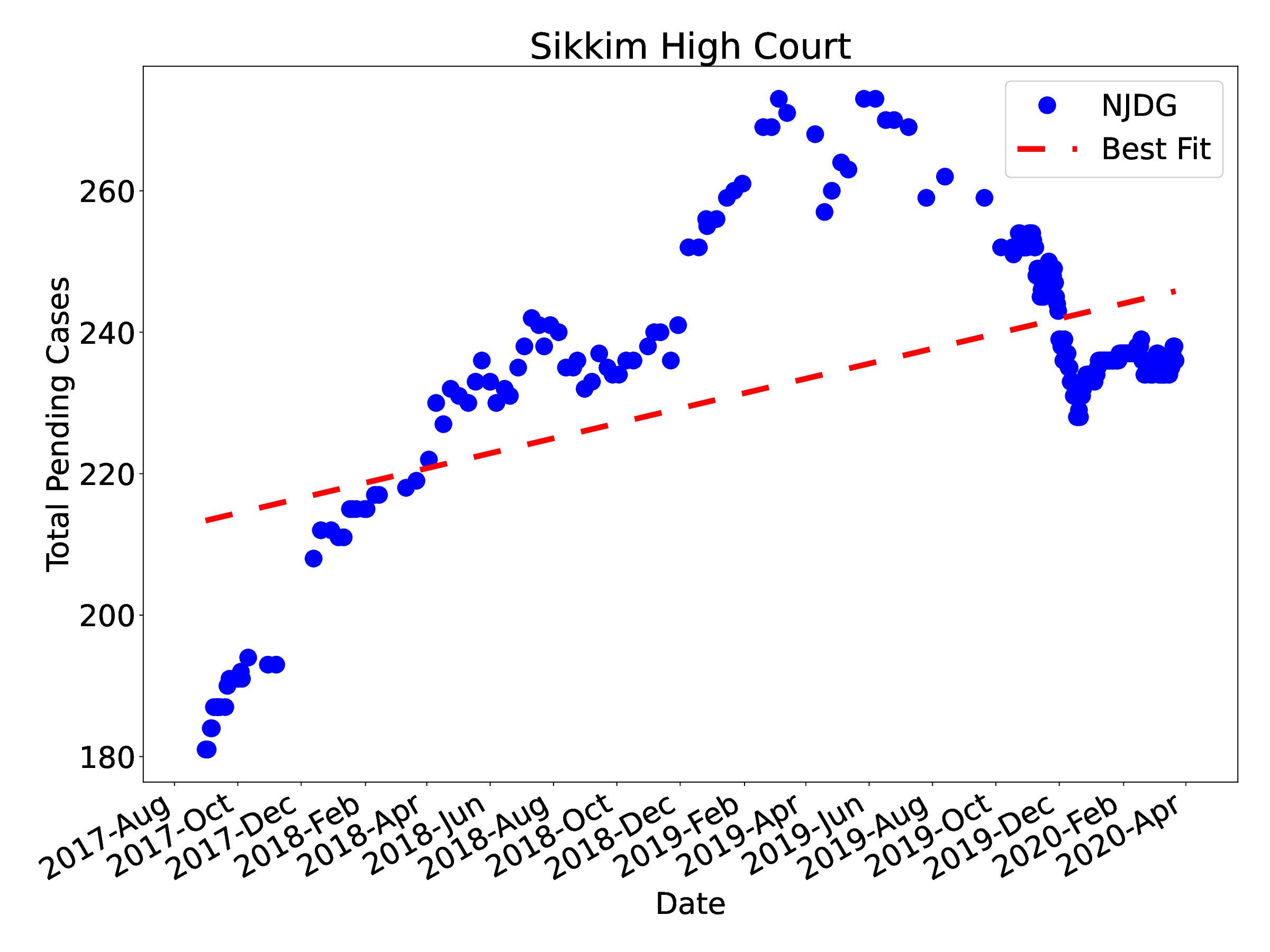}
\includegraphics[width=4.45cm]{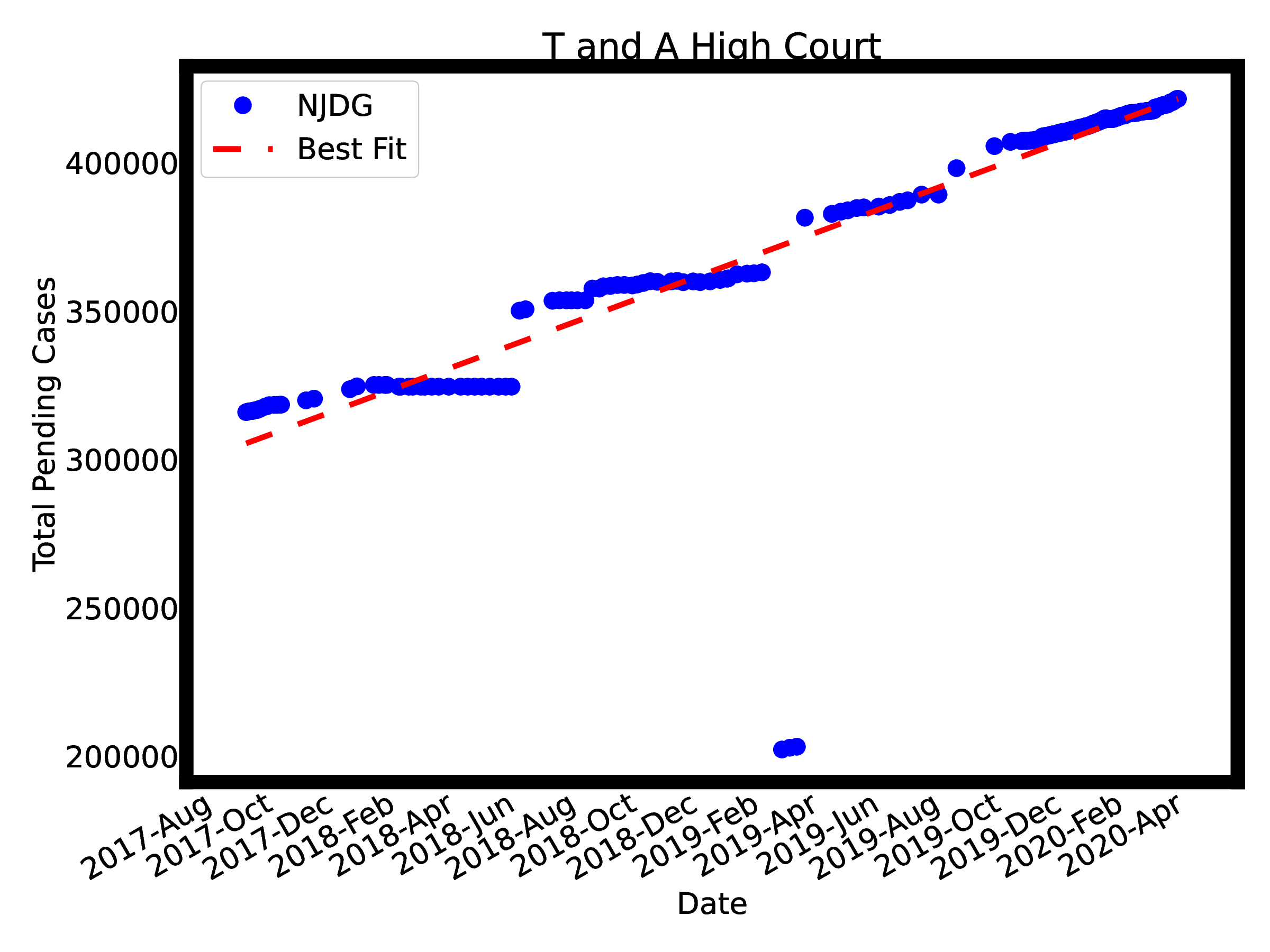}
\includegraphics[width=4.45cm]{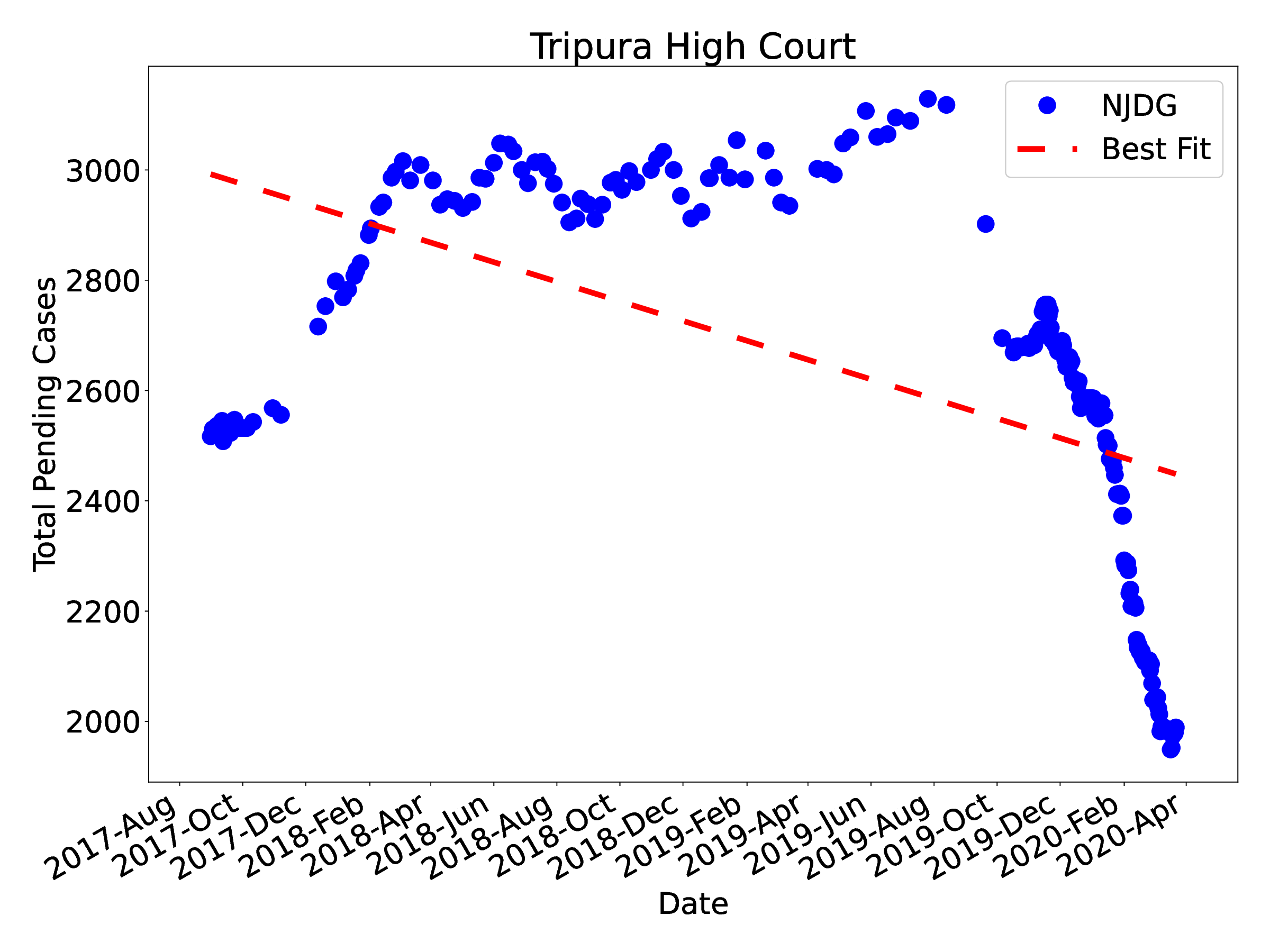}
\includegraphics[width=4.45cm]{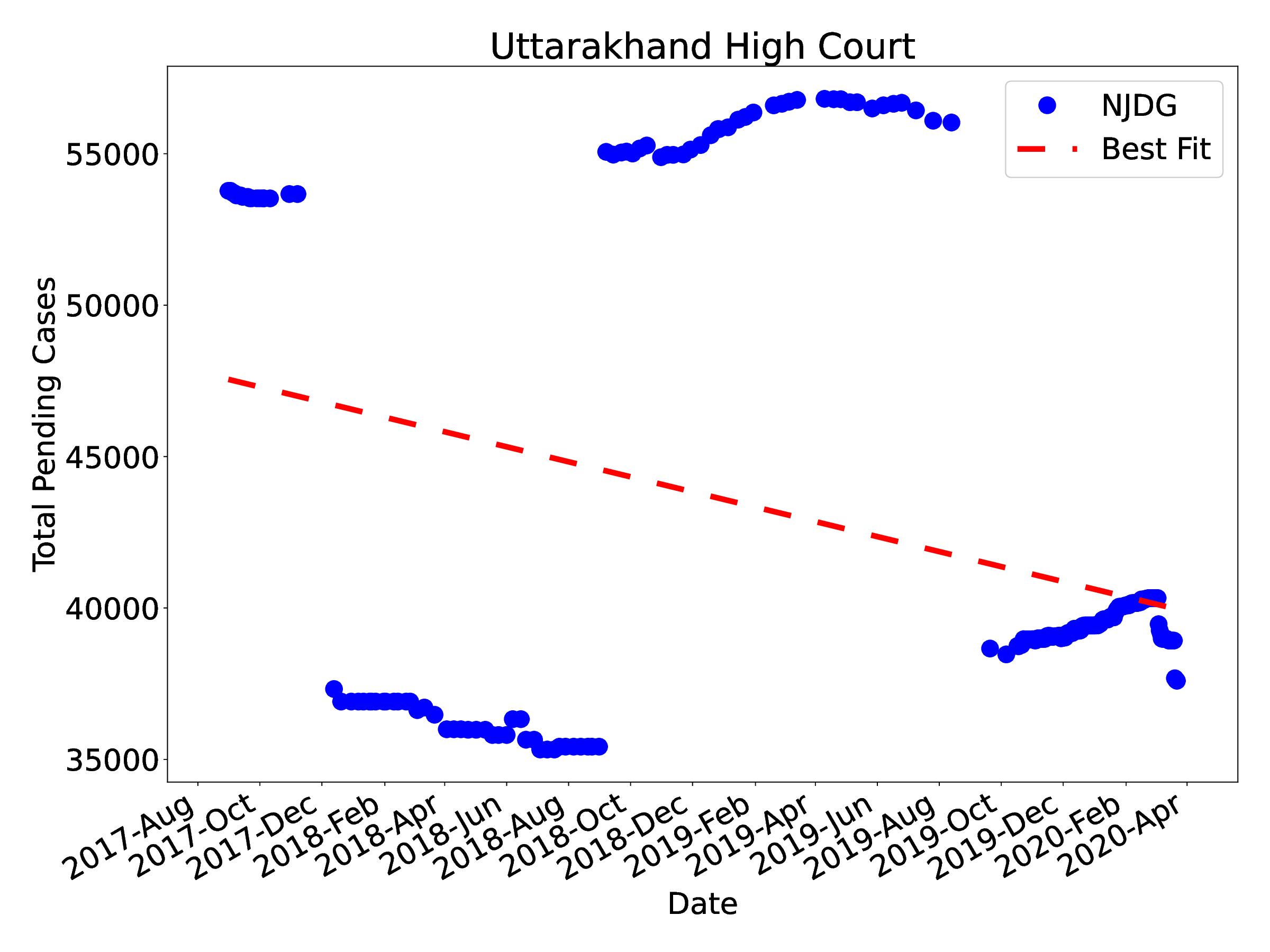}
\caption{Pending cases for individual high courts as plotted for the whole data collection duration, i.e., from August 31, 2017 to March 22, 2020. Dotted red line is the best fit straight line according to the least squares loss. For most of the high courts we use this figure to compute the rate of increase pendency, to be precise, such high courts are: Allahabad, Chhattisgarh, Delhi, Gauhati, Gujarat, Himachal Pradesh, Karnataka, Kerala, Madhya Pradesh, Madras, Meghalaya, Patna, Punjab and Haryana, Rajasthan and Telangana and Andhra. To emphasize, the frames of the plot of all these high courts have been made thicker compared to the frames of the high courts whose rate of increase of pendency is taken from other sources.}
\label{fig:np_hc1}
\end{figure*}

\begin{figure*}[h!]
\includegraphics[width=4.45cm]{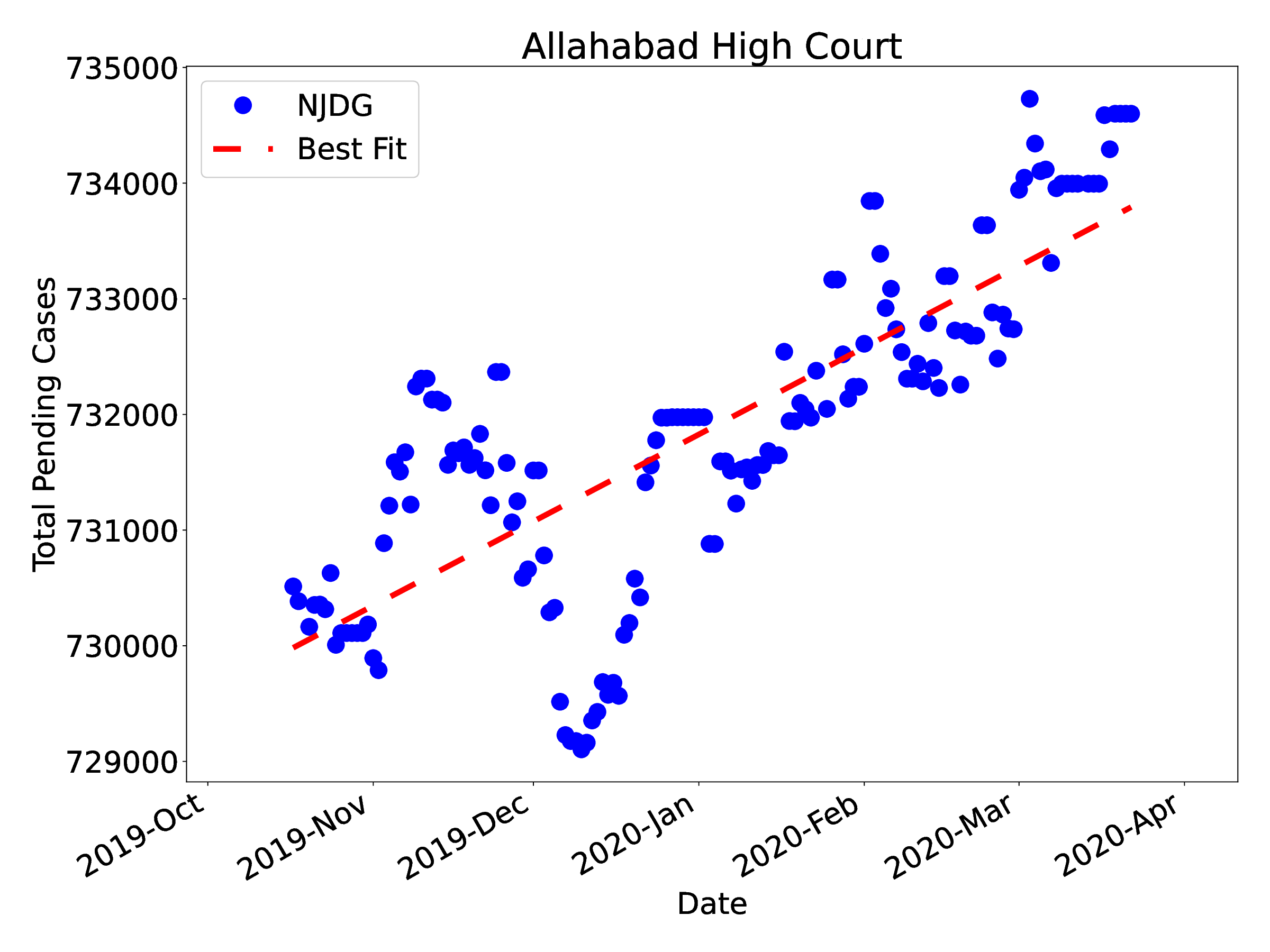}
\includegraphics[width=4.45cm]{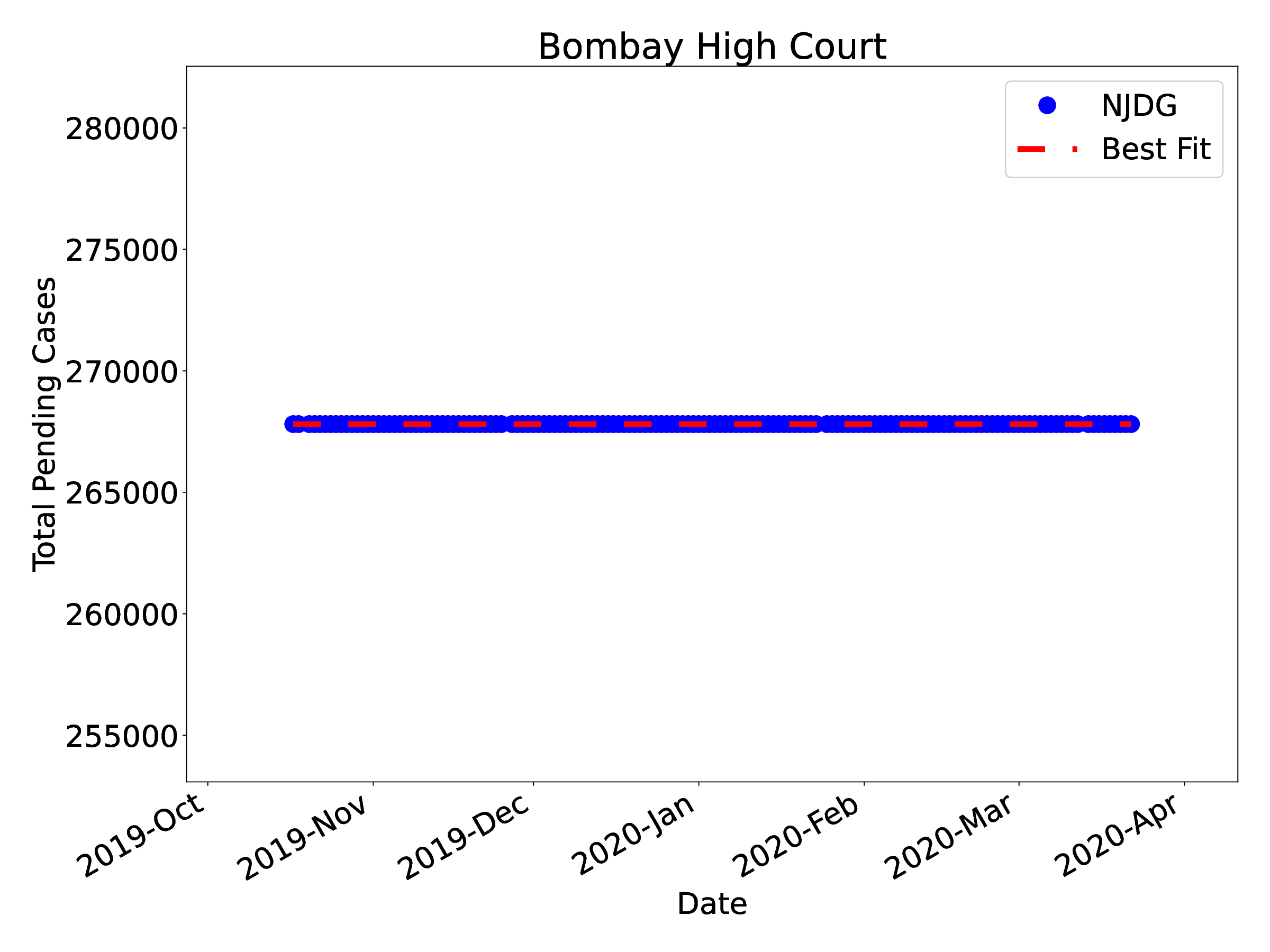}
\includegraphics[width=4.45cm]{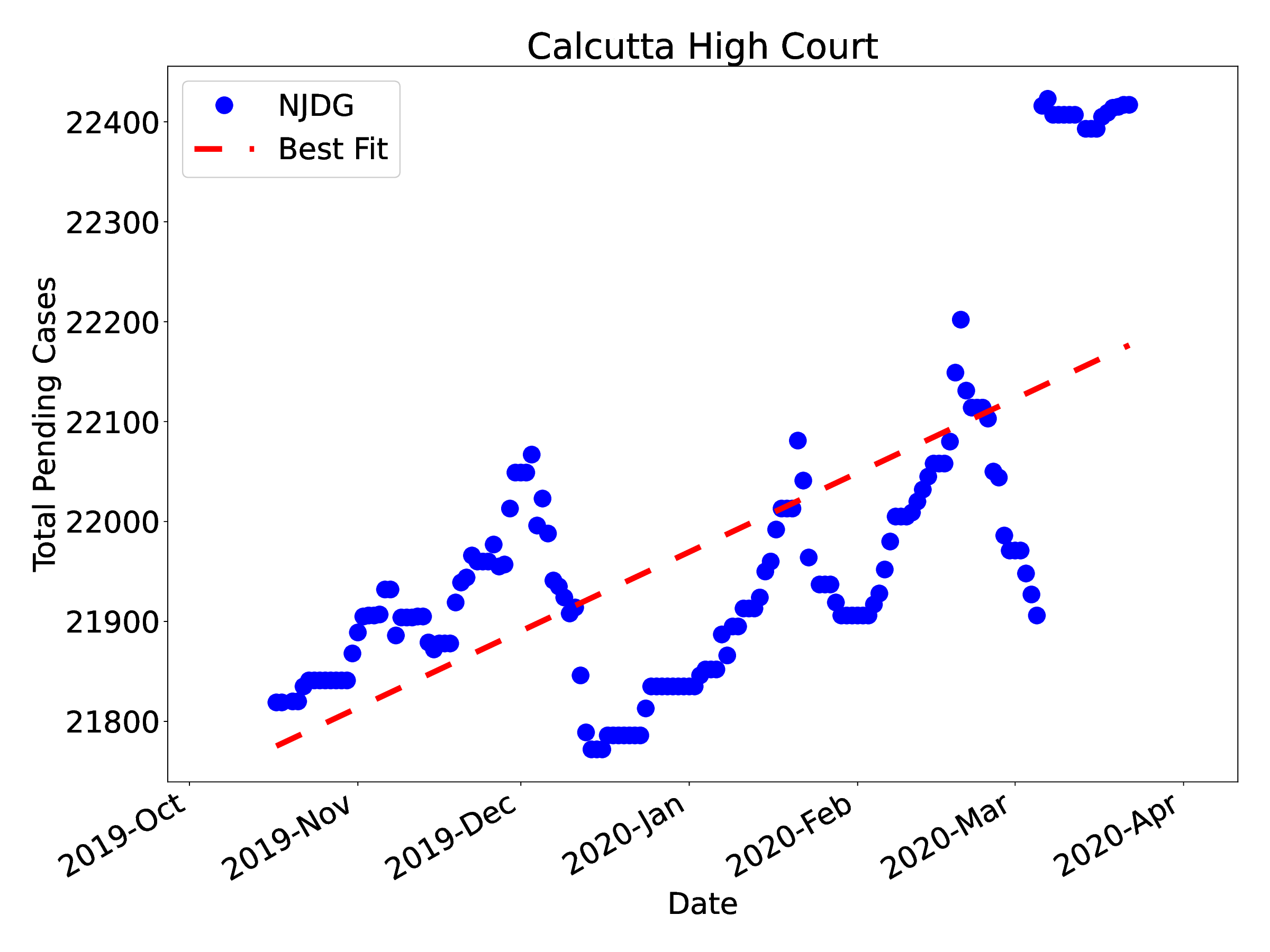}
\includegraphics[width=4.45cm]{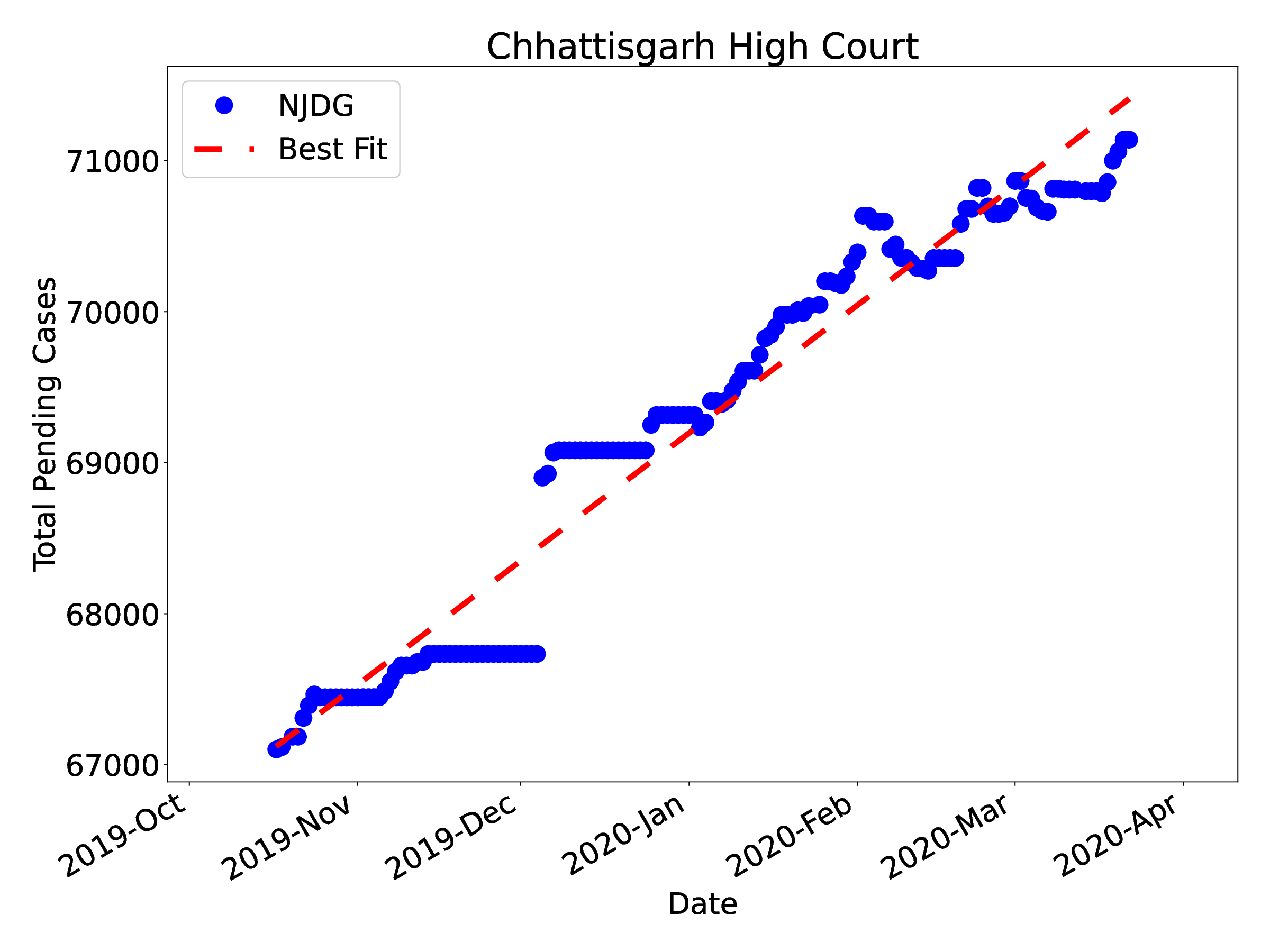}
\includegraphics[width=4.45cm]{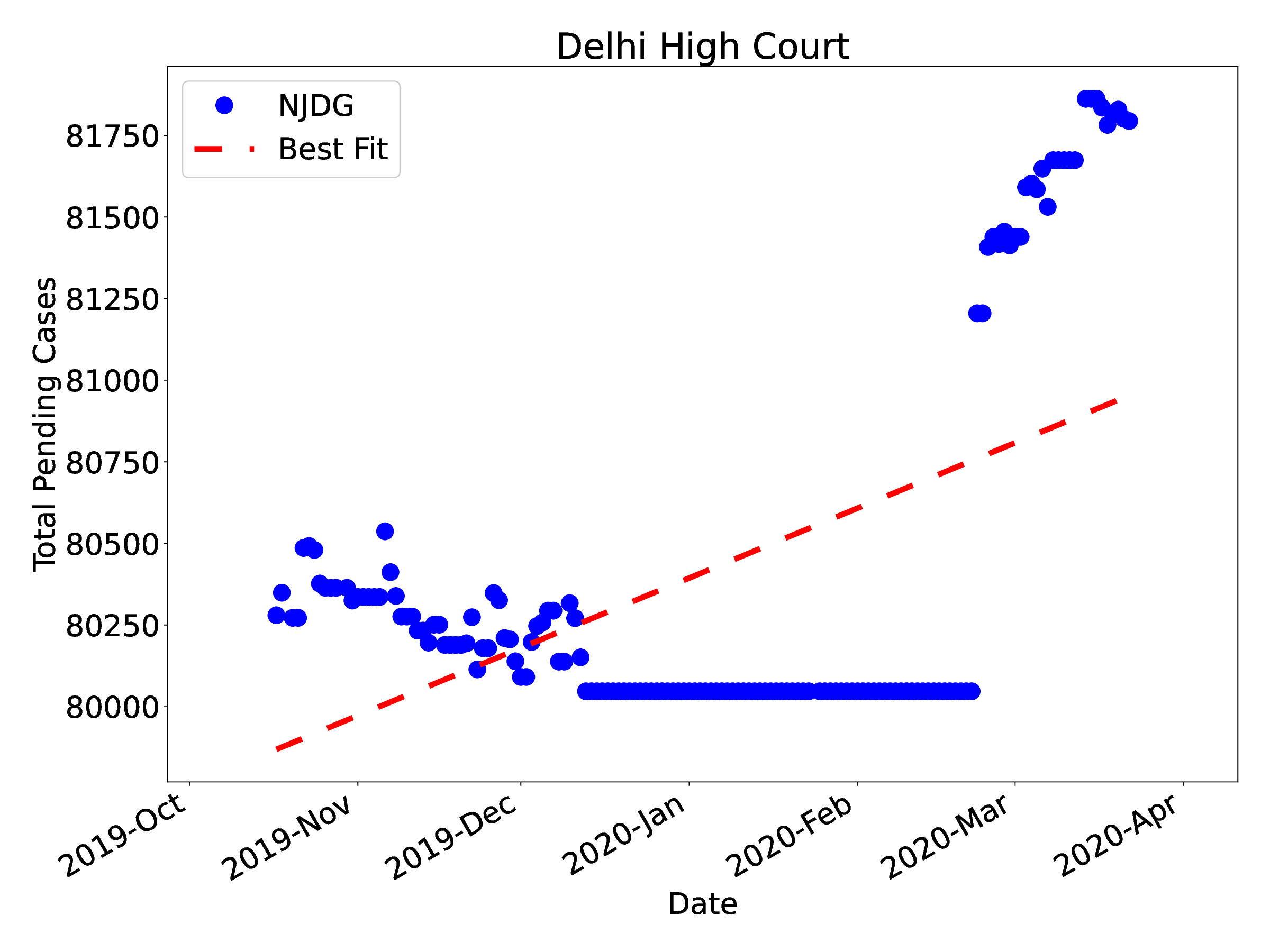}
\includegraphics[width=4.45cm]{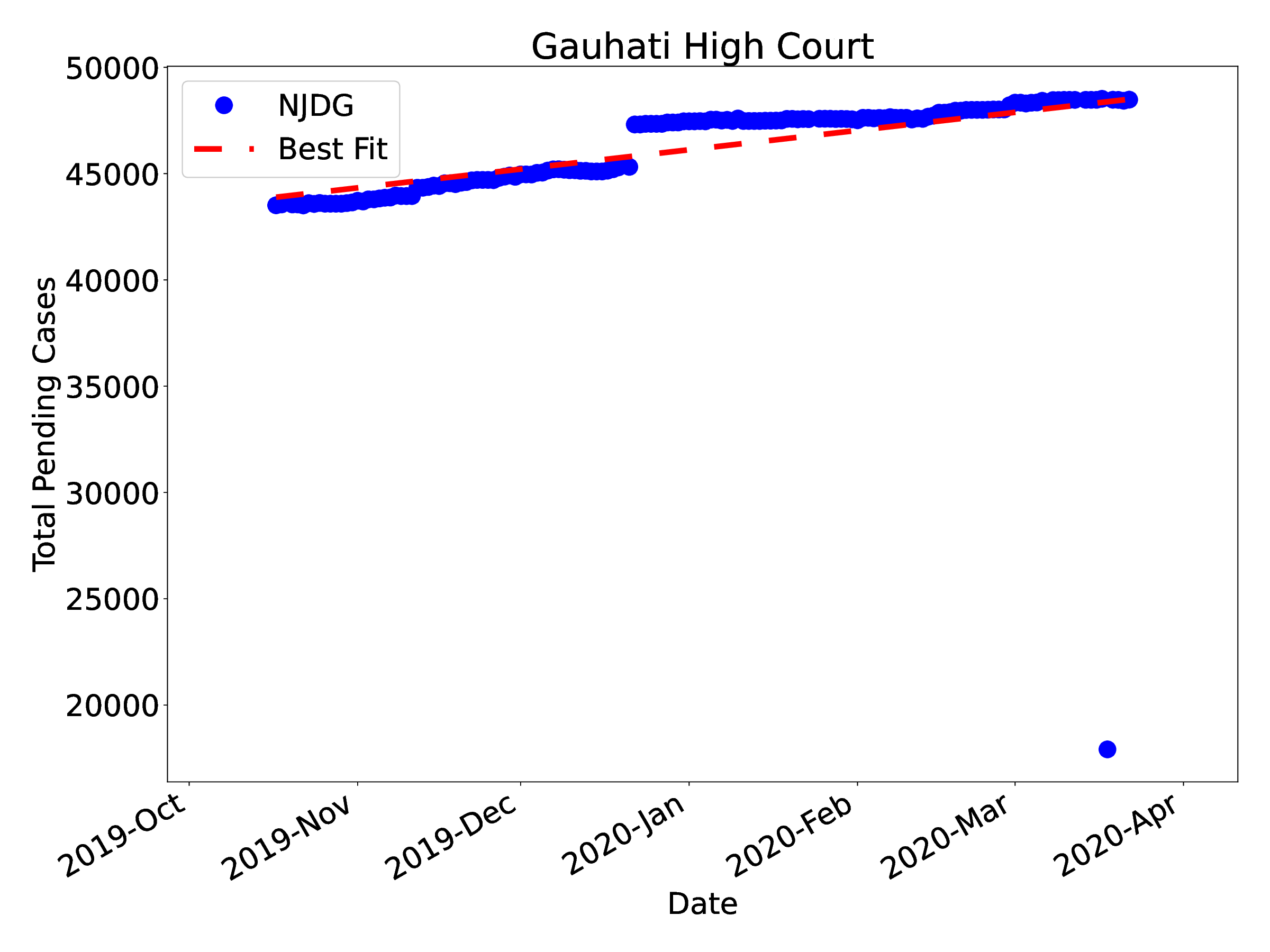}
\includegraphics[width=4.45cm]{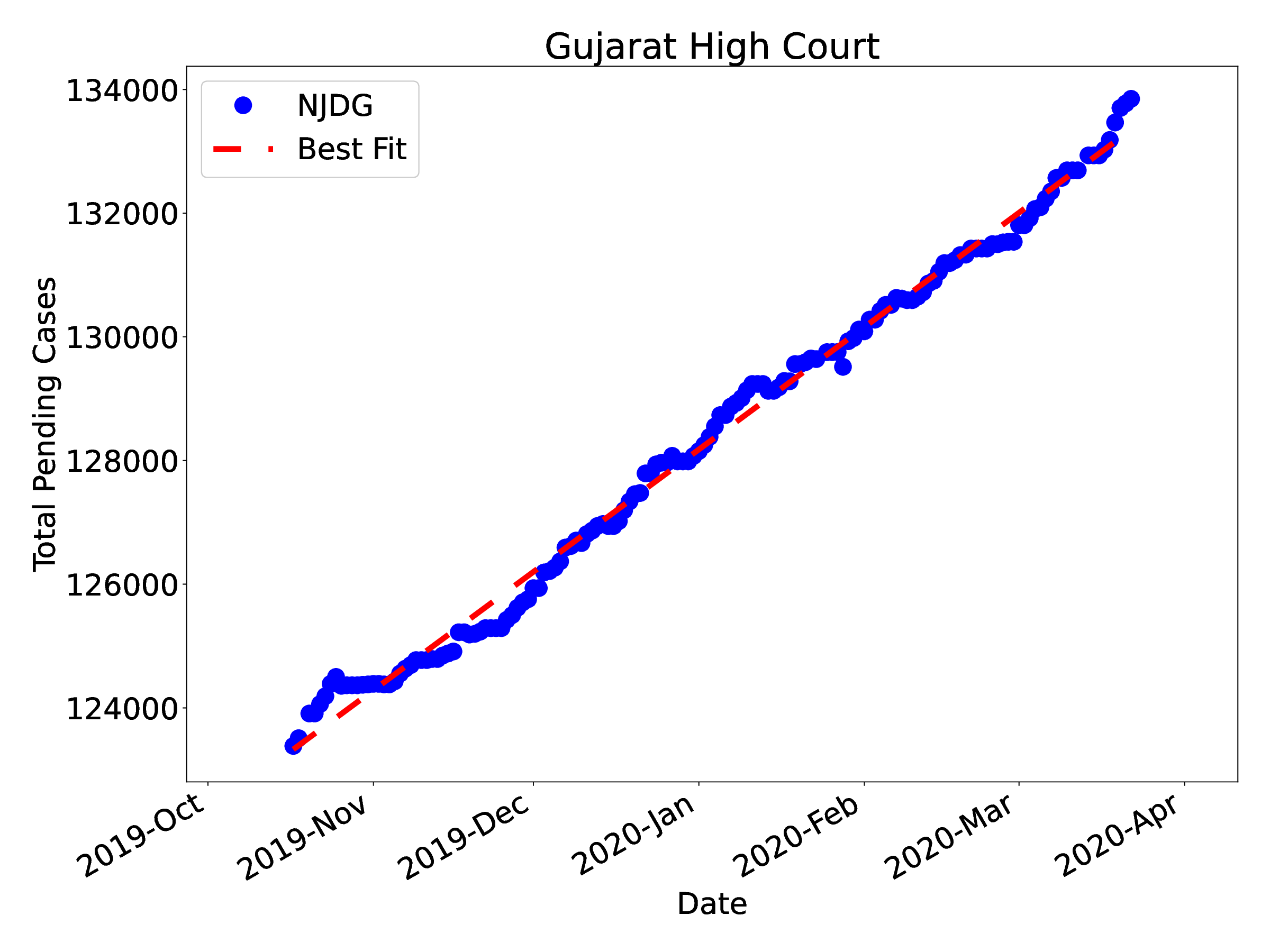}
\includegraphics[width=4.45cm]{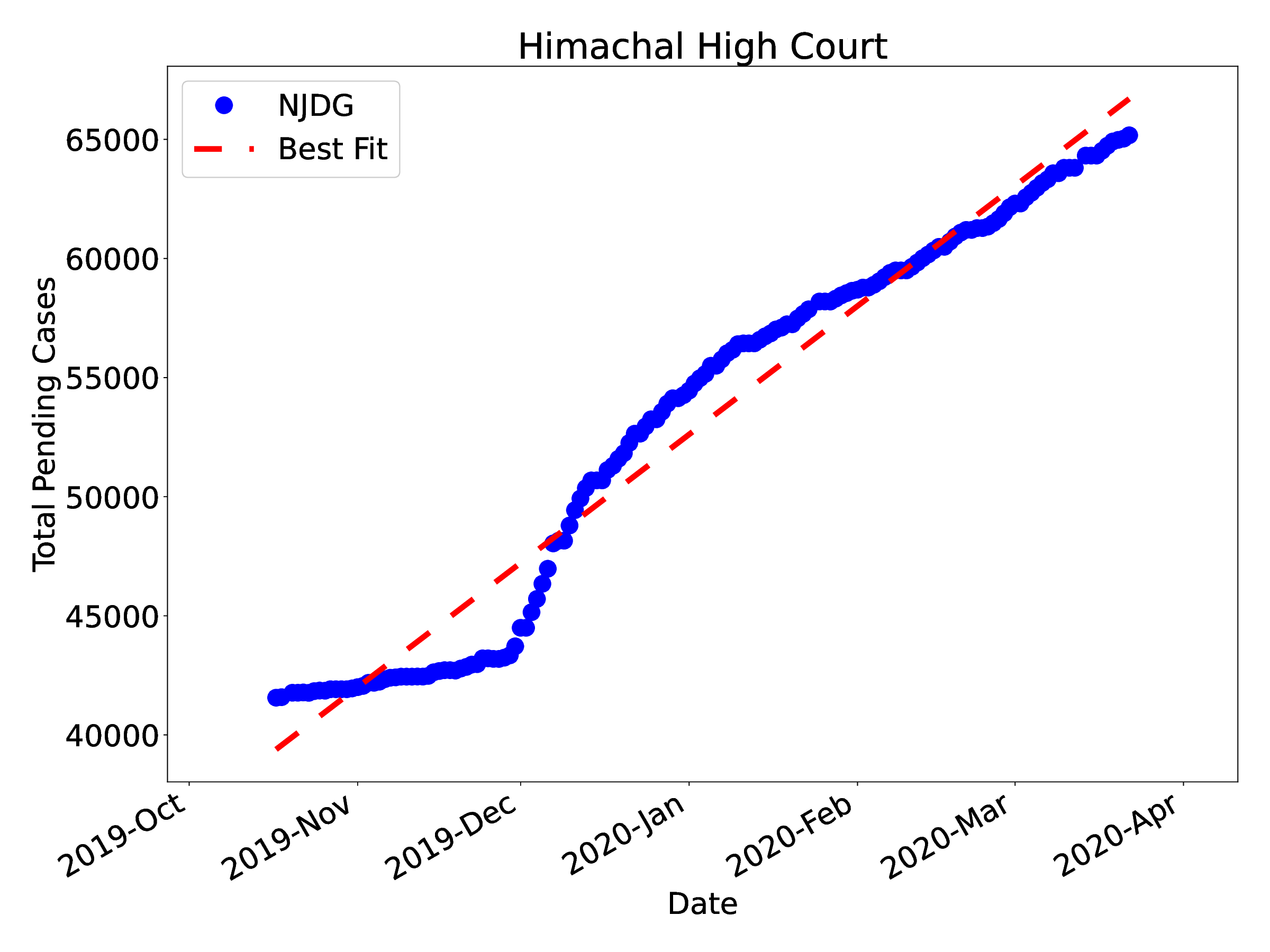}
\includegraphics[width=4.45cm]{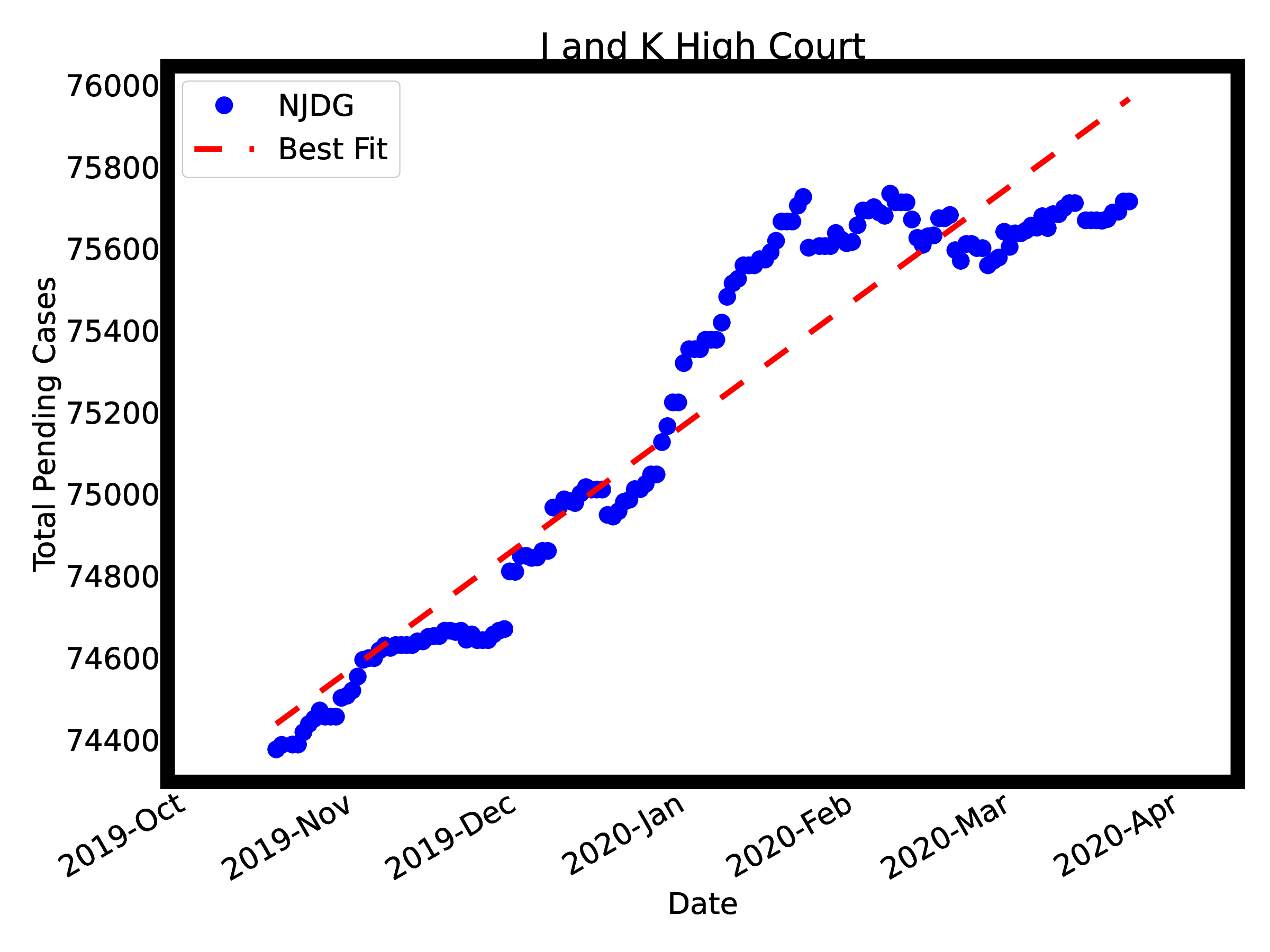}
\includegraphics[width=4.45cm]{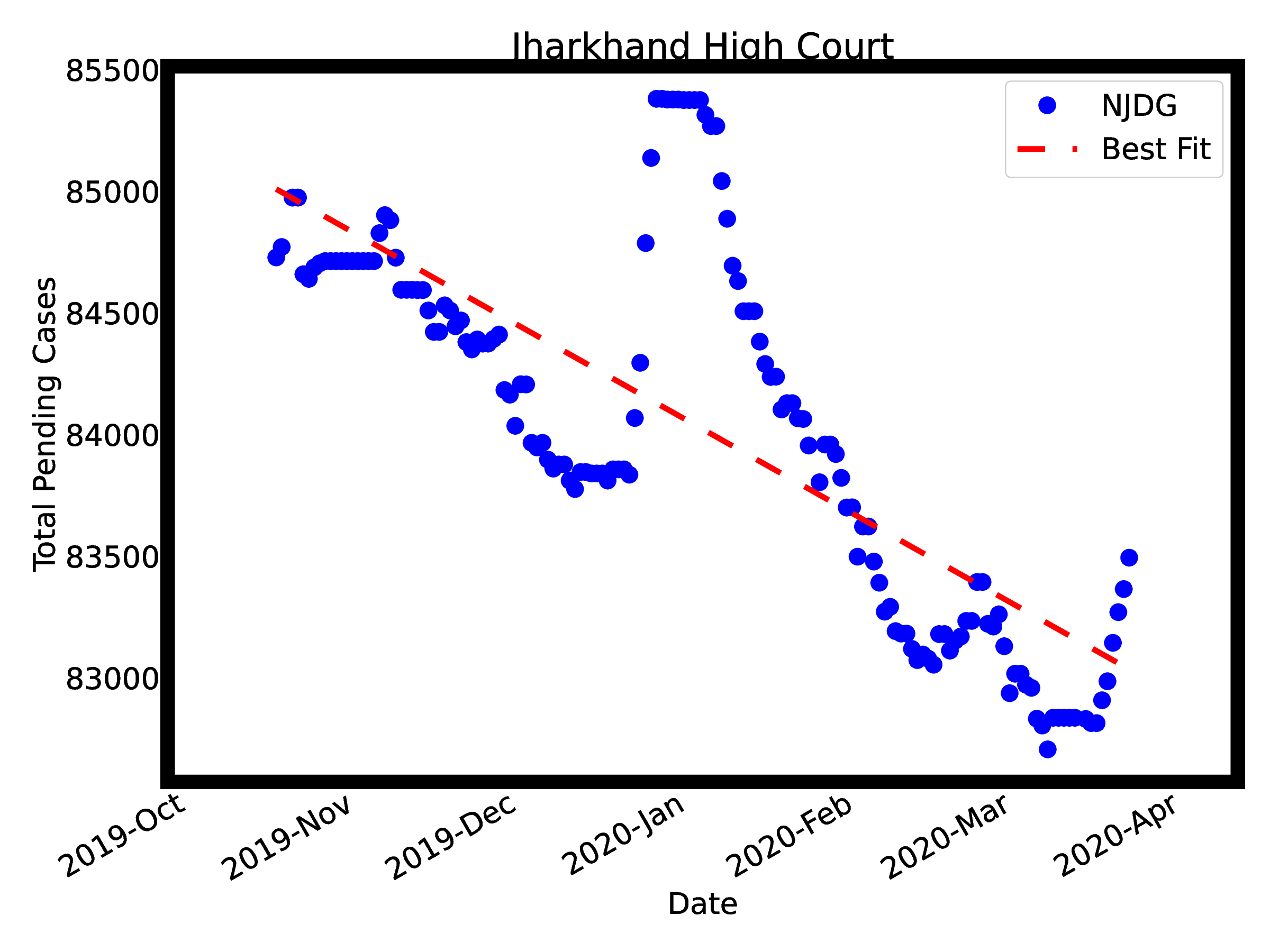}
\includegraphics[width=4.45cm]{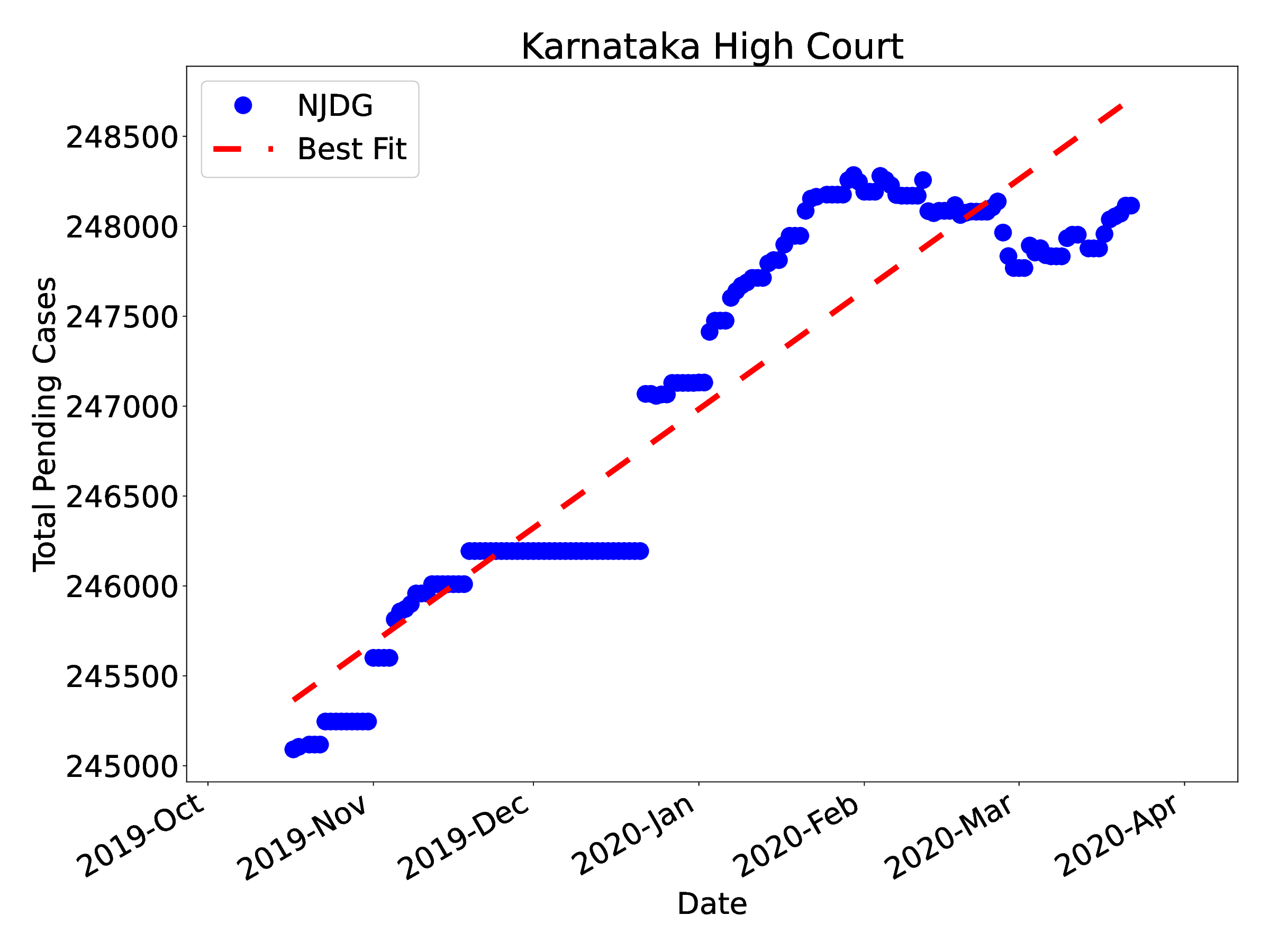}
\includegraphics[width=4.45cm]{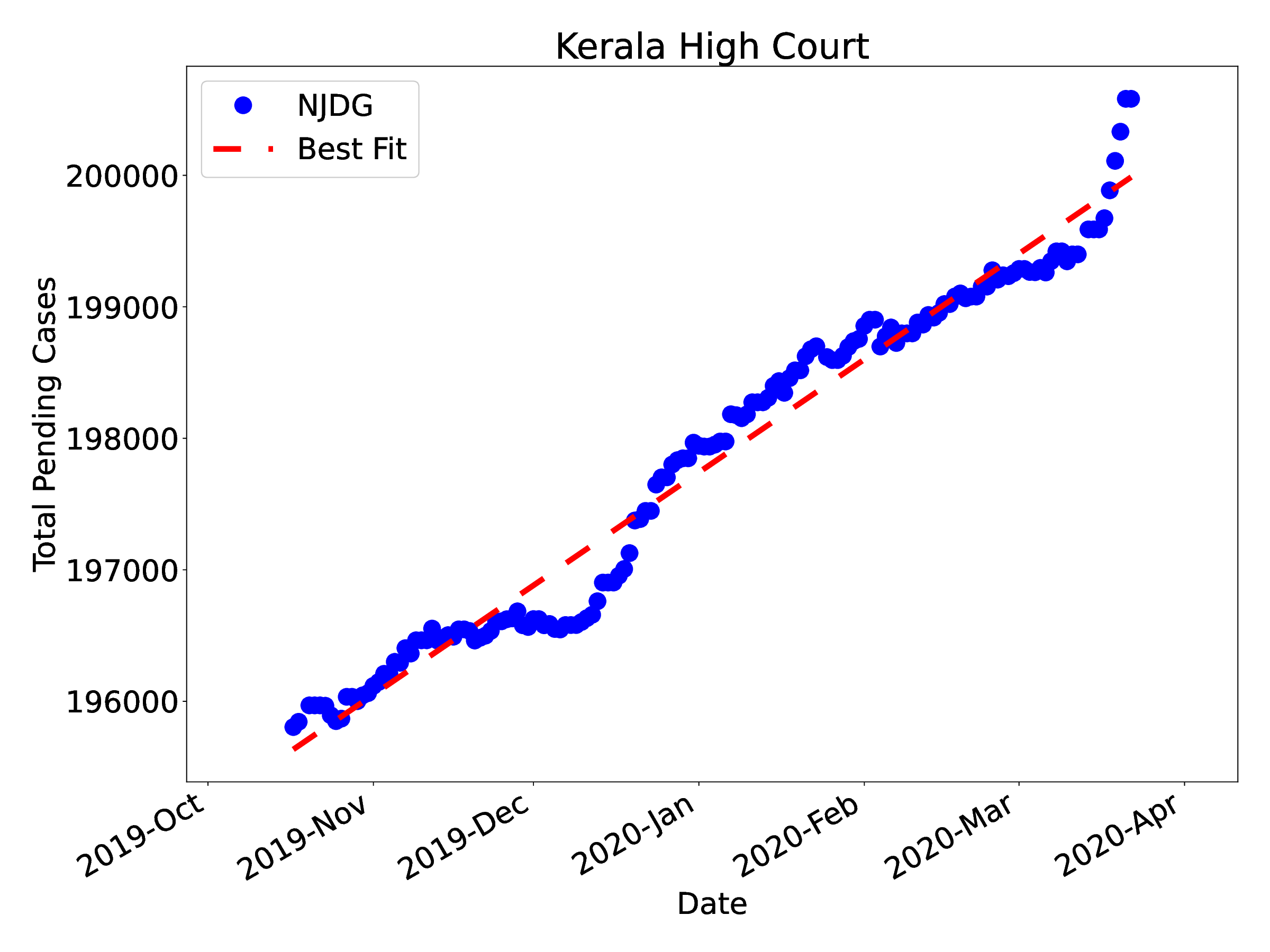}
\includegraphics[width=4.45cm]{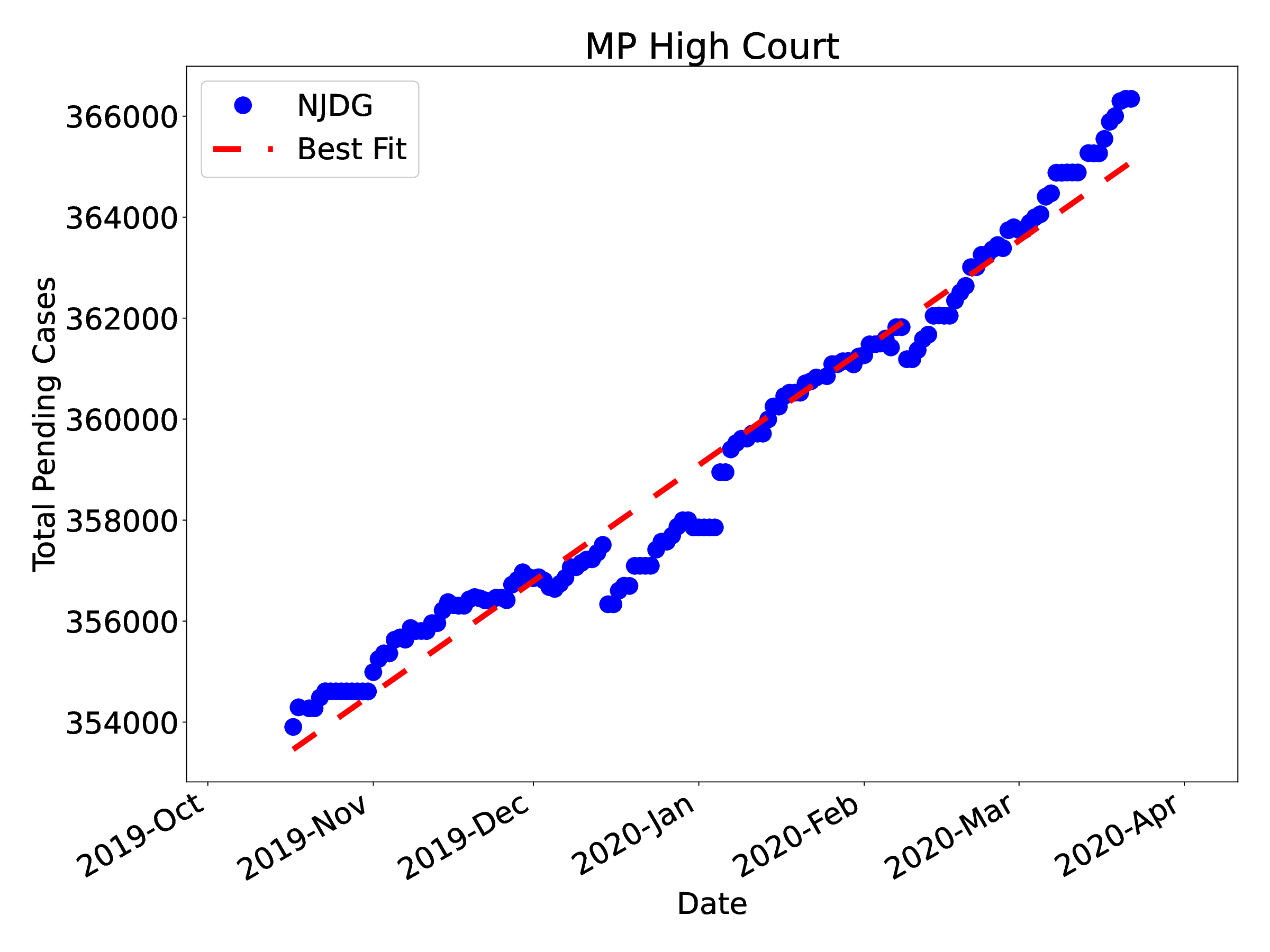}
\includegraphics[width=4.45cm]{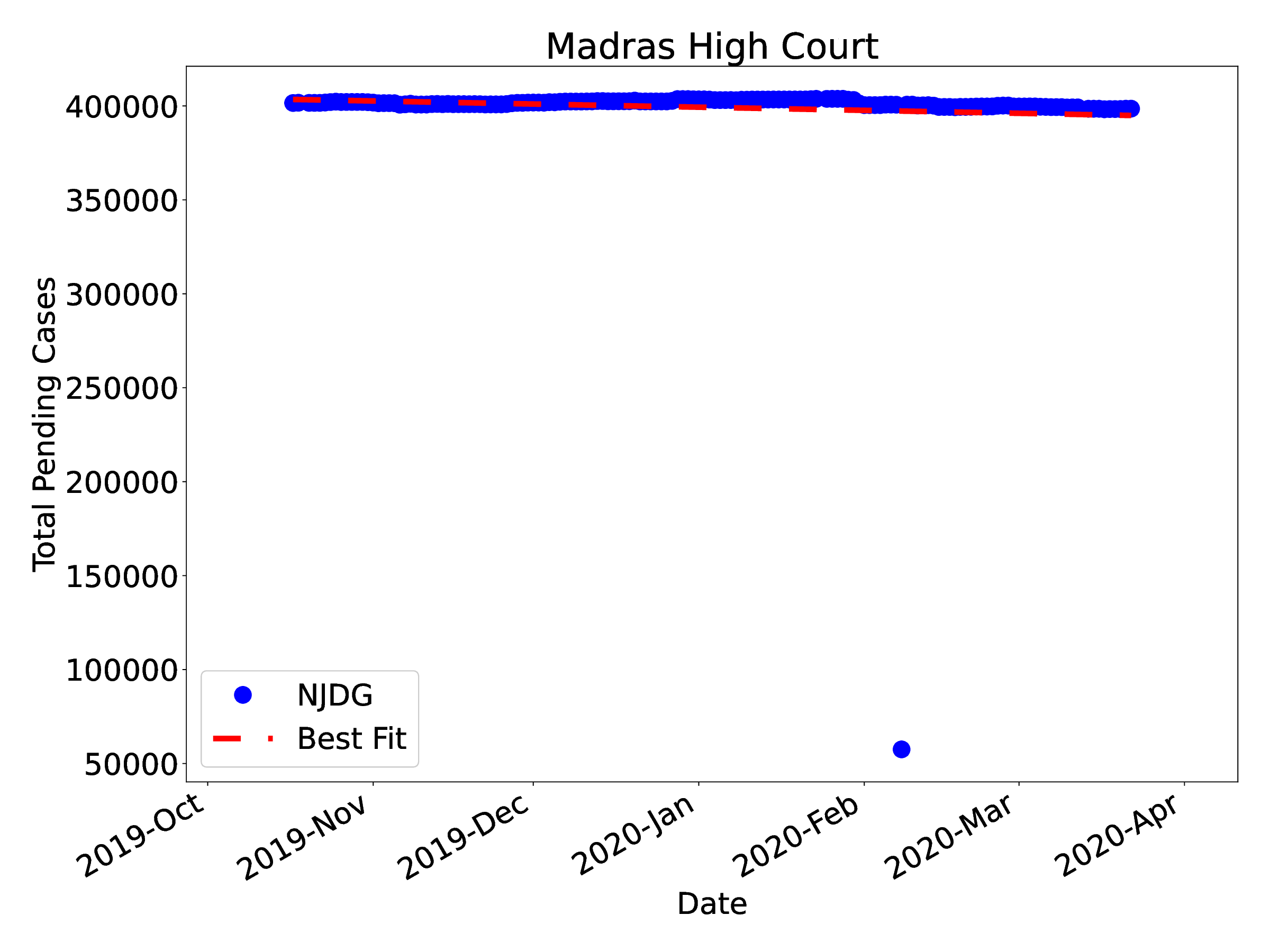}
\includegraphics[width=4.45cm]{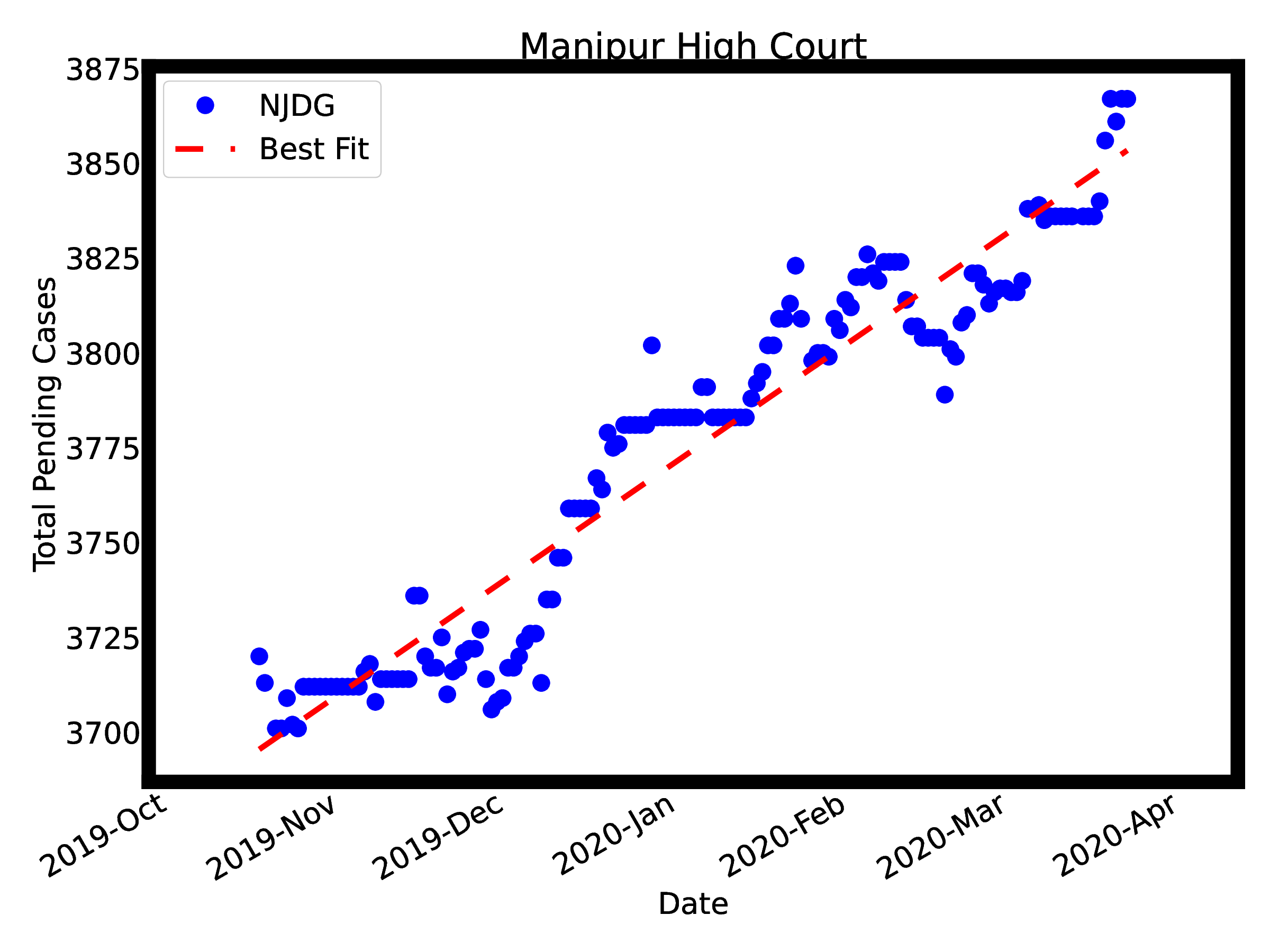}
\includegraphics[width=4.45cm]{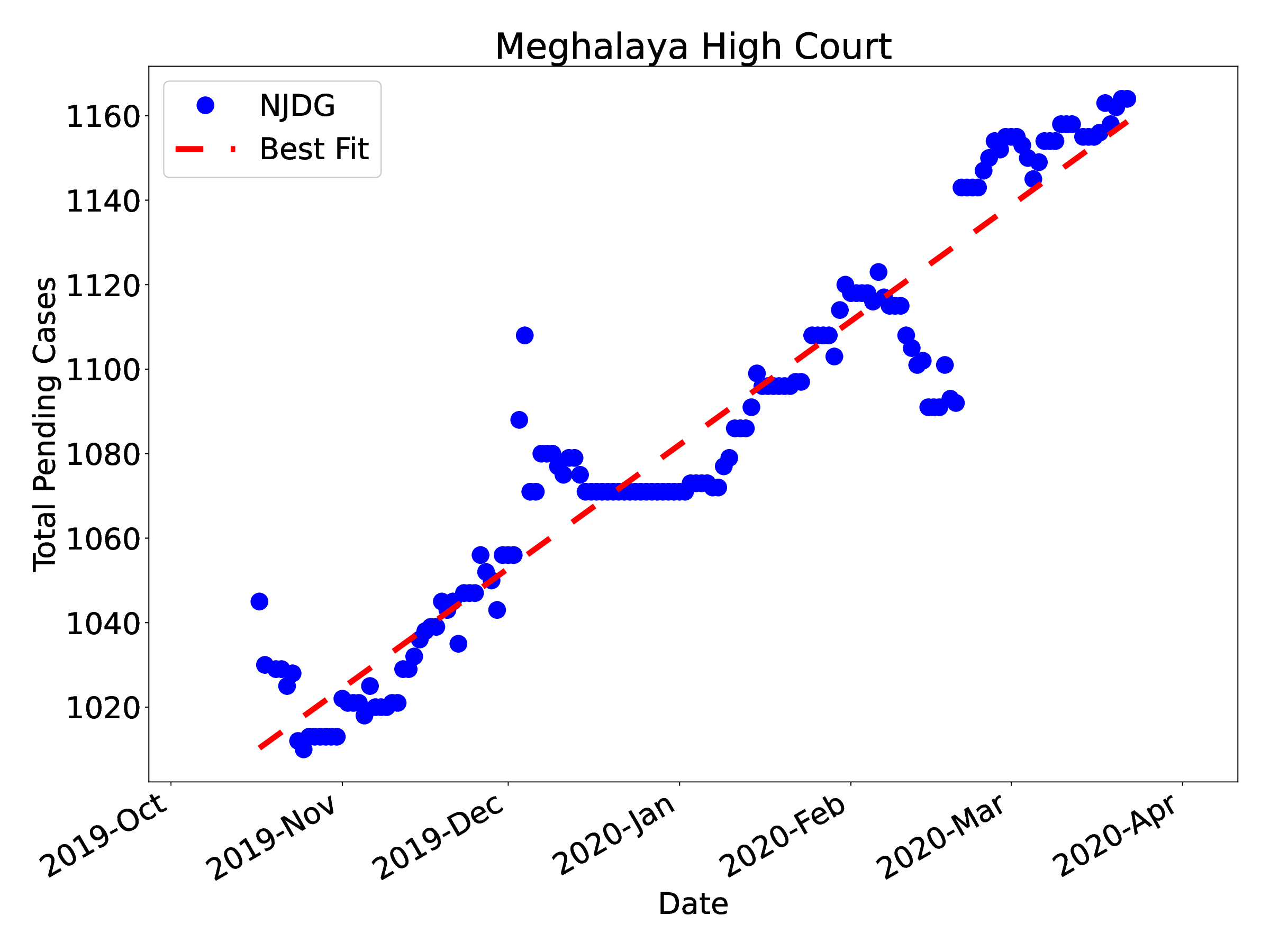}
\includegraphics[width=4.45cm]{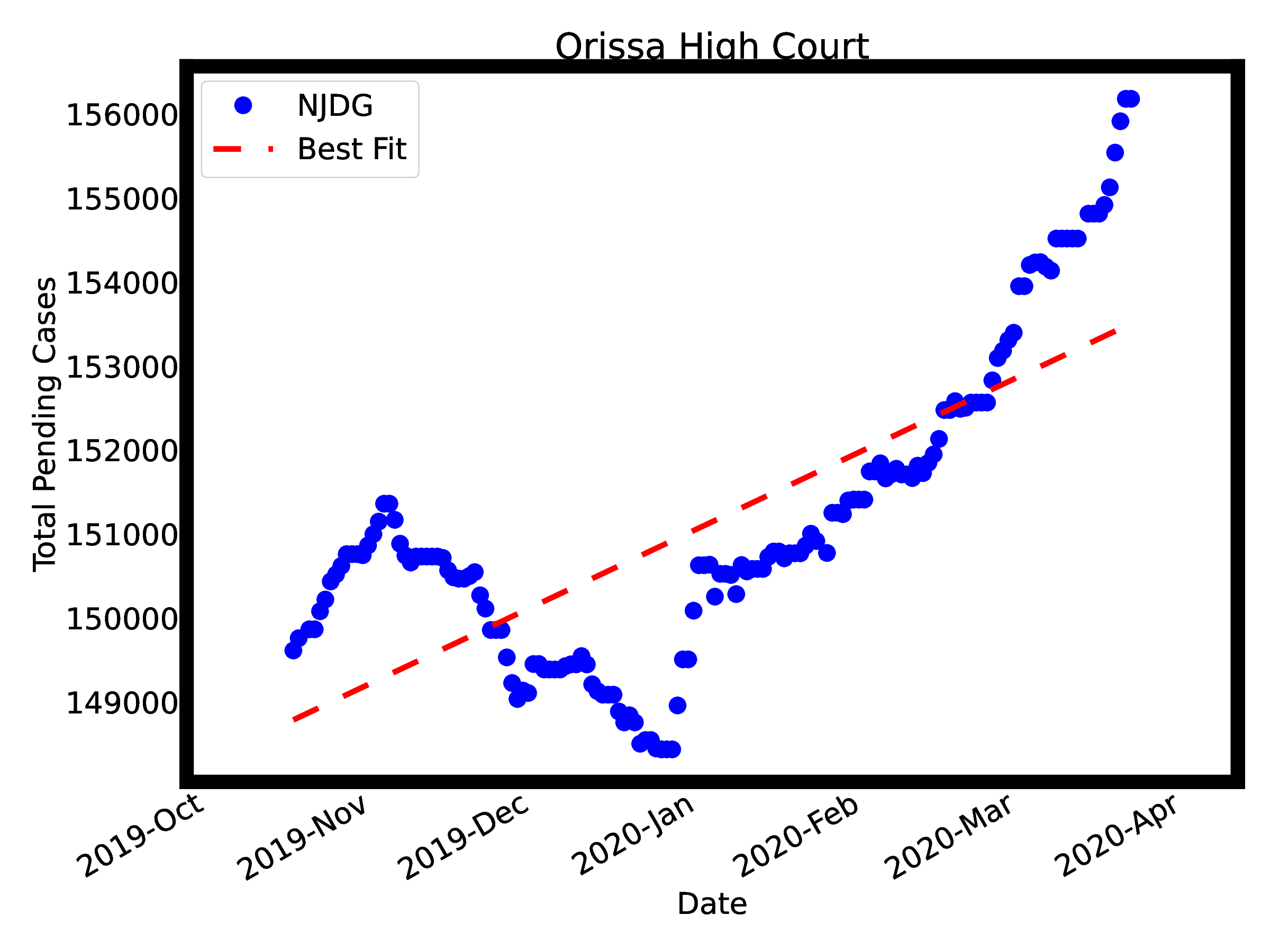}
\includegraphics[width=4.45cm]{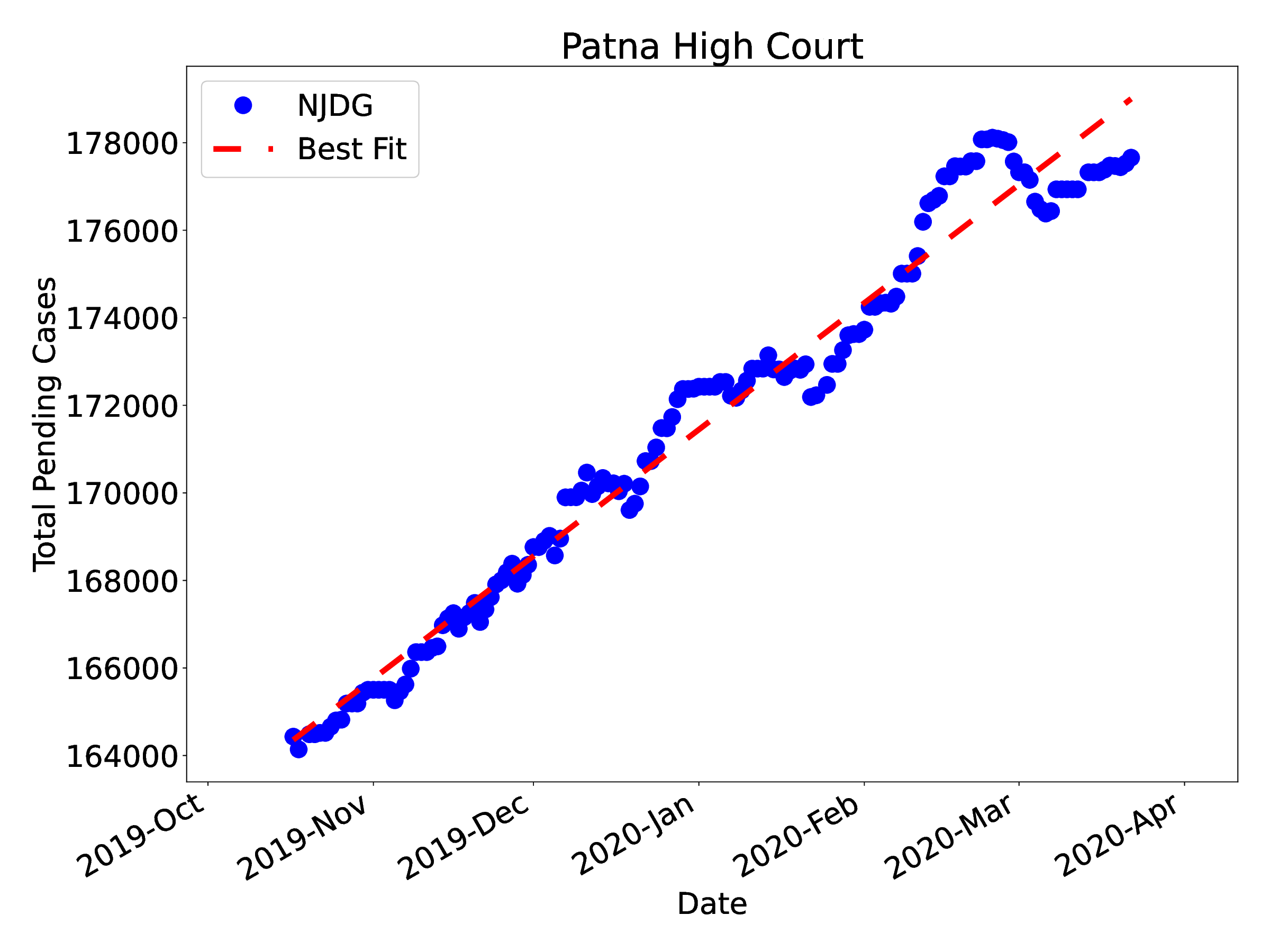}
\includegraphics[width=4.45cm]{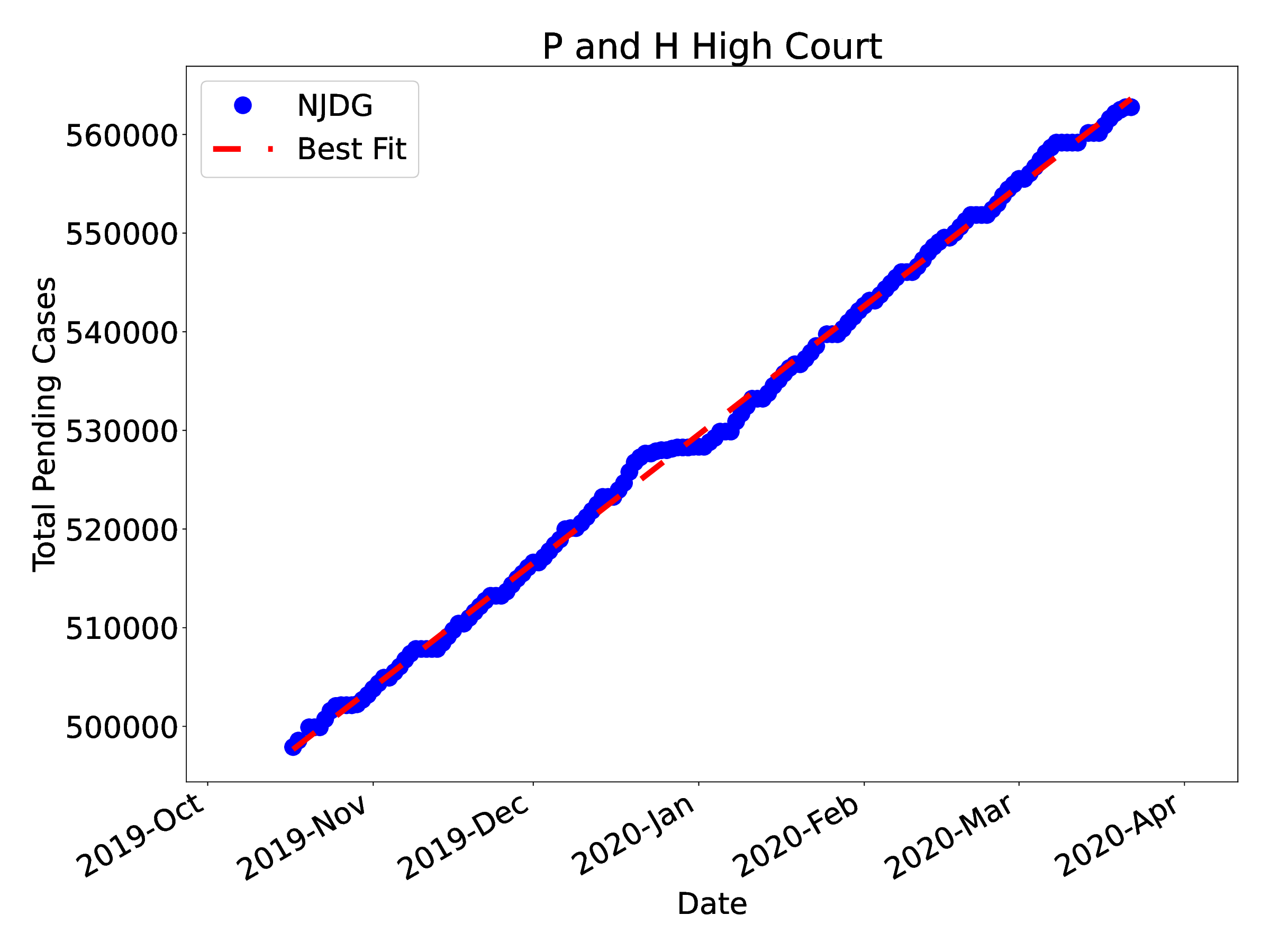}
\includegraphics[width=4.45cm]{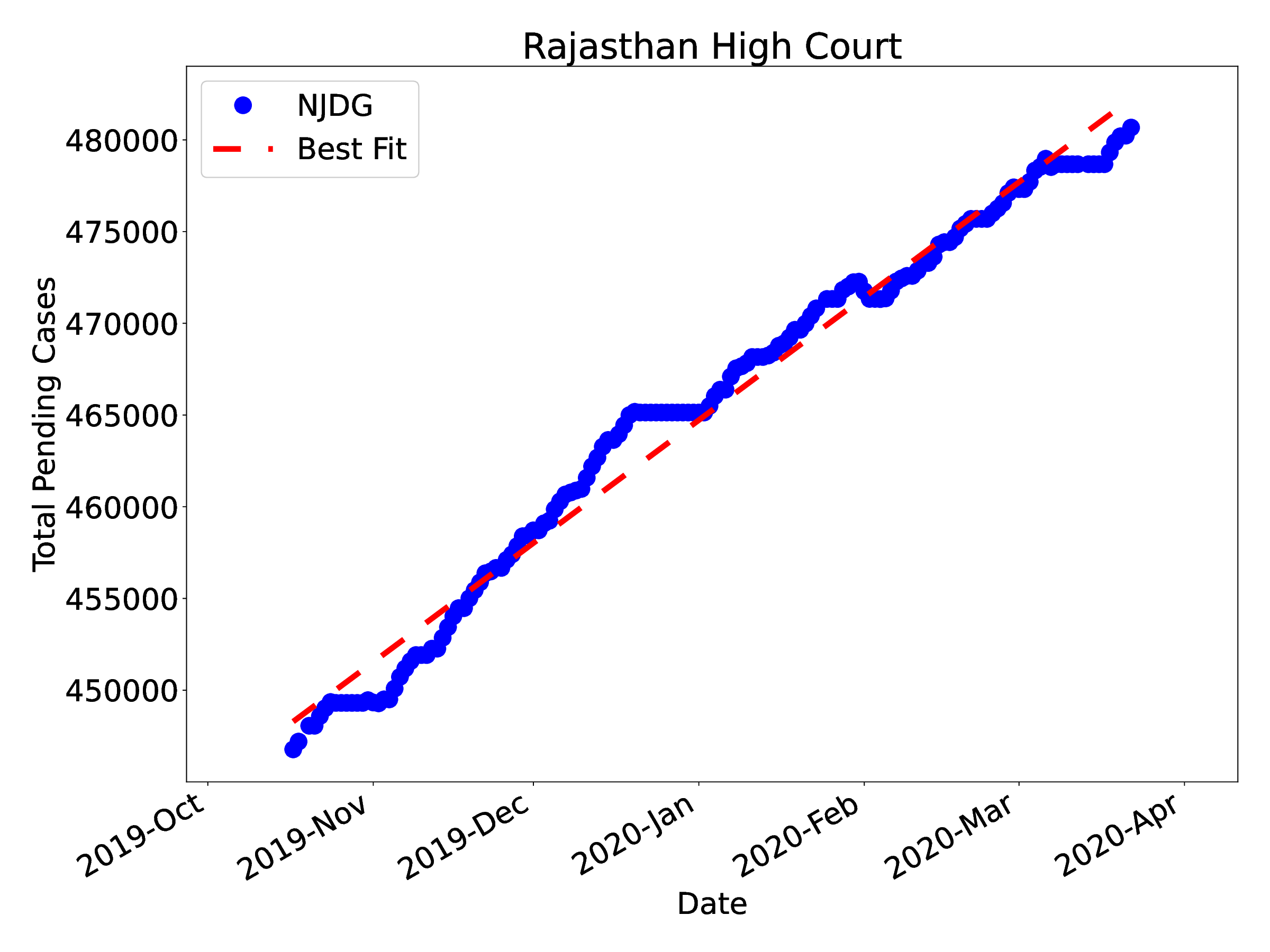}
\includegraphics[width=4.45cm]{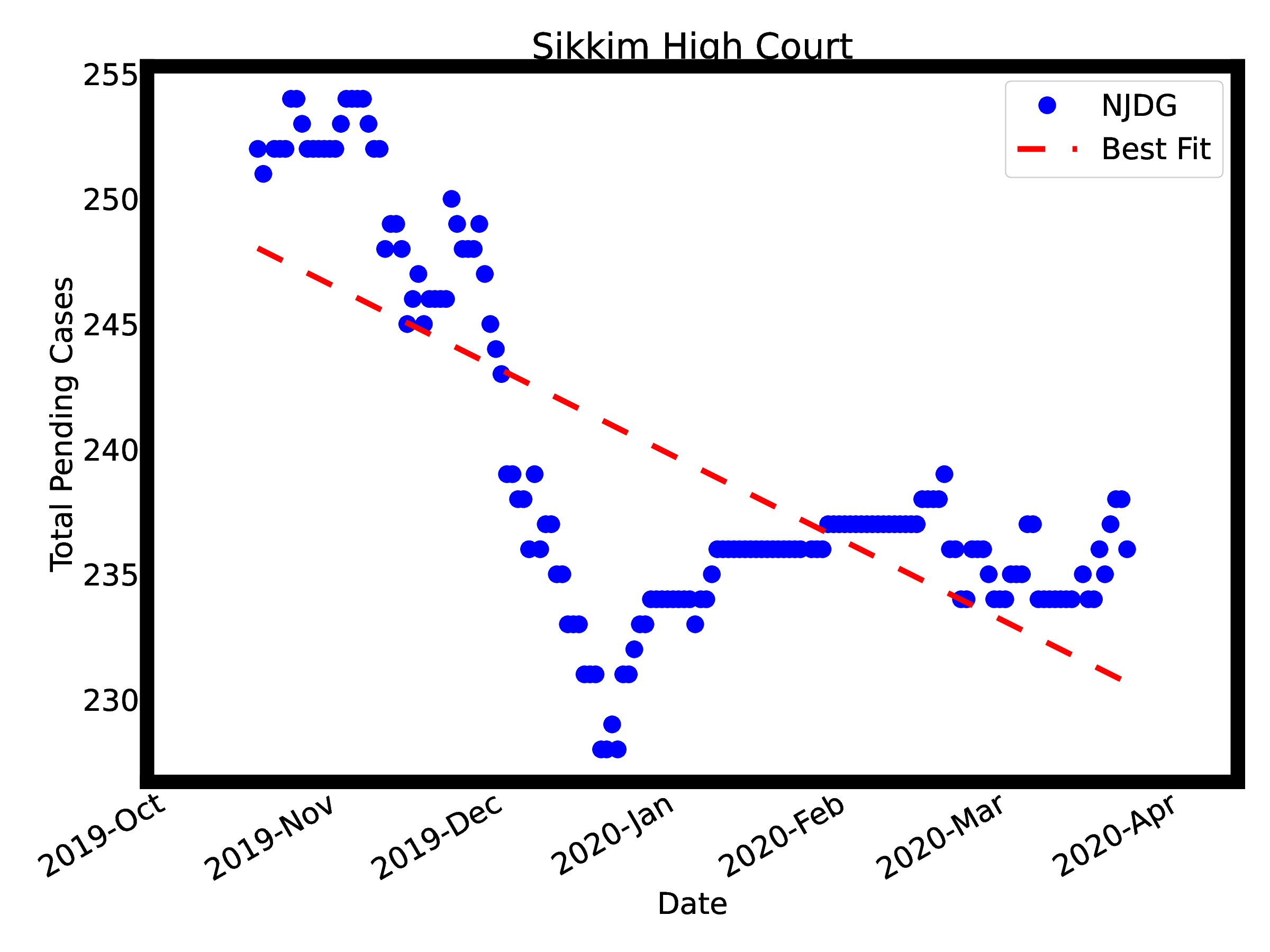}
\includegraphics[width=4.45cm]{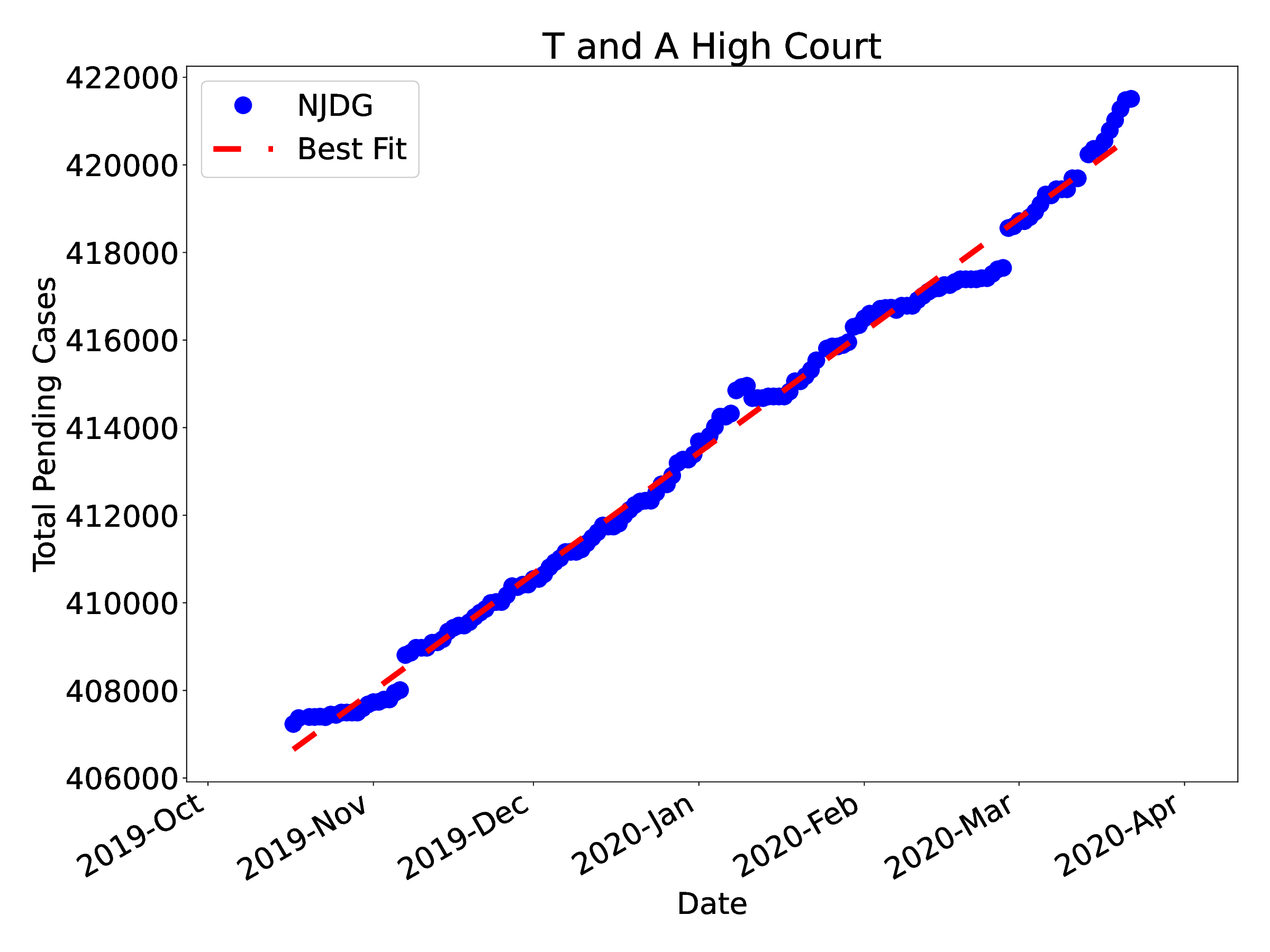}
\includegraphics[width=4.45cm]{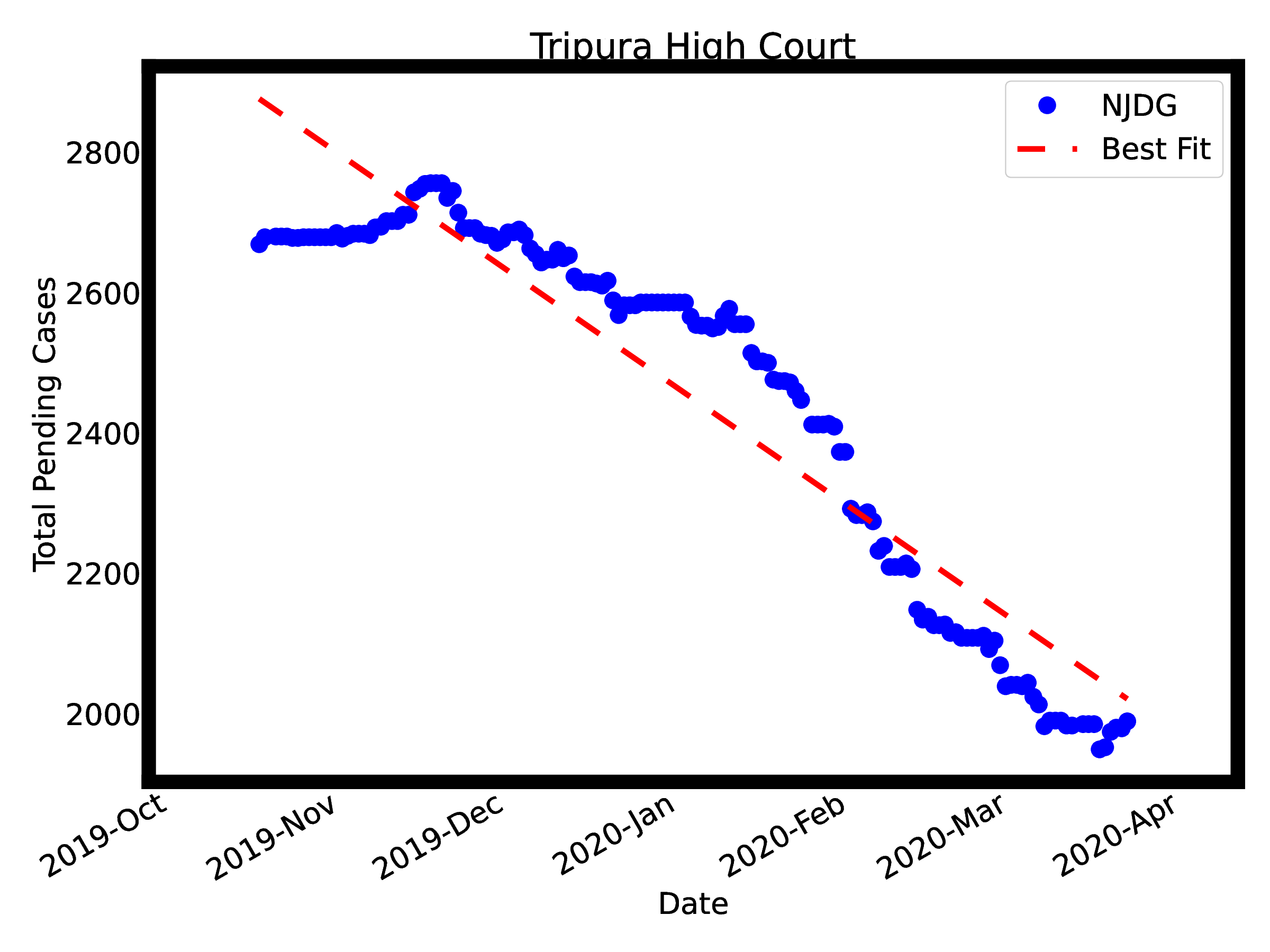}
\includegraphics[width=4.45cm]{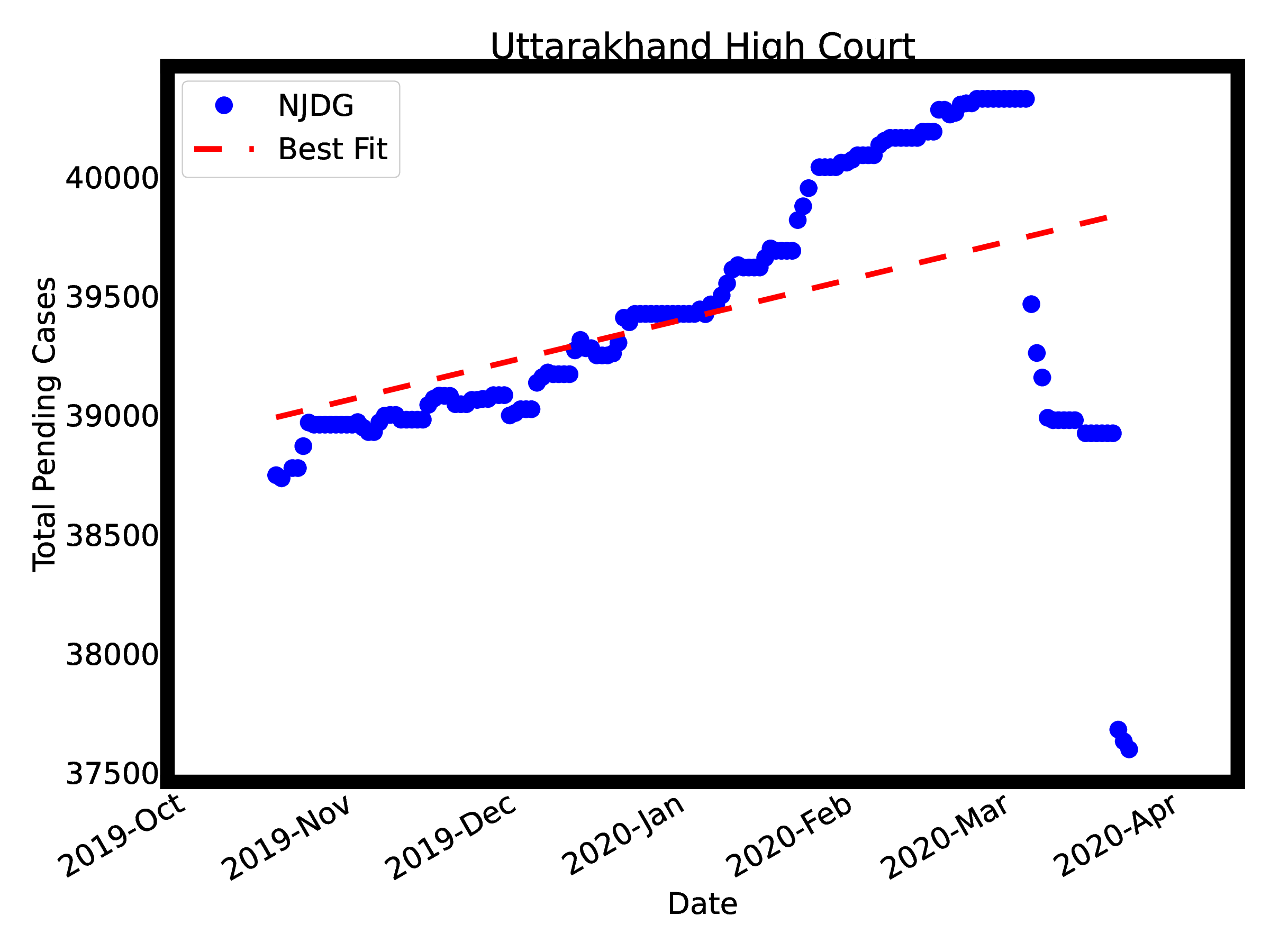}
\caption{Pending cases for individual high courts as plotted from October 2019 to March 2020. Dotted red line is the best fit straight line according to least squares loss. The updates on HC-NJDG have improved a lot during the above chosen period compared to the full duration of data collection. For many high courts we had to do discard the data collected during the whole duration and resort to the recently collected data. Such high courts include: Jammu and Kashmir, Jharkhand, Manipur, Orissa, Sikkim, Tripura and Uttarakhand. To emphasize, the frames of the plot of all these high courts have been made thicker compared to the frames of the high courts whose rate of increase of pendency is taken from other sources.}
\label{fig:np_hc2_p2}
\end{figure*}


\fref{fig:hcpendency} shows the aggregate number of pending cases in all the high courts of India. We have plotted the total number of pending cases in the high courts in India as obtained from the HC-NJDG portal in our data set. The blue dots are the data collected from the HC-NJDG and the dashed red line is the best fit straight line to the data minimizing the mean squared error cost function. It can be clearly seen that the data has few continuous clusters and few sudden jumps. While initial sudden jumps can be explained by the fact that few high courts have joined NJDG late and they may be taking time to converge to report stable number, the overall graph does not represent a healthy update culture until around March 2019. However, since April 2019 the updates have been smooth and barring a few outliers, the updates have been consistent. This is already a good news. This means that commendable efforts have been made to make data on HC-NJDG more reliable. 

\fref{fig:sanctioned_working} compares the working strength of the high courts in comparison with the sanctioned strength. The average working strength has been computed for the period June 2018 to March 2020, i.e., 22 months. The data is collected from the vacancy document available on the website of the Department of Justice \cite{doj_vacancy}. We see that on an average, around 38\% seats of judges in high courts remain vacant.




\fref{fig:np_hc1} and \fref{fig:np_hc2_p2} show the pendency data for the individual high courts. We also plot a linear regression best fit line (the red dashed line) to estimate the trends in pendency rather than depending on just one day of data. The difference between the two figures is that \fref{fig:np_hc1} plots the data collected during the whole duration and \fref{fig:np_hc2_p2} shows the data since October 2019. Barring a few High Courts, \fref{fig:np_hc2_p2} is much better in terms of regular updates than \fref{fig:np_hc1}. In both the plots, the high courts appear in the lexicographic order of their names. 

In \fref{fig:np_hc1}, the graph of Allahabad High Court depicts a decent update culture. There is not much diversion from the best fit line (using linear regression) either. It can be seen that the number of pending cases in Allahabad High Court has been increasing linearly with time. 

Be it \fref{fig:np_hc1} or \fref{fig:np_hc2_p2} Bombay High Court has the poorest record of data update on HC-NJDG among all the high courts. In the whole data collection period, the data has been updated only once. Due to this reason nothing can be said about Bombay High Court reliably. A similar case is with Calcutta High Court. Even though the updates have been frequent, wrong data was uploaded on the portal. The total number of pending cases in Calcutta High Court is more than 250 thousand but the graph shows a different number. Hence, nothing can be said reliably about Calcutta High Court either. For Calcutta High Court, the subsequent analysis is done based on the data available from Calcutta High Court website \cite{calcutta_portal}. For Bombay High Court, the data is not available on its website either, so we had to resort to the Supreme Court annual reports \cite{SCAR}. These are the only two high courts whose data is not taken from HC-NJDG. 

In \fref{fig:np_hc1}, data from Chhattisgarh High Court follows a nice update trend and the best fit straight line looks representative of the increase. Hence, it can be deduced that the number of pending cases in Chhattisgarh High Court has been increasing linearly. The Delhi High Court also has a nice update culture almost throughout the data collection period and the pendency is increasing linearly for this high court too. The updates in Gujarat High Court were not frequent until August 2018. However, after that, the number of cases have been increasing linearly. We consider the best fit line drawn in the figures as representing the rate of increase from this graph. Himachal Pradesh High Court data has seen a surge in the number of cases from October 2019 to March 2020, the best fit line, is with positive slope and hence, the number of pending cases are increasing with time for Himachal Pradesh High Court as well. 

In \fref{fig:np_hc2_p2} represents much better updates of NJDG in the Common High Court for the UT of Jammu \& Kashmir and UT of Ladhakh as well as for Jharkhand High Court. Initial updates of the HC-NJDG data seem erroneous and hence we use \fref{fig:np_hc2_p2} for these high courts. The number of cases in the Common High Court for the UT of Jammu \& Kashmir and UT of Ladhakh follow a linear increase whereas the number of pending cases in Jharkhand High Court are decreasing linearly.

In \fref{fig:np_hc2_p2}, barring a few erroneous updates, Karnataka High Court and Kerala High Court have regularly updated data on HC-NJDG. Again, the best fit straight lines looks quite a good representative of the increase. The pendency for both these high courts is increasing too. \fref{fig:np_hc1} depicts that the champion of updating data on HC-NJDG is Madhya Pradesh High Court. Not even a single outlier. The best fit straight line almost coincides with the data. The number of pending cases at this high court is increasing as well. The next comes Madras High Court, which, doesn't seem to have a good update culture. However, it is good enough to be consistent and is increasing which is the current expected trend in most of the high courts. 

Manipur High Court seems to have reconciled data and hence, we do not consider the data for the whole data collection period but only for the last six months as presented in \fref{fig:np_hc2_p2}. The trend of increasing pendency, however little, can be seen for Manipur High Court as well. 

In \fref{fig:np_hc1}, Meghalaya High Court also seems to have an increasing rate of pending cases. The data points may look erroneous on the first look, however, there is a variation of just 250 cases on the whole scale. So such updates are realistically possible. 

Some kinds of reconciliation seems to have taken place for Orissa High Court as well. Hence, we take the trend from \fref{fig:np_hc2_p2}, which again shows an increasing trend in pendency. 

The updates for Patna High Court, Punjab and Haryana High Court and Rajasthan High Court look reasonable and the best fit seems to be representative of the trend that the pendency is increasing. We take the rate of increase from \fref{fig:np_hc1}. 

Sikkim High Court has very low number of pending cases. So taking last six months of trend may be more beneficial. We see that the pendency is decreasing. So we use \fref{fig:np_hc2_p2} for computing the best fit line, which shows a decrease in the number of pending cases. 

In \fref{fig:np_hc1}, for the hypothetical aggregate of Telangana and Andhra High Court the trend is again a linear increase in the pendency. 

Tripura High Court has done very well since October 2019. There is a close to perfect linear decrease in the number of pending cases \fref{fig:np_hc2_p2}. For Uttarakhand High Court, we consider the best fit line for computing the increase in pendency from \fref{fig:np_hc2_p2}. The figure for Uttarakhand High Court in \fref{fig:np_hc1} is quite unreliable.

Hence, from the above analysis, we can deduce that we have enough data for computing the rate of increase of pendency reliably for 21 high courts, as well as the hypothetical aggregate of Telangana and Andhra High Court. However, the updates for two high courts, viz., Bombay and Calcutta are too unreliable on HC-NJDG to be able to make any conclusions about their pendency statistics. Other sources have been used for inferring their data. 

\begin{figure}[t]
\includegraphics[width=9.0cm]{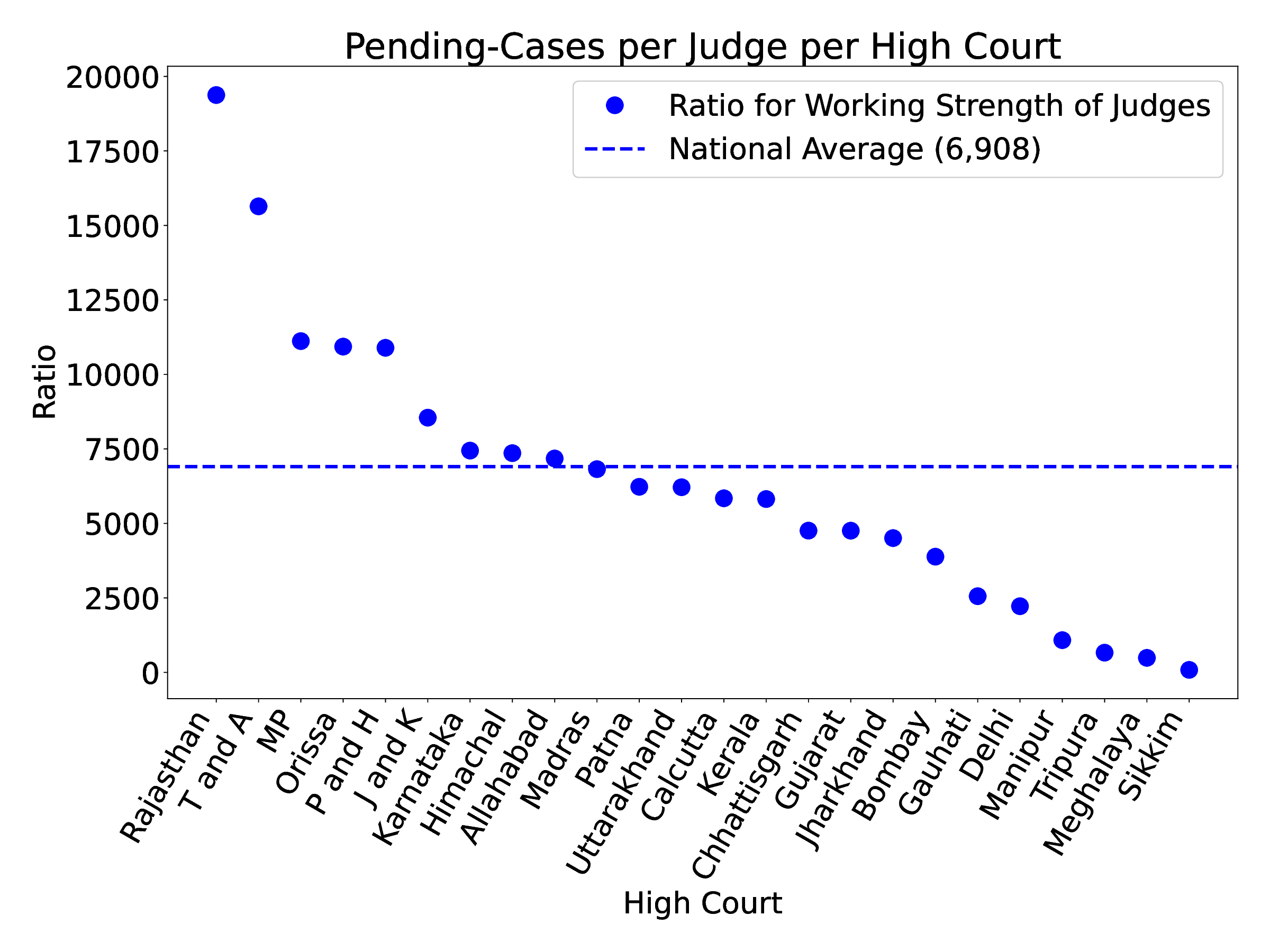}
\caption{Ratio of pending cases to the judges in High Courts, i.e., the average number of pending cases per judge per high court.}
\label{fig:pend_date_r}
\end{figure}

%

\subsection{Ratio of Pendency to Judges}

The total pendency, in itself, does not provide any information until the number of judges in the respective high court is also taken into account. This subsection considers the ratio of \emph{pending cases/number of judges} as a parameter for each high court.

\fref{fig:pend_date_r} plots the ratio \emph{pending cases/number of judges} for each high court. The number of pending cases is calculated by taking the pendency on the last day of data collected. The number of judges, however, are taken from \fref{fig:sanctioned_working}. The results are plotted in the descending order of the ratio so calculated. This graph provides the distribution of workload on each high court and judges thereof. The blue dots show the ratio \emph{pending cases/working strength of judges} for the average working strength of each high court. For example, Rajasthan High Court has the maximum value of 19,374 pending cases for each judge whereas Sikkim High Court has the minimum ratio which is 78. Hence, statistically we can say that a judge in Rajasthan High Court has almost 250 times more load than a judge in Sikkim High Court. It can be seen that the situation is similar for most of the high courts. The mean of this ratio is 6908, i.e., the national average of the number of pending cases per judge. It signifies that on an average each sitting judge of the high courts in India needs to dispose 6908 cases to reduce pendency to zero, provided no more cases are filed. It also means that the judges in some of the high courts are insanely overburdened. In the interest of justice, urgent appointments are required so that the case load may be shared. Hence, the number of pending cases per judge is huge and the numbers are so high that it would not be unfair to state that they are simply beyond the capacity of the current number of working judges.

\begin{figure}[t]
\includegraphics[width=9.0cm]{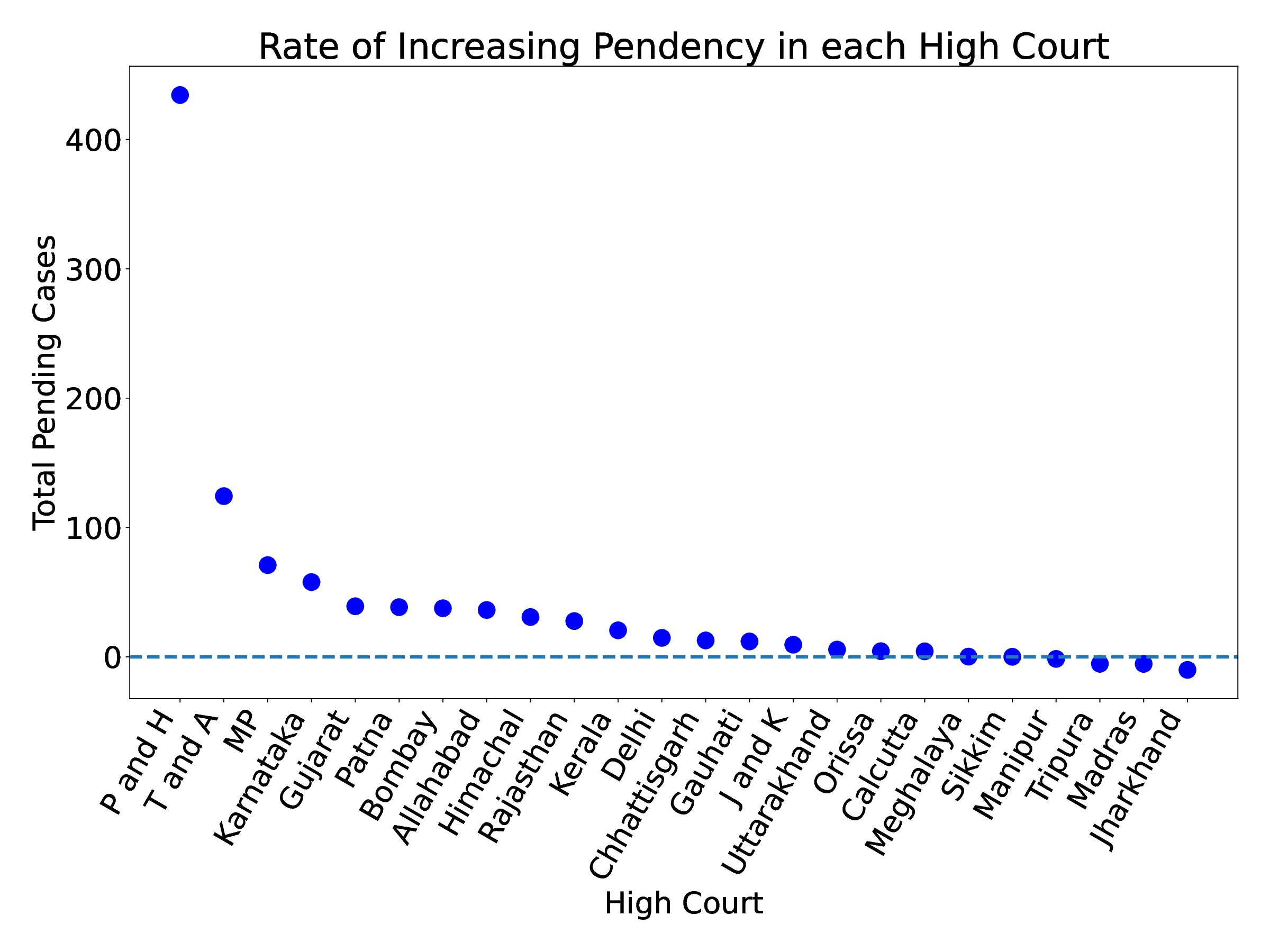}
\caption{Average rate of daily increase in the number of pending cases for each high court.}
\label{fig:hcrate_all}
\end{figure}


\section{Estimating Time Required to Combat Pendency}
\label{sec:combat}

Having discussed the trend of pending cases and the number of cases per judges in the high courts, we turn our attention to computing the time required to clear the pendency of the cases in high courts of India. Our attempt is the first -- to the best of our knowledge -- to be based on extremely rich statistical data to answer the question, ``How long will it take to reduce the pendency in the high courts to zero?".

\subsection{Rate of increase of pendency}
\fref{fig:hcpendency} presents increase in the number of total pending cases from August 31, 2017 to March 22, 2020. 


%

We observe that the number of pending cases in the high courts in India is increasing at a rate of approximately 1135 cases per day. It is basically the slope of the best fit line in \fref{fig:hcpendency}. We have plotted a similar best fit line for all the high courts in \fref{fig:np_hc1} and \fref{fig:np_hc2_p2}. The slope for various high courts is taken as rate of increase of pendency. For most of the high courts it means that the pendency will never get over rather increase with time. 

\fref{fig:hcrate_all} shows the rate of increasing pendency for each high court as computed from the slopes of the best fit lines in \fref{fig:np_hc1} and \fref{fig:np_hc2_p2}. It is quite expected that Rajasthan High Court, whose ratio of pending cases to judges is very high, has the highest rate of increase of pendency. A similar observation may be made for several other high courts whose ratio of pending cases to judges is very high.

\subsection{Towards Computing Time to Combat Pendency}

\begin{figure}[t]
  \includegraphics[width=8cm]{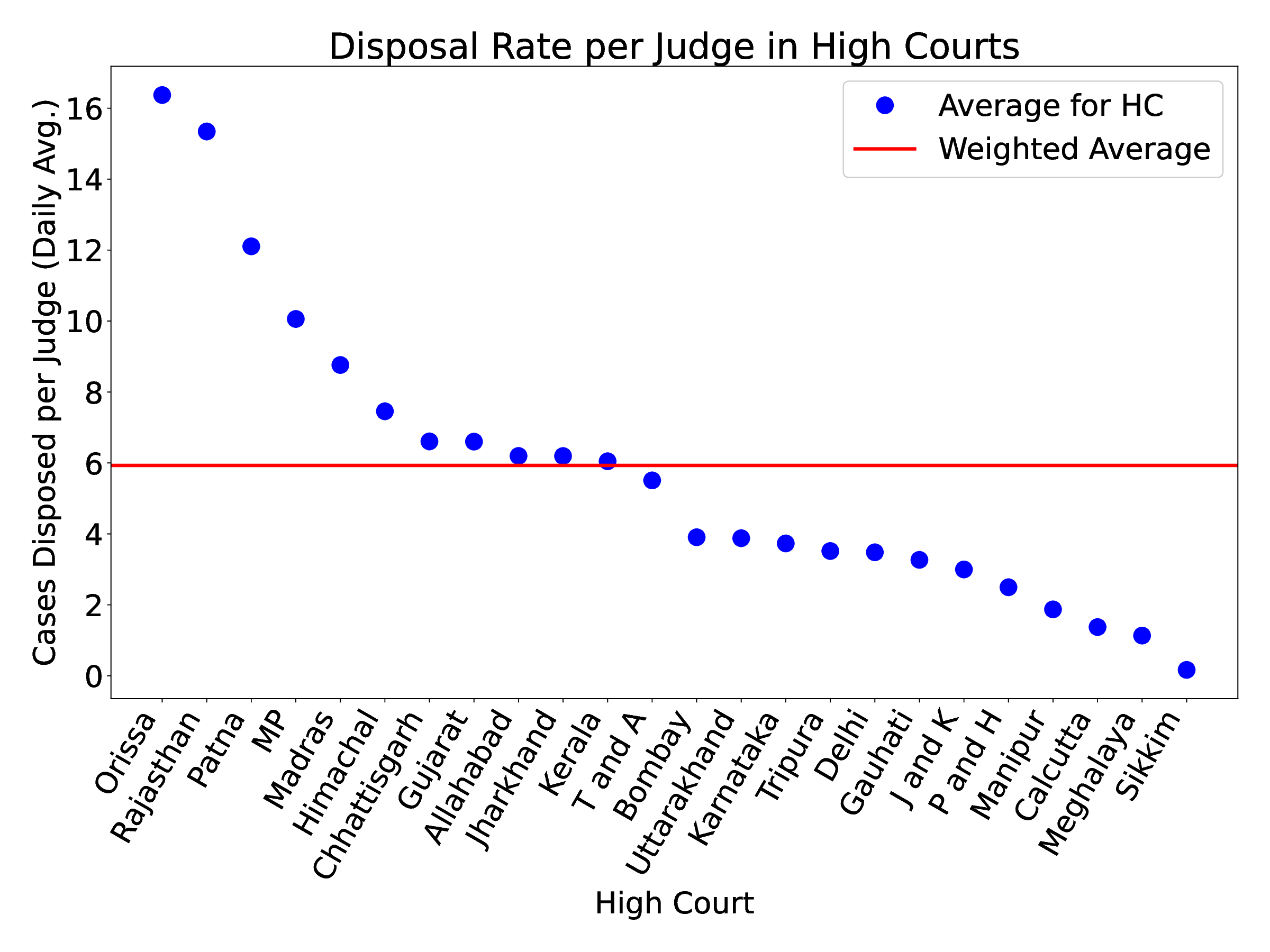}  
  \caption{Average case disposed per day per judge in High Courts. The average number of cases disposed across all the high courts is 5.93, i.e., around 6. } 
  \label{fig:avg_disp_judge}  
\end{figure}
\begin{figure}[t]
\includegraphics[width=9.0cm]{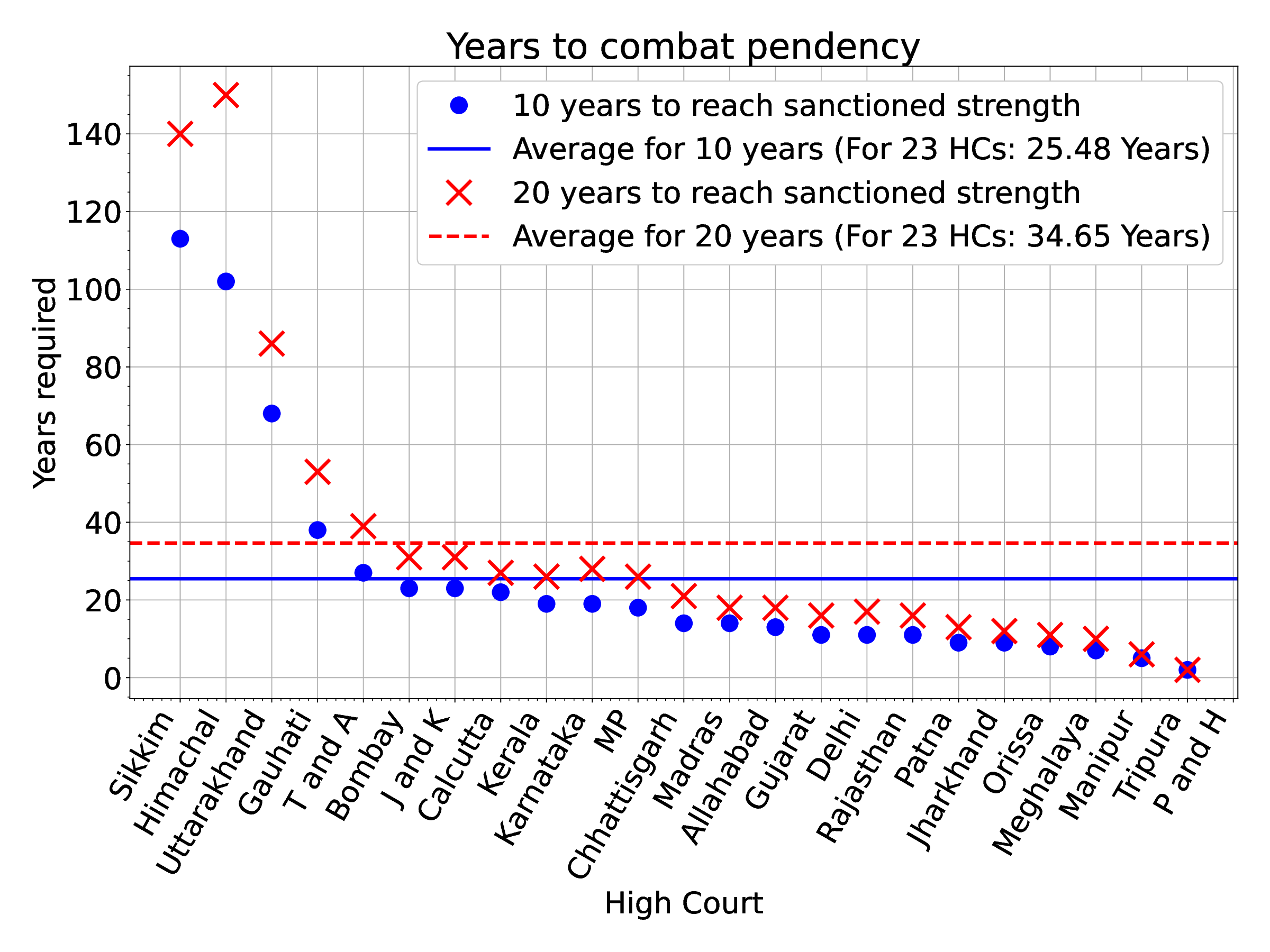}
\caption{Years required to nullify pendency if the working strength of each high court is assumed to reach its sanctioned strength in ten and twenty years. The rate of disposal of cases per day per judge for each high court is taken as reported in \fref{fig:avg_disp_judge}.}
\label{fig:years_combat_real}
\end{figure}

We use our analysis of NJDG data to find out answers to the following questions:
\begin{enumerate}
\item What is the rate of disposal of cases per day per judge in high courts? (\fref{fig:avg_disp_judge})
\item If the number of judges in high courts increase linearly and reach their sanctioned strength in ten or twenty years from now, and the average disposal rate used for a judge is as provided in \fref{fig:avg_disp_judge}, then how many years are required to reduce the pendency of cases to zero? (\fref{fig:years_combat_real})
\end{enumerate}





Disposal related statistics are provided on NJDG portal on a monthly basis. Thus, we have divided the number by 30 to get the daily figure. In \fref{fig:avg_disp_judge}, we plot the number of cases disposed per judge per day for each high court. The national average is 5.93. This figure provides the average number of cases disposed by each high court judge in a day. We use these results to estimate the time required to nullify the pendency in different high courts in India.

We have enough information to compute the time required to nullify the pendency in high courts. We are assuming that the number of judges increase linearly every year. We define the following variables:
\begin{enumerate}
\item Assumed to be constant, disposal rate per judge per year, denoted as $d$, of a high court (extrapolated from \fref{fig:avg_disp_judge}),
\item Pendency $p_t$ at the start of any given year $t$ in a high court,
\item Working strength $w_t$ of a high court during any given year $t$,
\item Yearly rate of increase ($r_t$) of pendency for a high court when average working strength is $w_t$ (extrapolated from \fref{fig:hcrate_all}),
\end{enumerate}
Then the following holds:
\begin{equation}
p_t = p_{t-1} + r_{t-1}
\label{eq:1}
\end{equation}
\begin{equation}
r_t = r_{t-1} - d\cdot(w_t-w_{t-1})
\label{eq:2}
\end{equation}

where $p_0$ and $r_0$ are taken as the values of pendency on the last day of our data collection and from \fref{fig:hcrate_all} respectively. The later values are updated according to Eq. \ref{eq:1} and \ref{eq:2} to compute $t$ for which $p_t\leq 0$. 

\begin{figure}[t]
\includegraphics[width=9.0cm]{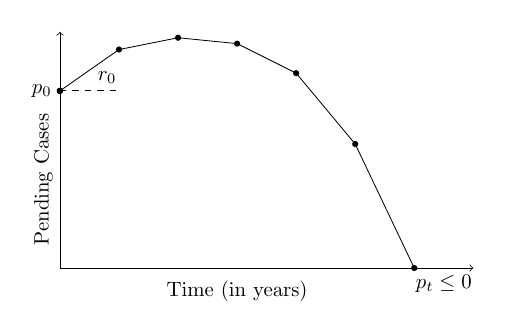}
\caption{A visual representation of how eventually the pendency may be reduced to zero with the help of Eq \ref{eq:1} and \ref{eq:2}. Initially, the rate of increasing pendency is high. As we increase the number of judges, the efficiency increases and the rate of increasing pendency decreases, soon to become negative. }
\label{fig:tikz}
\end{figure}

\fref{fig:tikz} shows how the pendency may be decreasing. If we begin from $p_0$ at the rate $r_0$, then we reach at pendency $p_1$ at the end of the first year. Since the number of judges will increase at the end of the year, more number of cases will be disposed and the rate of increase of the cases will be lesser than the previous rate. This will continue until the rate of increase becomes zero and eventually becomes negative making pendency to hit 0 at some point in time. 

\fref{fig:years_combat_real} shows the number of years required to nullify pendency in the high courts. It presents results assuming that the sanctioned strength of high courts are reached in ten years and twenty years respectively. We also assume that the rate of increase of the judges in both the cases is linear. We can see that there is a huge gap between the years taken to clear the pendency in the two cases. If we assume that the vacancy of judges in the high courts is to be filled in twenty years, then Himachal High Court and Madras High Court may take 150 and 113 years respectively. However, if we assume that the working strength of the high courts reach their sanctioned strength in ten years then the numbers for both the high courts mentioned above are 102 and 83 respectively, i.e., an improvement of 48 and 30 years respectively. On the other side of the spectrum we see Tripura High Court and Sikkim High Court that will take 2 years and 6 years respectively, irrespective of whether it takes ten or twenty years to fill the vacancy in these high courts. Thanks to the number of the pending cases, rate of decrease of pendency and the sufficient number of judges to handle that. Another extreme case is the Punjab and Haryana High Court. There is no plot against that high court in either case because the sanctioned strength, no matter whether reached in ten or twenty years, the rate of increase of pendency will still be positive rather than negative. We also see that the majority of the high courts will take more than twenty years if the sanctioned strength is reached in twenty years and more than 14 years if the sanctioned strength is reached in ten years. Hence, if only ten years are taken to fill the vacancy in high courts then substantially lesser number of years are required to clear the pendency. This is also reflected in the average number of years taken. For ten years to fill vacancy, on an average, it will take 25.3 years and for twenty years to fill vacancy, on an average it will take 35.35 years to clear the pendency. Hence, filling the vacancies in the high courts is a key to clearing the pendency. More details on the numbers used to plot \fref{fig:years_combat_real} is provided in Table \ref{tab:years} in the Appendix.

\section{Policy to Improve Access to Justice}
\label{sec:policy}
In this section, we comment on some of the required fundamental changes in various customs and enactment of laws to increase the number of judges in the high courts. In order to reduce pendency in the future, many policy level changes have been proposed by various studies. We focus on the number of judges required in the high courts. We advocate filling of the vacancy of judges in the high courts \cite{appointment_of_judges} \cite{judicial_manpower} and while filling the vacancies, age should be considered \cite{age_for_elevation}. Young judges should be elevated so that the retirement rate of the judges in a high court decrease. We:

\begin{figure*}[ht!]
\begin{subfigure}{.33\textwidth}
  \centering
  \includegraphics[width=5.8cm]{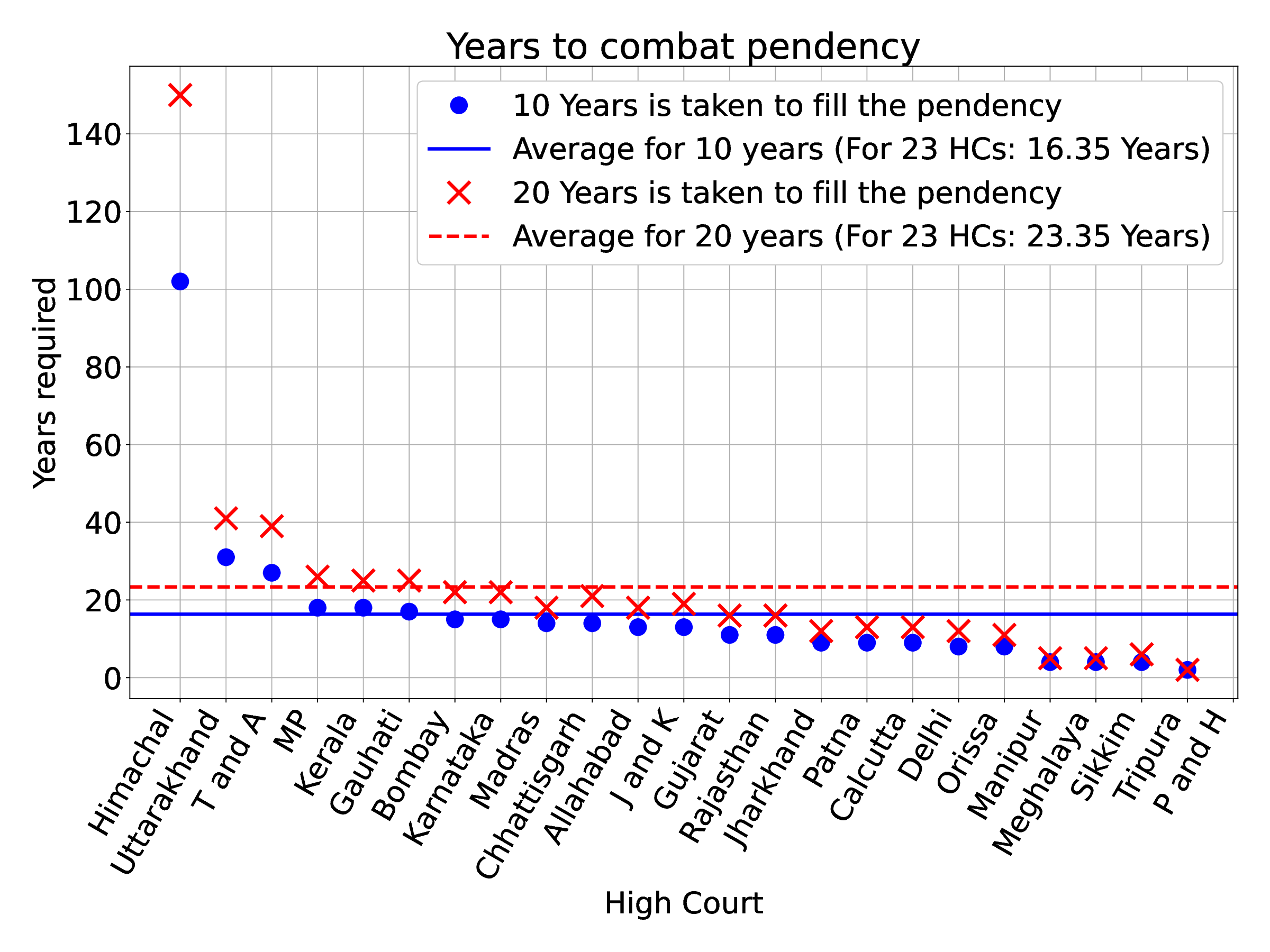}  
\caption{Years required if the minimum rate of disposal of cases per day per judge is fixed at a minimum of 5.93 in each high court.}
\label{fig:years_combat_national_avg_lowest}
\end{subfigure}
\begin{subfigure}{.33\textwidth}
  \centering
  \includegraphics[width=5.8cm]{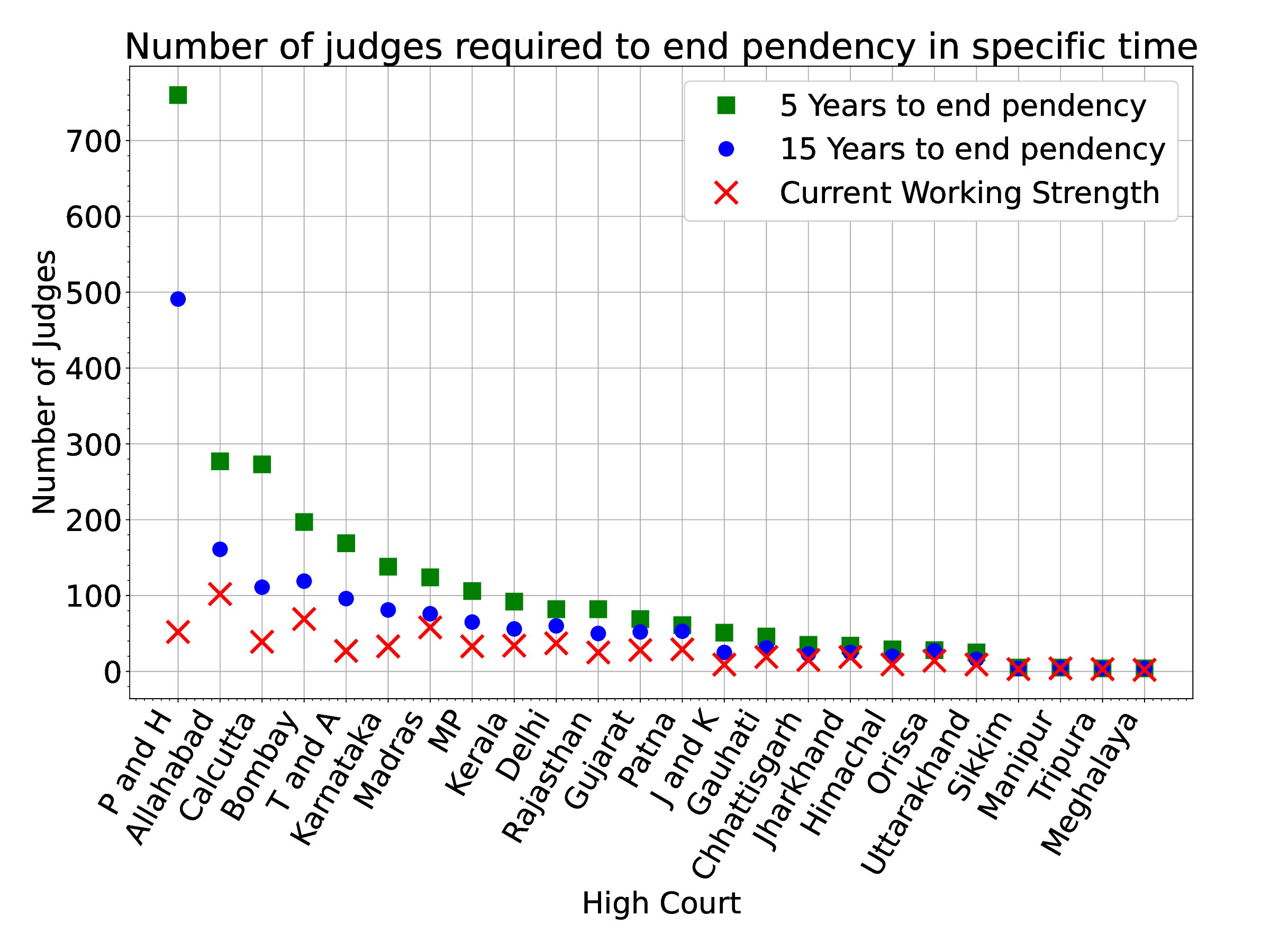}
\caption{Number of judges required in each high court to clear the pending cases in five and fifteen years respectively.}
\label{fig:req_individual}
\end{subfigure}
\begin{subfigure}{.33\textwidth}
  \centering
  \includegraphics[width=5.8cm]{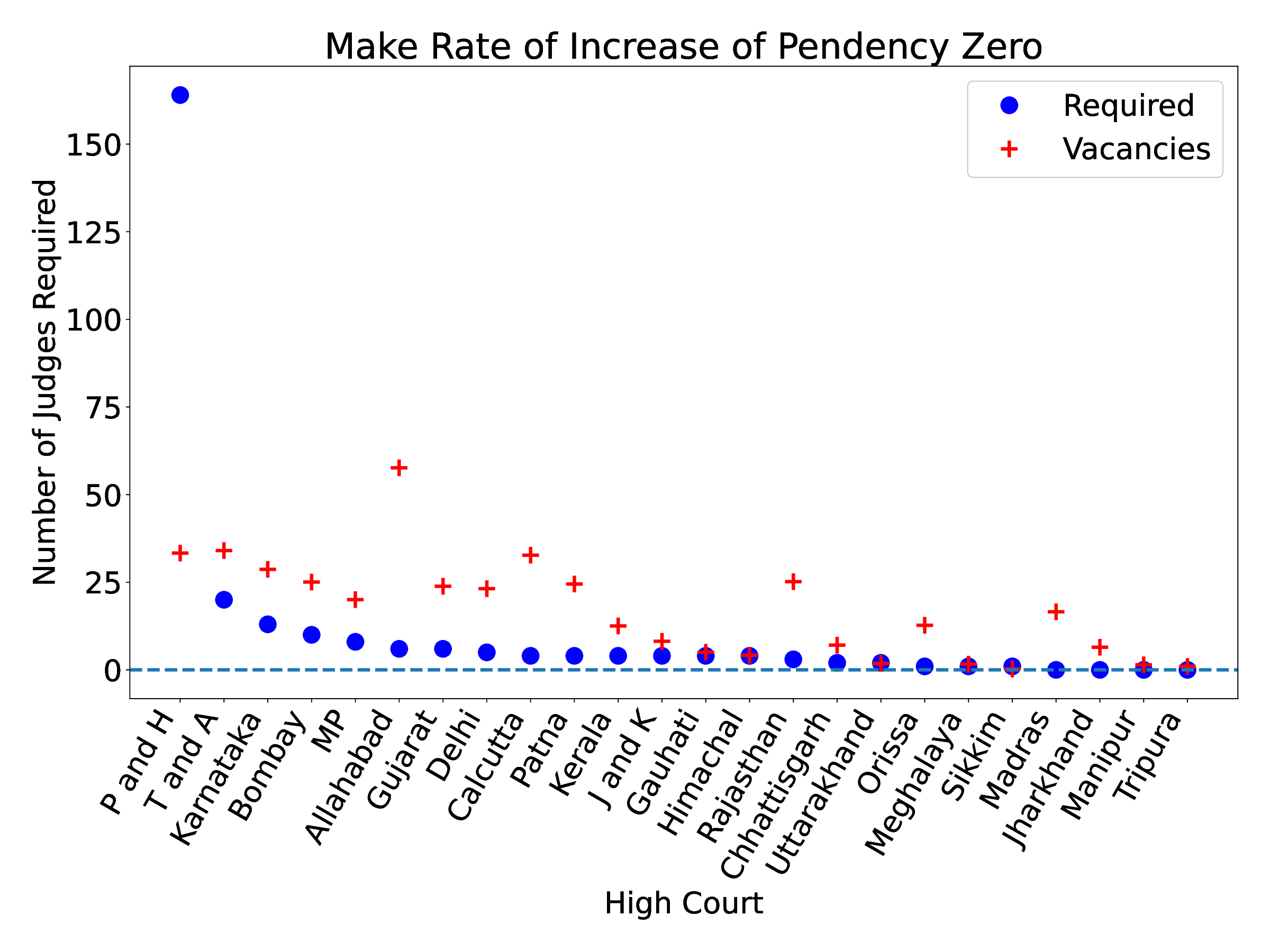}  
\caption{Number of judges required to make the rate of increase of pendency to zero, i.e., the pendency neither increases nor decreases.}
\label{fig:increase_judges}
\end{subfigure}
\caption{Some insights to drive policies to improve access to justice in the high courts in India.}
\label{fig:fig}
\end{figure*}

\begin{enumerate}
\item argue the impact of increasing the efficiency of judges on pendency in high courts \fref{fig:years_combat_national_avg_lowest}. This may be increased by providing more staff and ICT infrastructure.
\item reason for the proposed sanctioned number of judges in high courts depending on the targets set to reduce pendency to zero.
\end{enumerate}

\fref{fig:years_combat_national_avg_lowest} presents the number of years to nullify pendency if the disposal rate of those high courts is increased to 5.93 for which it is lesser than that. In other words, we hypothetically increase the number of cases disposed per judge per day to 5.93, if it is lesser than that, unchanged otherwise. This figure represents the number of years corresponding to a very ambitious case in which we assume the minimum disposal of cases per high court judge per day. We see that there is a significant improvement in the average number of years required to nullify the pendency. The average has come down from 25.3 years to 20.61 years if it takes ten years to reach to the sanctioned strength and from 35.35 years to 29.48 years if it takes twenty years to reach the sanctioned strength. In this work, we have not tried to estimate the optimal number of cases a judge may dispose on an average in a day, without compromising on the quality of justice. However, if we are given such a number then we are in a position to deduce its impact on the pendency of cases in the high courts.

\fref{fig:req_individual} shows the number of judges required if the pendency in high courts is to be nullified in five or fifteen years. We also assume that the number of judges as well as the pendency increase linearly in high courts and that the proposed sanctioned strength reaches in five or fifteen years. To provide a comparison, we have also plotted the current working strength of judges in the high court. Thus, we insist that the only way to substantially reduce the pendency in the high courts is to increase the number of judges. All the other factors like introduction of technology, etc will have a role to play but the scarcity of the judges and supporting staff is the primary reason for pendency. The numbers are very high compared to the current working strength of the high courts and it is in sharp contrast with the fact that the number of judges have not changed much during the data collection period. The rate of appointment of judges is roughly canceled by the rate of retirement, leaving the average number of judges unchanged in the high courts. Elevation of younger judges may help solve this problem. Since the number of judges required is very high compared to the current working strength of judges, the whole purpose of this graph is to provide an estimate on how aggressively the judges should be elevated to the high courts. However, such large number of judges may not be required once the pendency is cleared. For completeness, we have also provided the details about the numbers in this figure in Table \ref{tab:time} in the Appendix.

\fref{fig:increase_judges} provides an insight on the number of judges required if the rate of increase of pendency is to be made zero, i.e., the pending number of cases should neither increase nor decrease. This provides a good sign for most of the high courts as the number of judges required to make the rate of increase equal to zero is less than the vacancy in that particular high court according to the current sanctioned strength. Note that only Punjab and Haryana High Court has the required number greater than the vacancy. This means that once the pendency is taken care of, the current sanctioned strength is capable of handling the volume of fresh cases that are instituted in the high courts. A similar finding is also reported in \cite{jud_history}.

From the above discussion, we can see that the the government will have to be a bit innovative to be able to aggressively clear the pendency. So without increasing the sanctioned strength of the permanent judges in the high courts, government may, through appropriate legislation, increase the number of judges in high courts by elevating judges purely for clearing the pendency. Once pendency is cleared, the current sanctioned strength of the high courts is sufficient to take care of the newly instituted cases.

A due analysis of the cost and the infrastructure required has to be done which is beyond the scope of the current paper.


\section{Conclusion}
\label{sec:conclusion}

The problem of pending cases in India has taken an unimaginable form. In high courts alone, close to 4.5 million cases are pending of which around 20\% are pending for more than 10 years. We use the data collected from HC-NJDG portal for a period of more than two and a half years to study the trends in the pendency in the high courts. We realize that the pending cases are increasing for almost every high court. We use linear regression to capture the rate of increase of pendency in the high courts. We also make use of the data on disposed cases from the HC-NJDG portal to compute the number of cases disposed by each high court judge per day and use these statistics to estimate the number of years required to clear the pendency in the high courts. The number of pending cases is well beyond the capacity of the number of judges currently working in the high courts. Hence, the number of judges should be increased by taking necessary legislative measures.

The energy and efforts put in e-Courts project, NJDG in particular, must continue for few more years, if not decades, to see the real impact. Hence, more aid of ICT in judiciary must be sought to reduce the pendency of millions of cases and the use of artificial intelligence should be more than just welcome.

\bibliographystyle{IEEEtran}
\bibliography{/home/kshitizv/Dropbox/law/LVI/2017/legal_informatics}


\section{Appendix}
We present \fref{fig:years_combat_real} and \fref{fig:req_individual} in the form of tables. 
\begin{table}[h!]
\centering
\footnotesize
    \begin{tabular}{ | l | c | c | c | c | c |}
    \hline
    & High Court & 10 Years & 20 Years \\ \hline
    1& Himachal & 102  & 150 \\ \hline
    2& Madras & 83 & 113 \\ \hline
    3& Uttarakhand & 64 & 81 \\ \hline
    4& Rajasthan & 39 & 59 \\ \hline
    5& Gauhati & 38 & 53 \\ \hline
    6& T and A & 27 & 39 \\ \hline
    7& J and K &24 &32 \\ \hline
    8& Bombay &23 &31 \\ \hline
    9& Calcutta &22 &27 \\ \hline
    10& Kerala &19 &26 \\ \hline
    11& Karnataka &19 &28 \\ \hline
    12& MP &18 &26 \\ \hline
    13& Chhattisgarh &14 & 21  \\ \hline
    14& Manipur & 14 &22  \\ \hline
    15& Allahabad &13 &18 \\ \hline
    16& Delhi &11 &17 \\ \hline
    17& Gujarat &10 &14 \\ \hline
    18& Orissa &9 &14 \\ \hline
    19& Patna &9 &13 \\ \hline
    20& Jharkhand &9 &11 \\ \hline
    21& Meghalaya &7 &10 \\ \hline
    22& Sikkim &6 &6 \\ \hline
    23& Tripura &2 &2 \\ \hline
    24& P and H & - & - \\ \hline
    \end{tabular}
    \caption{Used to create \fref{fig:years_combat_real}. The number of years required to clear the backlog if the sanctioned strength of the high courts is reached in ten and twenty years. Note that for Punjab and Haryana High Court, the backlog can never be cleared if the sanctioned strength is not increased. }
    \label{tab:years}
\end{table}

\begin{table}[t]
\centering
\footnotesize
    \begin{tabular}{ | l | c | c | c | c | c |}
    \hline
    & High Court & 5 Years & 15 Years & Sanctioned & Working \\ \hline
    1& P and H & 549 & 310 & 85 & 52 \\ \hline
    2& Allahabad & 277 & 161 & 160 & 102 \\ \hline
    3& Calcutta & 273 &111 & 72 & 39 \\ \hline
    4& Bombay & 197 & 119 & 94 & 69\\ \hline
    5& T and A & 169 & 96 & 61 & 27 \\ \hline
    6& Madras & 160 & 107 & 75 & 58\\ \hline
    7& Karnataka & 138 & 81 & 62 & 33 \\ \hline
    8& Rajasthan & 124 & 80 & 50 & 25 \\ \hline
    9& MP &106 & 65 & 53 & 33\\ \hline
    10& Kerala & 92 & 56 & 47 & 34 \\ \hline
    11& Delhi & 82 & 60 & 60 & 37 \\ \hline
    12& Gujarat & 64 & 52 & 52 & 28 \\ \hline
    13& Patna & 61 & 53 & 53 & 29 \\ \hline
    14& J and K & 51 & 26 & 17 & 9 \\ \hline
    15& Gauhati & 46 & 31 & 24 & 19 \\ \hline
    16& Chhattisgarh &35 & 23 & 22 &15  \\ \hline
    17& Orissa & 33 & 27 & 27 & 14\\ \hline
    18& Jharkhand & 32 & 25 & 25 & 19 \\ \hline
    19& Himachal & 29  & 20 & 13 & 9\\ \hline
    20& Uttarakhand & 25 & 15 & 11 & 9 \\ \hline
    21& Manipur &7 & 5 & 5 & 4  \\ \hline
    22& Meghalaya &4 & 4 & 4 & 2 \\ \hline
    23& Tripura &4 & 4 & 4 & 3 \\ \hline 
    24& Sikkim &3 &3 &3 & 3\\ \hline \hline
    25& Total & 2561 & 1534 & 1079 & 672 \\ \hline
    \end{tabular}
    \caption{Used to create \fref{fig:req_individual}. The number of judges required in each high court to clear the backlog in five or fifteen years. The third column is the current sanctioned strength of the high courts and the last column shows the average number of working judges in the high courts for 22 months starting June 2018 to March 2020.}
    \label{tab:time}
\end{table}

\end{document}